\documentclass{aa}  
\usepackage{amsmath, epsfig}
\usepackage{graphicx}
\usepackage{caption}
\usepackage[varg]{txfonts}
\usepackage{gensymb}
\usepackage{longtable, lscape}
\usepackage{natbib}
\usepackage{lineno}
\usepackage{xcolor}
\usepackage[countmax]{subfloat}
\usepackage{booktabs}
\usepackage{float}
\usepackage{placeins}
\usepackage{mwe}
\newcommand{\RNum}[1]{\uppercase\expandafter{\romannumeral #1\relax}}
\def\pg{\object{PG\,1553+113}}

\usepackage{hyperref} 
\hypersetup{backref=true,       
    pagebackref=true,               
    hyperindex=true,                
    colorlinks=true,                
    breaklinks=true,                
    urlcolor= black,                
    linkcolor= blue,                
    bookmarks=true,                 
    bookmarksopen=false,
    filecolor=black,
    citecolor=blue,
    linkbordercolor=blue}

\begin{document}

\title{Multi-band behaviour of the TeV blazar \pg\ in optical range on diverse timescales}

\subtitle{Flux and spectral variations\thanks{Calibrated $BVRI$ light curves of \pg\ are only available in electronic form
at the CDS via anonymous ftp to cdsarc.u-strasbg.fr (130.79.128.5)
or via http://cdsweb.u-strasbg.fr/cgi-bin/qcat?J/A+A/}}
   	
\author{A. Agarwal\inst{1,2}
          \and
        B. Mihov\inst{3}\fnmsep\thanks{Corresponding author.}
          \and
        I. Andruchow\inst{4,5}
          \and
        Sergio A. Cellone\inst{5,6}
          \and
        G. C. Anupama\inst{2}
          \and
        V. Agrawal\inst{7}
          \and
        S. Zola\inst{8,9}
          \and
        L. Slavcheva-Mihova\inst{3}
          \and
        Aykut \"Ozd\"onmez\inst{11}
          \and
        Erg\"un Ege\inst{12}
           \and
        Ashish Raj\inst{10}
          \and
        Luis Mammana\inst{5,6}
          \and
        L. Zibecchi\inst{4,5}
          \and
        E. Fern\'andez-Laj\'us\inst{4,5}
        }
          
\authorrunning{Agarwal et al.}
\titlerunning{Multi-band behaviour of \pg\ in optical range}

\institute{Raman Research Institute, Vyalikaval, Lower
            Palace Orchards, Bengaluru, India 560003\\
            \email{aditi.agarwal@rri.res.in}
         \and
            Indian Institute of Astrophysics, Block II,  Koramangala, Bangalore, India, 560034
       	 \and
       	    Institute of Astronomy and NAO, Bulgarian
       		Academy of Sciences, 72 Tsarigradsko Chaussee 
       		Blvd., 1784 Sofia, Bulgaria\\
       		\email{bmihov@astro.bas.bg}
         \and
            Instituto de Astrof\'isica de La Plata (CCT La Plata-CONICET-UNLP), La Plata, Argentina
       	 \and
            Facultad de Ciencias Astron\'omicas y Geof\'isicas, Universidad Nacional de La Plata, Paseo del Bosque, B1900FWA, La Plata, Argentina
        \and
            Complejo Astron\'omico ``El Leoncito'' (CASLEO), CONICET-UNLP-UNC-UNSJ, San Juan, Argentina
        \and    
            Embibe, Diamond District, Old Airport Road, Bengaluru - 56008
        \and    
            Astronomical Observatory, Jagiellonian University, ul. Orla 171, 30-244 Krakow, Poland
        \and
            Mt. Suhora Observatory, Pedagogical University, ul. Podchorazych 2, 30-084 Krakow, Poland
        \and
            Dept. of Physics \& Astrophysics, University Road, University Enclave, Delhi, India, 110007
        \and    
            Ataturk University, Faculty of Science,  Department of Astronomy and Space Science, Yakutiye, Erzurum
        \and
            Istanbul University, Faculty of Science, Department of Astronomy and Space Sciences, 34116, Beyazıt, Istanbul, Turkey
            }

\date{Received xxx; accepted xxx}
 
\abstract
{The TeV BL Lac object \pg\ is one of the primary candidates for a binary supermassive black hole system.}
{We study the flux and spectral variability of \pg\ on intra-night to long-term timescales using (i) $BVRI$ data collected over 76 nights from January 2016 to August 2019 involving nine optical telescopes and (ii) historical $VR$ data (including ours) obtained for the period from 2005 to 2019.}
{We analysed the light curves using various statistical tests, fitting and cross-correlation techniques, and methods for the search for periodicity. We examined the colour-magnitude diagrams before and after the corresponding light curves were corrected for the long-term variations.}
{Our intra-night monitoring, supplemented with literature data, result in a low duty cycle of $\sim$(10--18)\%. In April 2019, we recorded a flare, which marks the brightest state of \pg\ for the period from 2005 to 2019: $R \simeq 13.2\,\rm mag$. 
This flare is found to show a clockwise spectral hysteresis loop on its $VR$ colour-magnitude diagram and a time lag in the sense that the $V$-band variations lead the $R$-band ones. We obtain estimates of the radius, the magnetic field strength, and the electron energy that characterize the emission region related to the flare. We find a median period of $(2.21 \pm 0.04)$\,years using the historical light curves. In addition, we detect a secondary period of about 210\,days using the historical light curves corrected for the long-term variations. We briefly discuss the possible origin of this period.}
{}

   
\maketitle

\section{Introduction}
\label{sect:intro}

Blazars are a sub-class of active galactic nuclei, which are characterized by intense non-thermal radiation dominating their spectral energy distribution (SED) from radio to very high energy (VHE) passbands.
This radiation is strongly beamed, originating in a relativistic jet pointing towards the Earth at a small angle with the line of sight.
From their optical spectra, blazars have been classified into two categories, viz. BL\,Lac objects (having featureless optical spectra) and flat spectrum radio quasars (FSRQ, with broad emission lines in their optical spectra).  

Blazars have SEDs showing two peaks: one at low energies (from radio to optical and X-rays) due to synchrotron emission and another one at high energies (X- to $\gamma$-rays) due to inverse-Compton scattering of soft photons off the same relativistic electrons, which are responsible for the synchrotron radiation \citep{1981ApJ...243..700K}.

The frequency at which the synchrotron emission peaks is used for further blazar classification. \citet{1995ApJ...444..567P} classified BL\,Lac objects into low-energy peaked BL\,Lacs (LBLs) and high-energy peaked BL\,Lacs (HBLs). This classification was then generalized by \citet{2010ApJ...716...30A} to include FSRQs: classes become low, intermediate, and high synchrotron peaked blazars. In particular, the low synchrotron peaked blazar class includes LBLs and virtually all FSRQs \citep{2010ApJ...716...30A}.

Blazars show variability in all passbands, from radio to VHE $\gamma$-rays, as well as strong radio and optical polarization and superluminal motions.
Since the relativistic jet is supposed to dominate the SED the strong variability could either be caused by internal processes within the jet itself or by accretion disk instabilities triggering changes in it. At radio frequencies the interstellar scintillation may also play a role \citep[e.g.][]{1987AJ.....94.1493H}. 

Variability in blazars can be detected on diverse timescales.
We can broadly divide blazar variability into three temporal classes, viz. intra-night variability (INV, or intra-day variability, or
microvariability) on timescales from minutes to hours \citep[e.g.][]{1995ARA&A..33..163W, 1975IAUS...67..573K, 2003AJ....126...37C, 2003AJ....126...47R, 2004MNRAS.348..831X}, short-term variability (STV) on timescales from several days to a few months, and long-term variability \citep[LTV, e.g.][]{2008AJ....136.2359G,
2017MNRAS.469..813A} on timescales
from several months to many years; in the latter two classes the amplitude of the optical variability can often exceed even 5\,mag \citep[e.g.][]{2018BlgAJ..28...22B}.
Variability studies are a powerful tool to probe the nature of emission
processes occurring in blazars, their magnetic field geometry, dominant particle
acceleration mechanisms, etc.

Along with flux variability studies, analyses of colour trends accompanying brightness changes have begun to be implemented in the last decade through the colour-magnitude diagrams (CMDs). 
Three types of CMD behaviour could be discerned: redder-when-brighter (RWB), bluer-when-brighter (BWB), and achromatic. The RWB chromatism is most frequently associated with FSRQs, for which the contribution of the accretion disk to the total emission could be significant. 
The BWB behaviour is thought to be related to processes associated to the relativistic jet, such as particle acceleration and cooling in the framework of the shock-in-jet model \citep[e.g.][]{1985ApJ...298..114M,1998A&A...333..452K}.
Alternatively, the BWB chromatism could arise from a Doppler factor variation on a convex spectrum \citep{2004A&A...421..103V,2007A&A...470..857P}. Finally, the achromatic behaviour is most frequently interpreted as being due to the variations of the Doppler factor, which are most likely explained in the framework of the geometric scenario \citep[e.g.][]{2002A&A...390..407V}.

\pg\ was discovered by the Palomar-Green survey of UV-excess stellar
objects \citep{1986ApJS...61..305G} and was classified as a BL\,Lac object
due to its featureless optical spectrum \citep{1983BAAS...15..957M} and significant optical variability \citep{1988ESASP.281b.303M}. Furthermore, it was classified as an HBL, as its synchrotron emission peak falls in the UV and X-ray frequency range \citep{1990PASP..102.1120F} and was detected at TeV energies \citep{2006A&A...448L..19A}.
Due to its featureless optical spectrum and its host galaxy remaining undetected, the redshift of \pg\ has remained uncertain \citep[e.g.][]{2007A&A...473L..17T}.
The recent detection of its putative galaxy group would set this object's redshift at $z=0.433$ \citep{TZCA2017RMxAC, JMC2019}; we shall thus use this value throughout the paper.

\pg\ is among the few blazars whose flux variability is claimed to be periodic. The first hint for periodicity comes from \citet{2015ApJ...813L..41A}~-- they found a $(2.18 \pm 0.08)$\,years period analysing the $Fermi$/LAT $\gamma$-ray light curve (LC) of the source.
This period was further confirmed by \citet{2017MNRAS.471.3036P}, \citet{2018ApJ...854...11T}, \citet{2018A&A...615A.118S},
\citet{2020ApJ...895..122C}, and
\citet{2020ApJ...896..134P}\footnote{The authors presented the most comprehensive search for periodicities at $\gamma$-rays using ten different methods along with the corresponding techniques for significance estimation of the periods found.}
at $\gamma$-rays and by \citet{2016agnt.confE..58C}, \citet{2018A&A...615A.118S}, and \citet{2020ApJ...895..122C} in the $R$-band. 
On the other hand, \citet{2019MNRAS.482.1270C} and \citet{2020A&A...634A.120A} expressed some caution about the significance of the year-long period detected at $\gamma$-rays.
\citet{2018A&A...620A.185N} did not find the claimed year-long period in their $R$-band LC.
In addition, \citet{2020A&A...634A..87L} found no signs for clear periodic variability at the radio frequencies.

The caution expressed by \citet{2019MNRAS.482.1270C} and \citet{2020A&A...634A.120A} is based on the low global significance of the periods they found. \citet{2018A&A...615A.118S} also reported modest global significance of the periods for both high energy and optical LCs of \pg. 
According to the authors, however, the presence of one and the same period in both $\gamma$-ray and $R$-band LCs means that this period could be real in spite of the modest individual global significance. Finally, \citet{2020ApJ...895..122C} found that the addition of a periodic signal to the modelling of the \pg\ LCs improves their statistical description; the modelling itself is done by means of Gaussian processes.

In this paper, we present the results from intensive intra-night monitoring (INM) of \pg\ along with the analysis of its historical $VR$-band LCs.
The paper is organized as follows.
We describe the observations and data reduction techniques in Sect.\,\ref{sect:obs}.
In Sect.\,\ref{sect:res} 
we report the results of flux and colour variability of the source from intra-night to longer timescales, while,
finally, Sect.\,\ref{sect:disc} presents a discussion of our results followed by the conclusions.

\begin{table*}
\caption{Details about the telescopes and instruments used.}
\label{tab:telescopes} 
\centering
\label{tab:telFGI}
\resizebox{1\textwidth}{!}{
\begin{tabular}{ccccccc} \hline\hline\noalign{\smallskip}
Telescope & C1 & C2 & F & G & H & I \\
 & 2.0\,m RC & 2.0\,m RC & 1.3\,m RC & 1.0\,m RC & 60\,cm RC & 50\,cm Cassegrain \\ \noalign{\smallskip}
\hline\noalign{\smallskip}
CCD model         & VersArray:1300B & iKon-L & 2k $\times $ 4k UKATC      &SI 1100 Cryo & Andor iKon-L  & Apogee Alta     \\
 & & & &  UV, AR, BI & 936 BEX2-DD & U42 \\

Chip size         & $1340 \times 1300$ px & $2048 \times 2048$ px & $2048 \times 4096$ px  & $4096 \times 4096$ px & $2048 \times 2048$ px  & $2048 \times 2048$ px \\
Pixel size        & $20 \times 20\,\mu\rm m$ & $13.5 \times 13.5\,\mu\rm m$ & $13\times 13\,\mu$m         & $15\times 15\,\mu$m   & $13.5\times 13.5\,\mu$m      & $13 \times 13\,\mu$m \\
Scale             & $ 0\farcs258$ px$^{-1}$ &$ 0\farcs497$ px$^{-1}$ & $0\farcs30$ px$^{-1}$        & $0\farcs31$ px$^{-1}$    & $0\farcs456$ px$^{-1}$    &$0\farcs83$ px$^{-1}$  \\
Field             & $5\farcm8\times5\farcm6$ & $17\arcmin\times17\arcmin$  & $10\arcmin\times20\arcmin$ &$21\farcm5\times21\farcm5$ &$15\farcm6\times15\farcm6$ &$14\farcm2\times14\farcm2$ \\
Gain              & 1.0\,$\rm e^-$ ADU$^{-1}$ & 1.0/1.1\,$\rm e^-$ ADU$^{-1}$ & 0.74\,$\rm e^-$ ADU$^{-1}$             & 0.57\,$\rm e^-$ ADU$^{-1}$   & 1.1\,$\rm e^-$ ADU$^{-1}$    & 1.28\,$\rm e^-$ ADU$^{-1}$  \\
Read Out Noise    & 2.0\,$\rm e^-$ rms  & 6.7/6.9\,$\rm e^-$ rms & 4.2\,$\rm e^-$ rms           & 4.19\,$\rm e^-$ rms & 6.9\,$\rm e^-$ rms    & 8.3\,$\rm e^-$ rms    \\
Binning used      & 1$\times$1   & 1$\times$1 & 1$\times$1           &  2$\times$2    & 1$\times$1    &  2$\times$2   \\
Typical seeing    & 1\farcs5 to 2\farcs5 & 1\farcs5 to 2\farcs5 & 1$\arcsec$ to 3$\arcsec$  & 1$\arcsec$ to 3$\arcsec$ & 1\farcs5 to 3$\arcsec$ & 2$\arcsec$ to 3$\arcsec$ \\
\hline
\end{tabular}}
\tablefoot{Configuration C1: 2.0\,m RC telescope at Rozhen NAO, Bulgaria. Configuration C2: 2.0\,m RC telescope + FoReRo-2 focal reducer at Rozhen NAO. The gain and read-out noise
are reported for the blue and red channels, respectively.
Telescope F: 1.3\,m JC Bhattacharya telescope (JCBT), Kavalur, India.
Telescope G: 1.0\,m RC telescope,  Turkey.
Telescope H:  60\,cm RC robotic telescope, Turkey.
Telescope I: 50\,cm Cassegrain telescope, Krakow, Poland.}
\end{table*}

\begin{table*}
\caption{Log of photometric observations for the blazar \pg.}
\label{tab:obs_log} 
\centering
\begin{tabular}{ccrrrrccrrrr} 
\hline\hline\noalign{\smallskip}
Date of & Telescope & \multicolumn{4}{r}{Number of data points} & Date of & Telescope & \multicolumn{4}{r}{Number of data points} \\
observation & & & & & & observation & & & & & \\
(yyyy mm dd) & & $B$ & $V$ & $R$ & $I$ & (yyyy mm dd) & & $B$ & $V$ & $R$ & $I$ \\
\noalign{\smallskip}
\hline\noalign{\smallskip}
2016 01 12    & D  & 3 & 0 & 3 & 3  & 2018 09 10    & A  &2 & 0 & 2 & 0  \\
2016 05 28    & C1 & 2 & 2 & 2 & 1  & 2019 03 01    & A  &2 &12 &12 & 2  \\
2016 05 29    & C1 & 3 & 2 & 2 & 1  & 2019 03 02    & A  &1 & 4 & 1 & 1  \\
2016 06 10    & D  & 3 & 0 & 3 & 3  & 2019 03 11    & F  &1 & 1 & 1 & 1  \\
2016 06 11    & D  & 3 & 3 & 3 & 3  & 2019 03 12    & F  &1 & 1 &83 & 1  \\
2016 07 02    & D  & 2 & 2 & 2 & 2  & 2019 03 13    & F  &1 & 1 &84 & 1  \\
2016 07 03    & D  & 2 & 2 & 2 & 2  & 2019 04 01    & B  &2 & 4 & 4 & 4  \\
2016 07 03    & C1 & 2 & 3 & 5 & 3  & 2019 04 01    & A  &2 & 2 &110& 2  \\
2016 07 06    & C1 & 2 & 2 & 2 & 3  & 2019 04 04    & D  &35&35 &36 &35  \\
2016 07 07    & C  & 0 & 2 & 2 & 0  & 2019 04 05    & F  &1 & 1 & 1 & 1  \\
2016 07 10    & D  & 2 & 1 & 2 & 1  & 2019 04 06    & F  &1 & 1 &58 & 1  \\
2016 08 04    & C  & 0 & 0 & 2 & 0  & 2019 04 06    & A  &12&12 &12 &12  \\
2016 08 06    & C  & 0 & 2 & 2 & 0  & 2019 04 07    & F  &1 & 1 &50 & 1  \\
2016 08 11    & D  & 2 & 2 & 2 & 1  & 2019 04 07    & A  &9 & 9 & 9 & 9  \\
2017 03 23    & D  & 3 & 3 & 3 & 3  & 2019 04 08    & A  &1 & 2 &99 & 2  \\
2017 03 24    & D  & 3 & 2 & 3 & 2  & 2019 04 09    & A  &0 & 0 &60 & 0  \\
2017 03 27    & C  & 0 & 3 & 3 & 0  & 2019 04 09    & F  &0 & 0 &60 & 0  \\
2017 03 29    & C  & 0 & 3 & 3 & 0  & 2019 04 09    & E  &2 & 2 & 2 & 1  \\
2017 03 31    & D  & 3 & 3 & 2 & 2  & 2019 04 10    & A  &1 & 1 &37 & 1  \\
2017 04 24    & D  & 3 & 3 & 3 & 3  & 2019 04 10    & F  &1 & 1 &47 & 1  \\
2017 04 25    & D  & 2 & 3 & 3 & 3  & 2019 04 11    & A  &1 & 2 &76 & 2  \\
2017 06 22    & D  & 3 & 3 & 3 & 3  & 2019 04 12    & F  &2 & 2 & 2 & 2  \\
2017 06 23    & D  & 2 & 2 & 2 & 2  & 2019 04 14    & F  &2 & 2 & 2 & 1  \\
2018 03 09    & A  & 0 & 0 & 2 & 0  & 2019 04 25    & B  &1 & 1 & 4 & 1  \\
2018 03 10    & A  & 0 & 9 & 9 & 9  & 2019 06 21    & E  &1 & 1 &346& 1  \\
2018 05 07    & I  &17 &17 &17 &16  & 2019 07 06    & C2 &3 & 3 & 3 & 3  \\
2018 05 10    & I  &61 &61 &61 &61  & 2019 07 07    & C2 &3 & 3 & 2 & 3  \\
2018 05 12    & I  & 0 &67 &67 &67  & 2019 07 08    & C2 &3 & 3 & 3 & 2  \\
2018 05 13    & I  & 0 & 0 &40 & 0  & 2019 07 10    & E  &1 & 1 &121& 1  \\
2018 05 28    & I  &51 &54 &53 &53  & 2019 07 13    & G  &2 & 3 &111& 2  \\
2018 05 29    & I  &35 &32 &39 &37  & 2019 07 13    & H  &1 & 2 & 2 & 1  \\
2018 06 04    & I  &30 &34 &34 &34  & 2019 07 15    & H  &1 & 1 & 1 & 1  \\
2018 06 05    & I  &17 &17 &17 &17  & 2019 07 19    & G  &1 & 2 &55 & 2  \\
2018 07 07    & A  & 0 & 4 & 3 & 3  & 2019 07 20    & G  &2 & 2 &65 & 2  \\
2018 07 08    & A  & 0 & 4 & 5 & 5  & 2019 07 21    & A  &1 & 1 &11 & 0  \\
2018 07 14    & D  & 2 & 2 & 2 & 2  & 2019 07 21    & H  &1 & 2 & 2 & 2  \\
2018 07 15    & D  & 6 & 1 & 2 & 2  & 2019 07 22    & A  &1 & 1 &14 & 0  \\
2018 07 21    & D  &14 &18 &19 &21  & 2019 08 18    & A  &1 & 0 & 0 & 0 \\
\hline
\end{tabular}
\end{table*}

\section{Observations and data reduction}
\label{sect:obs}

The data presented here were obtained on 76 nights over a period from January 2016 to August 2019.
Observations were carried out in the optical $BVRI$-bands using nine different telescopes around the globe which are: 2.15\,m Jorge Sahade telescope (JS, telescope A) and 60\,cm Helen Sawyer Hogg telescope (HSH, telescope B), CASLEO, Argentina;
2.0\,m Ritchey-Chr\'etien (RC, telescope C) and 50/70\,cm Schmidt (telescope D) at the Rozhen National Astronomical Observatory (NAO), Bulgaria;
2.01\,m RC Himalayan Chandra Telescope (HCT, telescope E) at Indian Astronomical Observatory, Hanle, India;
1.3\,m JC Bhattacharya telescope (JCBT; telescope F) at the Vainu Bappu Observatory (VBO), India;
1.0\,m RC telescope (telescope G) and 60\,cm RC robotic telescope (telescope H) at TUBITAK National Observatory (TUG), Antalya, Turkey; 50\,cm Cassegrain telescope (telescope I) located at the Astronomical Observatory of the Jagiellonian
University in Krakow, Poland.

The main characteristics and technical details for telescopes A, B, C, D, and E can be found in Table 1 of \citet{2019MNRAS.488.4093A}.
In addition to these, for the present
work, we have used four other optical telescopes (F, G, H, I), which we describe below.

The JCBT telescope located at the VBO, Tamil Nadu, India, is equipped with a 1k $\times$ 1k proEM CCD and a 2k $\times$ 4k
regular CCD (the proEM CCD has a smaller field of view). As described later in this section, to perform differential photometry, suitable pairs of non-variable stars from the same observation frame are necessary.
Therefore, we chose the 2k $\times$ 4k CCD for our photometric observations,
whose central 2k $\times$ 2k region was used. 

The 1.0\,m RC telescope is located at the Bakirlitepe Mountain and is currently operated remotely from TUG in Antalya. The telescope is equipped with 4k $\times$ 4k CCD camera working at $-90^{\,\circ}$C.
Each observation was taken with $2\times2$ binning and consisted of several exposures through $R$ filter and 1-2 sets of $BVRI$ frames.

The 60\,cm RC robotic telescope is also located at TUG. The observations with this telescope are object-based and fully automatic. Observers are allocated a maximum of 3\,000\,sec for each day on this telescope. After observations, the bias, dark, and flat corrected frames by the TALON pipeline are sent to observers. The robotic telescope is equipped with a 2k $\times$ 2k CCD camera and with 12 standard filters. As long as the weather is suitable, the observation was taken daily with $1 \times 1$ binning and 1-2 sets of $BVRI$ frames.

In the period between May 7 and June 5, 2018, we performed
observations of \pg\ using the 50\,cm Cassegrain telescope
located at the Astronomical Observatory of the Jagiellonian
University in Krakow, Poland. The telescope is equipped with
an Apogee Alta U42 CCD and a set of broadband filters \citep{1990PASP..102.1181B}. The $BVRI$ data were taken with
$2 \times 2$ binning, with exposure times between 60\,sec and 100\,sec,
depending on the weather conditions and the filter used.
This camera accommodates an e2v, back-illuminated chip with
2048\,px of 13\,$\mu$m in size.
The calibration images were taken every night by collecting
series of bias, dark and sky flat-field frames.

Finally, the 2.0\,m telescope of the Rozhen NAO was used in two additional observing configurations compared to that reported in \citet{2019MNRAS.488.4093A}: the CCD camera VersArray:1300B attached directly to the RC focus and the two CCD cameras Andor iKon-L attached at the ports of the two-channel focal reducer FoReRo-2 \citep{2000KFNTS...3...13J,2011gfun.conf...89B}.

Additional information about the new telescopes and configurations can be found in Table\,\ref{tab:telescopes}.
The complete observations log corresponding to the data collected during our monitoring campaign is given in Table\,\ref{tab:obs_log}.

A detailed description of the data reduction procedure is given in \citet{2015MNRAS.450..541A}.
To summarize, the data reduction of all CCD frames, including bias
correction, flat-fielding, and cosmic-ray removal was performed
using IRAF\footnote{IRAF is distributed by the National Optical Astronomy Observatories, which are operated
by the Association of Universities for Research in Astronomy, Inc., under a cooperative agreement with the
National Science Foundation.} software. After correcting the
CCD frames, the Dominion Astronomical
Observatory Photometry (DAOPHOT II) software \citep{S1987PASP, S1992ASPC} was used to perform aperture photometry which gave us the instrumental magnitudes of the source and all the
stars in the CCD frames.

To build the instrumental differential LCs,
we used four steady comparison stars in the frame \citep{2015MNRAS.454..353R}. We finally selected those two stars that show smaller scatter of the star-star differential LC.
As pointed out by \citet{2007MNRAS.374..357C}, spurious INV could be detected if the
comparison stars differ much in brightness with respect to the
target. Therefore, we selected those comparison stars which had
magnitude and colour close to that of the target blazar.
The dispersion in the star-star LCs was checked for every night and various
apertures, starting from a value equal to the full width at
half maximum to four times this value. Finally, the aperture
with minimum dispersion in the star-star LCs was selected.
An aperture radius of 4\arcsec\ was thus used.
The calibration of instrumental magnitudes into the standard system was made using magnitudes listed in \citet{2015MNRAS.454..353R}.

\section{Results}
\label{sect:res}

\subsection{Intra-night variability}
\label{sect:res:inv}

The LCs derived during the 28 nights of INM are shown in Fig.\,\ref{LC_BL1}. The date of observation and the telescope used are indicated in each plot. We show only LCs having a minimum of 10 data points acquired during the particular night.

Considering the modest number of data points in a single filter, we used the $C$-test and the $F$-test, both set at a 99.5\% confidence level to assign variability status to the intra-night LCs. The tests are described in detail in \citet{2019MNRAS.488.4093A}.
Even if the comparison and control stars were appropriately selected, we applied the $\Gamma$ scaling factor \citep{1988AJ.....95..247H} to equalize the target and control error distributions, since this procedure has been shown to give the most reliable results \citep[e.g.][]{2017MNRAS.467..340Z,2020MNRAS.498.3013Z}.

After computing the parameters of each test with the original choice of stars ($C_1$, $F_1$), we then recalculated them with comparison and control stars interchanged ($C_2$, $F_2$) as a check for any issue in the LC. 
We have designated a light curve to be variable (Var) if both tests rejected the null hypothesis (i.e. non-variable LC) at a 99.5\% confidence level and non-variable (NV) if at least one of the tests failed to reject the null hypothesis.
The results from the INV tests are summarized in Table\,\ref{tab:var_res}.

\begin{figure}[t]
\centering
\includegraphics[width=\columnwidth,clip=true]{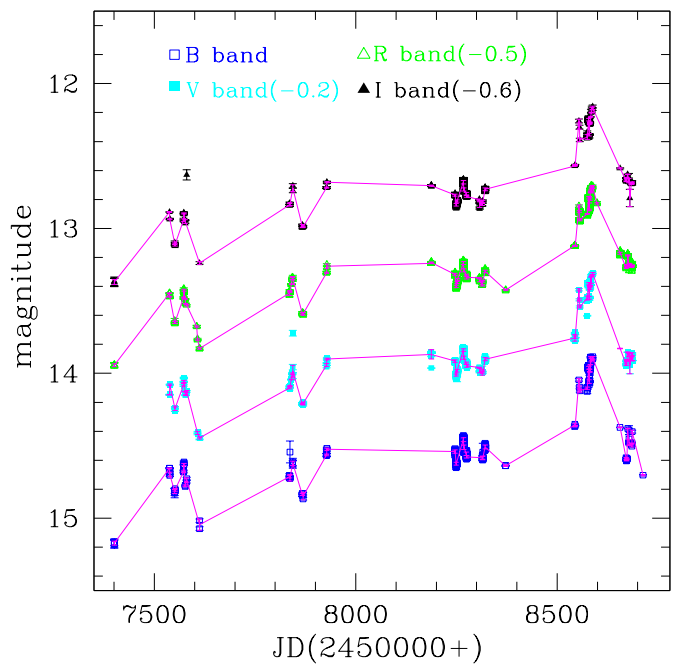}
  \caption{Long-term LC for \pg\ covering the entire monitoring duration.
  Different
colours and symbols denote data in different passbands as indicated in the plot.
In each passband the magenta solid lines connect the data points obtained after nightly binning of the respective LC.}
\label{fig:ourLC}
\end{figure}

\begin{table*}
\caption{Long-term variability characteristics.} 
\label{tab:ltv} 
\centering 
\begin{tabular}{ccccccc} 
\hline\hline  \noalign{\smallskip}
 Passband & Weighted mean & Faintest  & JD    & Brightest & JD    & $A$\tablefootmark{a} \\ 
     & Magnitude & Magnitude & & Magnitude & & [mag] \\ 
      (1) & (2) & (3) & (4) & (5) & (6) & (7) \\ \noalign{\smallskip}
\hline  \noalign{\smallskip}
$B$ & 14.501 $\pm$ 0.027 & 15.172 & 2457400.66179 & 13.893 & 2458588.36564 & 1.151 \\
$V$ & 14.156 $\pm$ 0.026 & 14.646 & 2457612.30060 & 13.511 & 2458588.36235 & 1.135 \\
$R$ & 13.790 $\pm$ 0.024 & 14.444 & 2457400.65818 & 13.212 & 2458588.35905 & 1.116 \\
$I$ & 13.240 $\pm$ 0.028 & 13.970 & 2457400.66539 & 12.768 & 2458588.36099 & 1.071 \\
\hline
\end{tabular}\\
\tablefoot{Table columns read: (1) Passband of observation. (2) Weighted mean magnitude.
(3) Faintest magnitude attained by the source. (4) JD corresponding to the faintest magnitude. (5) Brightest magnitude attained by the source. (6) JD corresponding to the brightest magnitude. (7) Variability amplitude expressed in magnitudes instead of percents (see Eq. 4 in \citealp{2019MNRAS.488.4093A}). \tablefoottext{a}{For $BRI$-bands the January 12, 2016 (MJD $\sim$7400.7) data point was not taken into account in calculation of $A$ as it has no counterpart in the $V$-band.}}
\end{table*}

\subsection{Duty cycle}
\label{sect:res:inv:dc}

We found no significant variability during 28 nights of INM. 
Thus, our INM campaign resulted in zero duty cycle \citep[DC; for details about the DC see][]{1999A&AS..135..477R}. In order to improve this estimate, we searched the literature for other cases of INM of \pg. 

The first INM of \pg\ to our knowledge was reported by \citet{2005MNRAS.356..607S}~-- they applied the $C$-test to the $R$-band LCs (total duration of about 10\,hr) and found the source to be variable in one night out of two.
Another two nights of INM about 13\,hr long were presented by \citet{2006AJ....132..873O}. They found no significant INV, but they did not present statistical tests. To quantify the conclusion of \citet{2006AJ....132..873O}, we used the $C$-test in the form $C=\sigma / \langle e \rangle$, where $\sigma$ is the standard deviation of the source LC and $\langle e \rangle$ the mean uncertainty of the source photometric measurements. We applied this test to the data listed in Table\,4 of \citet{2006AJ....132..873O} and found $C=1.52$ and $C=1.28$ for the respective nights, which means that \pg\ was non-variable.
\citet{2007BAAA...50..299A, 2011A&A...531A..38A} reported the results from the INM they carried out during four nights in the $VR$-bands (April 21 to 24, 2007) and five nights in the $BR$-bands (April 21 to 25, 2009), respectively. Applying the $C$-test, the authors found no significant variability.
\citet{2011MNRAS.416..101G} detected INV during three nights out of three using $C$- and $F$-tests (a total monitoring duration of 16\,hr in the $R$-band).
\citet{2012MNRAS.425.3002G} found no INV during six nights of monitoring in the $BR$-bands ($C$- and $F$-tests were applied).
\citet{2016MNRAS.458.1127G} presented the results from the INM during seven nights in the $R$-band (a total monitoring duration of about 26\,hr). The authors found the source to be variable in one night, non-variable in another, and probably variable in the remaining nights ($F$- and $\chi^2$ tests were applied). We, however, should point out that in the latter nights the variability amplitude seems to be quite close to the magnitude uncertainties. In such cases the usage of the $C$-test could be more appropriate \citep[see][]{2017MNRAS.467..340Z,2020MNRAS.498.3013Z}.
\citet{2019ApJ...871..192P} monitored \pg\ for eight nights in the $VR$-bands for 2--4\,hr each night. Employing enhanced $F$- and nested ANOVA tests, the authors found the source to vary on intra-night timescales for three nights.
Finally, \citet{2020MNRAS.492.1295P} found no INV in the $BVR$-bands during a single night of monitoring (duration of about 3.7\,hr, $F$-test employed).

It is worth mentioning the study of \citet{2018ApJS..237...30M}, which is focused mainly on the LTV of a sample of blazars including \pg. For this source the observations were performed for 28 nights through three intermediate-band filters. The authors did not find INV during each of the observing nights using four statistical tests. In some of the nights, however, the observing time span is insufficient for the INV testing~-- less than an hour. So, we re-analysed the \pg\ data selecting those nights for which the observing time span is $\Delta t > 2\,\rm hr$ and the number of the acquired data points is $N > 3$. In this way we got a total of eight nights with $2.2\,{\rm hr} \le \Delta t \le 4.6\,\rm hr$ and $7 \le N \le 11$. We then applied the $C$-test and assumed the source to be variable for the given night if $C\ge2.576$, at least in two of the passbands. We found this condition not met for the selected nights.

The literature data, combined with ours, resulted in a total of 74 nights of \pg\ INM. 
Despite the large number of nights of INM the blazar \pg\ was found to show INV during only eight nights; during another five nights, the source was classified as probably variable. 
Because of the lack of information about the total monitoring duration for some of the literature data, we computed the DC simply as $N_{\rm INV}/N_{\rm INM}$, where $N_{\rm INV}$ is the number of the nights with INV detected and $N_{\rm INM}$ the total number of nights of INM.
The so computed DC is 10.8\% if the probably variable cases are considered non-variable and 17.6\%~-- otherwise. Our result agrees with the DC of 11.6\% reported by \citet{2014RMxAC..44...95A} for a sample of 6 HBLs.

\subsection{Long-term variability}
\label{sect:res:ltv}

We based the study of the optical behaviour of \pg\ at long-term timescales on our observations along 76 nights from 2016 to 2019. In the course of this multi-wavelength campaign, we acquired a total of $\sim$3350 data points in $BVRI$-bands.

In order to remove effects due to dense intra-night sampling during some of the nights, we performed nightly binning of our data.
Firstly, we 
discarded some unreliable data points and then the magnitudes obtained during each given night were weight-averaged. 
Regarding the uncertainties of the weight-averaged magnitudes, we adopted the following conservative estimate: we took the larger between (i) the uncertainty of the weighted mean (which accounts only for the individual measurement uncertainties) and (ii) the weighted standard deviation about the weighted mean (which accounts mainly for the scatter of the individual data points with respect to the weighted mean). 
Finally, we took the median of the corresponding MJDs. The data points that give rise to uncertainties of the weight-averaged magnitudes larger than or equal to 0.1\,mag were removed and the binning was done again.

Our long-term $BVRI$-band LCs are presented in Fig.\,\ref{fig:ourLC}.
Both statistical tests applied to them revealed that LTV is highly significant at all four photometric passbands.

The most remarkable feature of the LCs is the flare that was detected in all passbands on April 14, 2019 (MJD $\sim$8588.4) with the $R$ magnitude reaching
13.212, which seems to be the highest level recorded for \pg\ \citep[see also the Erratum\footnote{http://www.astronomerstelegram.org/?read=12695} to][]{2019ATel12631....1J}.
On the other hand, the source attained the faintest state on January 12, 2016 (MJD $\sim$7400.7) with an $R$ magnitude of 14.444.
We observed a trend of a larger variability amplitude at higher frequencies on a long-term basis.
The LTV characteristics are given in Table \ref{tab:ltv}.
We also assembled the historical $VR$ LCs of \pg\ covering a time interval from 2005 to 2019 (Appendix\,\ref{app:hist}, Fig.\,\ref{fig:fullLC}).

\subsection{Colour-magnitude diagram}
\label{sect:res:ltv:cmd}

Multi-band monitoring campaigns in the optical $BVRI$-bands help us in studying spectral changes on various timescales.
The study of CMDs could shed some light on the physical processes causing the blazar flux variations.
In this context, we firstly considered our $BVRI$ data set and then the historical $VR$ one.
Generally, we calculated colour indices by coupling data taken during the same night. 
The coupling within the night is justified as the DC of the INV of \pg\ is low (see Sect.\,\ref{sect:res:inv:dc}).

\begin{figure}[t]
\centering
\includegraphics[width=\columnwidth,clip=true]{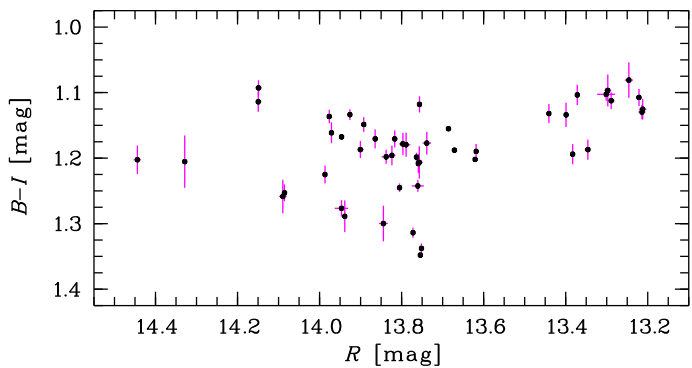}
\caption{Colour-magnitude diagram for our data only.}
\label{fig:bi_r}
\end{figure}

\begin{figure}[t]
\centering
\includegraphics[width=\columnwidth,clip=true]{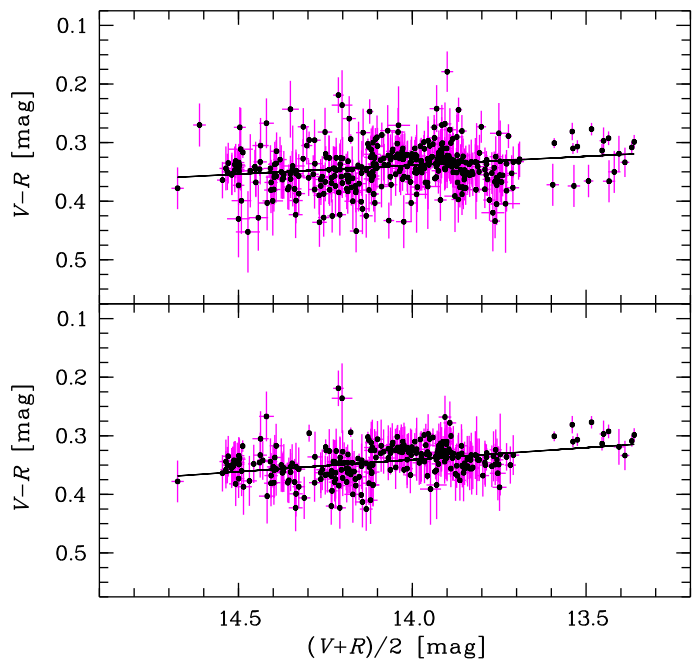}
\caption{Colour-magnitude diagram for the historical data set (upper panel, 373 data points plotted) and for our, Steward, and Kanata data sets (lower panel, 274 data points plotted). The linear least-squares fits are overplotted. For both panels the median uncertainties of the magnitudes and colour indices are $0.018$\,mag and $0.036$\,mag, respectively.}
\label{fig:vr_vr}
\end{figure}

\begin{figure}[t]
\centering
\includegraphics[width=\columnwidth,clip=true]{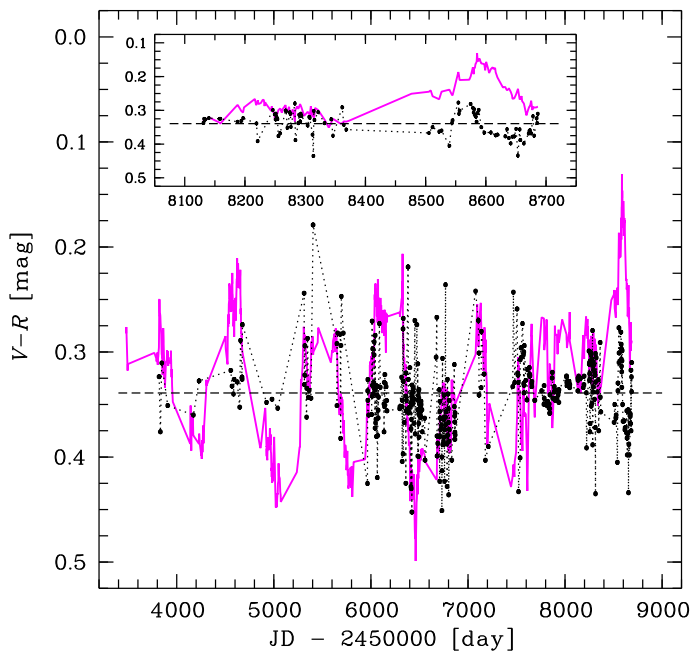}
\caption{Temporal evolution of the $V-R$ colour index (black dots). The scaled $R$-band LC is overplotted for comparison (magenta line, the flux increases along $+y$ axis).
The horizontal dashed line marks the weighted mean $V-R$ value of $(0.339 \pm 0.002)$\,mag with a weighted standard deviation of 0.030\,mag about the weighted mean. The insert shows the 2019 flare in details~-- the maximum of the spectral hardness (the minimum of $V-R$) precedes the flare by about 10\,days (see Sect.\,\ref{flare} for details).
}
\label{fig:vr_time}
\end{figure}

We plot in Fig.\,\ref{fig:bi_r} the colour index with the largest frequency base, viz. $B-I$, against the $R$ magnitude~-- we choose the $R$-band following the prescription of \citet{1996A&A...312..810M}. The presented CMD~-- from our data only~-- shows no clear correlation. 
Instead, one could trace a RWB chromatism when the source is faint and a BWB one when the source is bright. 
Similar behaviour was reported by \citet{2011PASJ...63..639I} using $V-J$ vs. $V$ CMD; the source was observed in the $VJK_{\rm S}$-bands from 2008 to 2010. The number of the data points used to build the above mentioned CMDs is relatively small, so,
any conclusions based on their analysis should be considered with caution. 
A denser sampling on $B-R$ vs. $R$ CMD for \pg\ was achieved by \citet{2015A&A...573A..69W}; the source was observed in $BR$-bands from 2007 to 2012. 
The authors reported a statistically non-significant RWB trend considering the whole data set
and both RWB and BWB trends on shorter time intervals.

Let us analyse the historical LCs constructed by us from the data sets presented in the Appendix\,\ref{app:hist}. Although the frequency base is not so large, the time span (from 2005 to 2019) and the total amplitude of variations are sufficient to look for
possible CMD peculiarities.
We plot in the upper panel of Fig.\,\ref{fig:vr_vr} the CMD for the historical data set: the colour indices are plotted against the mean of the $V$ and $R$ magnitudes. This mean could be considered as representative of the intermediate-band magnitude.
The overall trend is BWB with the following characteristics of the corresponding linear least-squares fit: a slope of $0.030 \pm 0.004$; $\chi^2_{\rm df}=2.8$; a linear Pearson correlation coefficient $r_{\rm CMD}=0.16$ with corresponding $p$-value of 0.002; a standard deviation about the fitted line of $0.037$\,mag. 
In the following, we shall consider a CMD trend (i) significant at 99\% confidence level if $r_{\rm CMD}\ge0.5$ and the null hypothesis probability is $p\le0.01$ \citep[e.g.][]{2016MNRAS.458.1127G} and (ii) achromatic if the absolute value of the slope is less than the corresponding fitted uncertainty. 
Therefore, the so obtained BWB trend is non-significant. In addition, the presented CMD shows a large scatter about the fitted line. This scatter is caused mainly by the data sets we used: they are obtained using various telescope+filter+CCD combinations and the differences in the resulting response curves were not taken into account during the inter-calibration process. In an attempt to minimize this effect we plot the CMD for our, Steward\footnote{In what follows, the Steward Observatory and the Kanata telescope will be referred to as Steward and Kanata, respectively, for short.}, and Kanata $VR$ data sets (Fig.\,\ref{fig:vr_vr}, lower panel); the other sets are single-band or had to be transformed to the $VR$-bands, so, their contribution to the scatter is more substantial.
Now, the BWB trend becomes clearer and has the following linear fit characteristics: a slope of $0.041 \pm 0.004$; $\chi^2_{\rm df}=3.6$; a linear Pearson correlation coefficient $r_{\rm CMD}=0.35$ with corresponding $p$-value less than $10^{-5}$; a standard deviation about the fitted line of $0.027$\,mag. This weak trend, however, is non-significant following the above criterion. 
The CMD shows three segments at different mean brightness levels in which local RWB trends could be followed.
Similar complicated behaviour~-- an overall BWB chromatism with local RWB trends~-- could be traced on the $B-R$ vs. $R$ CMD of \citet{2015MNRAS.454..353R} which has a smaller time span (about five months) but an excellent time coverage.

Furthermore, we plot in Fig.\,\ref{fig:vr_time} the temporal variations of the $V-R$ colour index compared to the scaled $R$-band LC. The comparison suggests no clear correlation. 

\begin{figure}[t]
\centering
\includegraphics[width=\columnwidth,clip=true]{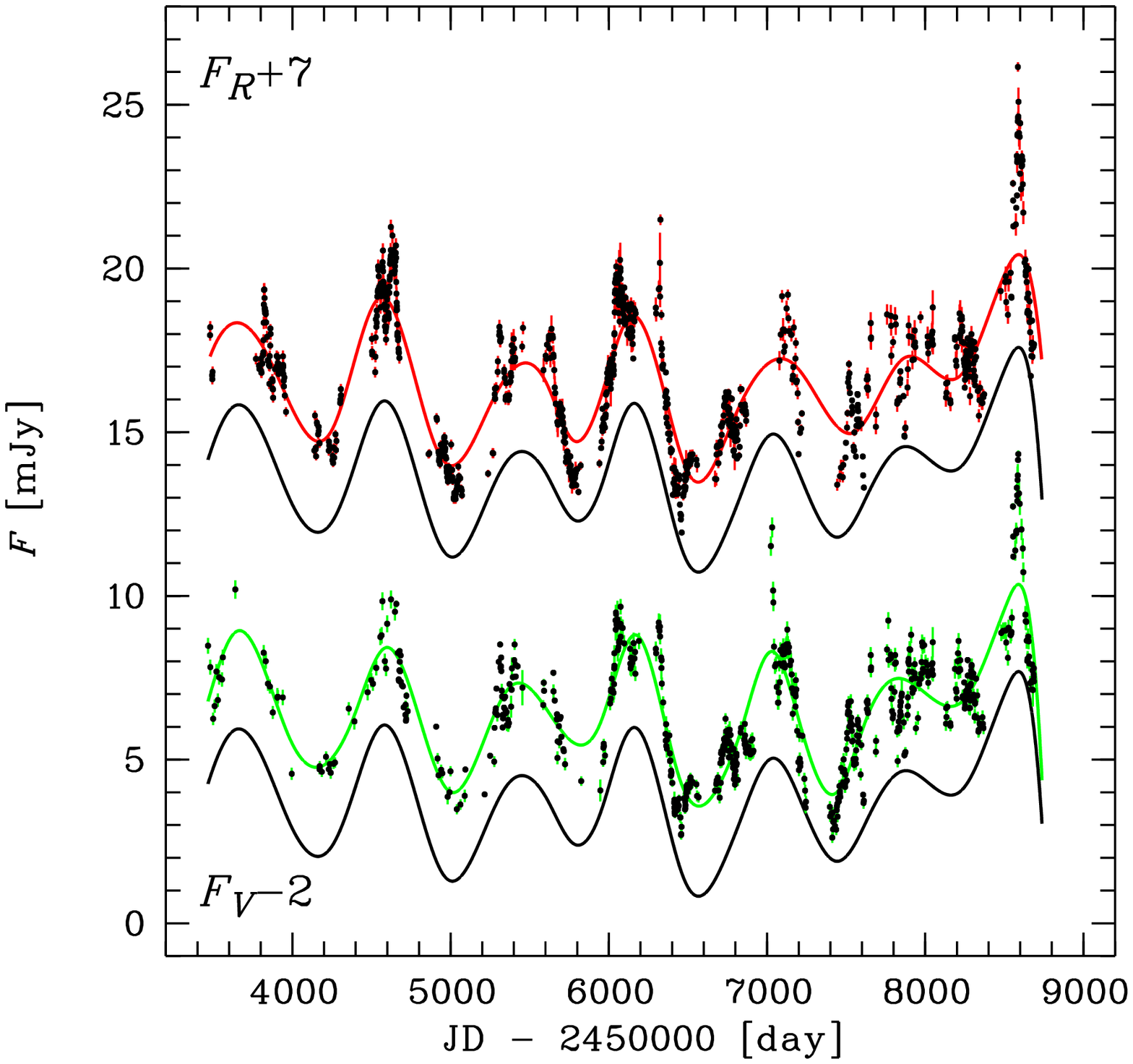}
\caption{Historical $VR$-band LCs (dots). The cubic splines, fitted through the 10-day binned LCs, are overplotted ($V$-band~-- green line; $R$-band~-- red line) along with the shifted mean spline representing the base levels for the $VR$-bands (black lines).}
\label{fig:base}
\vspace{0.5cm}
\centering
\includegraphics[width=\columnwidth,clip=true]{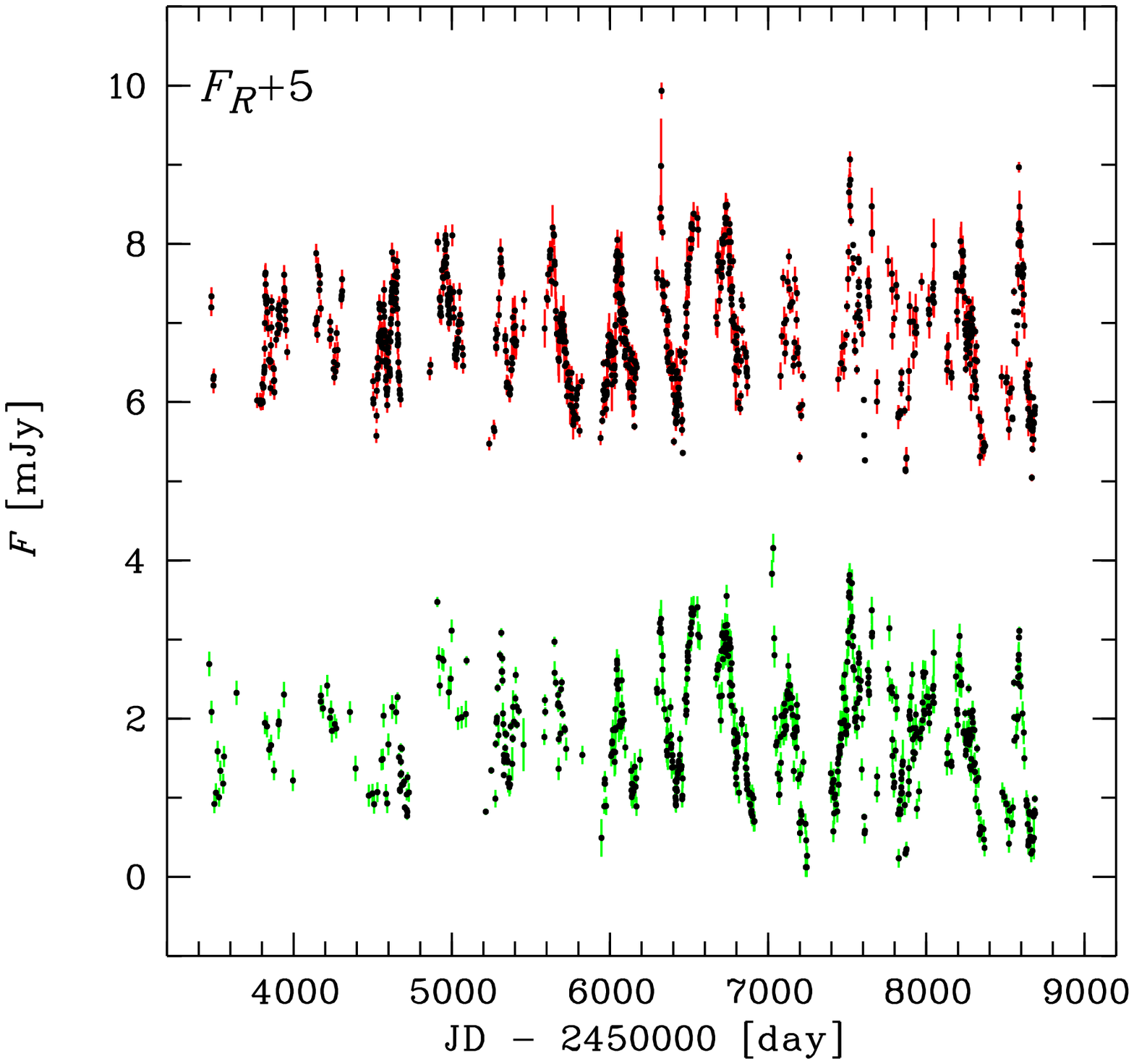}
\caption{Historical $VR$-band LCs corrected for the long-term variations.}
\label{fig:base_corr}
\end{figure}

\begin{figure}
\centering
\includegraphics[width=\columnwidth,clip=true]{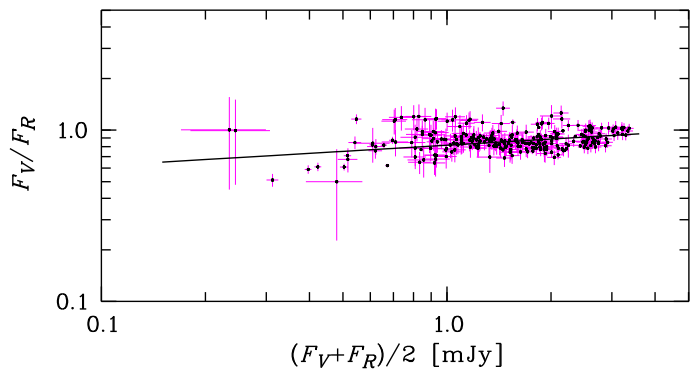}
\caption{Flux ratio $F_V/F_R$ plotted against $(F_V+F_R)/2$. Shown are our, Steward, and Kanata data sets corrected for the long-term variations.
The best fit linear model to the data is overplotted. The median uncertainties of the fluxes and flux ratios are $0.082$\,mJy and $0.087$\,mJy, respectively.}
\label{fig:vr_vr_part_corr_fx}
\end{figure}

To study the colour index behaviour on small timescales, we divided the LCs into separate segments. Considering Fig.\,\ref{fig:fullLC} one sees 5 distinct minima that divide the LCs into 6 flares; in addition, the 6$^{\rm th}$ flare can be divided into a pre-flare (around MJD of 8000) and a flare\footnote{This flare marks the brightest state of \pg\ recorded in our historical LC.}. 
Eventually, we divided the LCs into seven segments and built a CMD for each of them.
The most significant trend ($r_{\rm CMD}=0.49$ and $p=0.0007$) was found for the brightest flare~-- this flare will be analysed in details in Sect.\,\ref{flare}. Regarding the other segments, all of them show either achromatism or non-significant chromatism. In particular, the pre-flare could be considered achromatic~-- the absolute value of the CMD slope (viz. $-0.007 \pm 0.011$) is less than its uncertainty.

The inspection of Fig.\,\ref{fig:fullLC} suggests at least two variable components~-- a long-term one, responsible for the quasi-periodic behaviour of the source, and a short-term one producing the flux variations onto the top of the long-term component, that is, $F^{\rm (tot)}(t)=F^{\rm (long)}(t)+F^{\rm (short)}(t)$, where $F$ means flux. Therefore, to study the CMD behaviour on small scales, we have to remove the possible impact on their CMDs of the long-term variations. We shall assume that the long-term component is achromatic on the base of the above analysis.

To separate the components presumed, we used the approach of \citet{2002A&A...390..407V}. As a first step, we binned the $VR$-band LCs over a time interval of 10 days. Then, we transformed magnitudes into
fluxes\footnote{The magnitude-to-flux transformation was done as follows. Firstly, the calibrated $BVRI$ magnitudes of the blazar were corrected for the Galactic extinction using the values taken from the NASA/IPAC Extragalactic Database: $A_{B}$ = 0.188\,mag, $A_{V}$ = 0.142\,mag, $A_{R}$ = 0.113\,mag,
and $A_{I}$ =  0.078\,mag. The extinction corrected magnitudes were then converted into fluxes using the zero points of \citet{1998A&A...333..231B}.} and fitted a cubic spline to the binned LCs. 
The mean flux ratio of splines over the time, $\langle F^{\,\rm (spl)}_V(t)/F^{\,\rm (spl)}_R(t)\rangle_t$, is $0.89 \pm 0.06\,\rm(s.d.)$, which corresponds to a two-point spectral index $\alpha_{VR}^{\rm (spl)}=0.72 \pm 0.42$~-- these are the flux ratio and spectral index of the long-term component (or the base level). The spline fits are very similar, so, we averaged them for
further analysis. 
After that we subtracted from the $VR$-band LCs the mean spline, shifted in such a way, that the condition $[F(t)-e(t)]-F^{\rm (spl)}(t)>0$ is fulfilled for each MJD and both passbands; here $e$ is the uncertainty of the flux level $F$. In this way, we corrected the LCs for the variable long-term component. The spline fits and the adopted base levels are shown in Fig.\,\ref{fig:base}.

\begin{figure}[t]
\centering
\includegraphics[width=\columnwidth,clip=true]{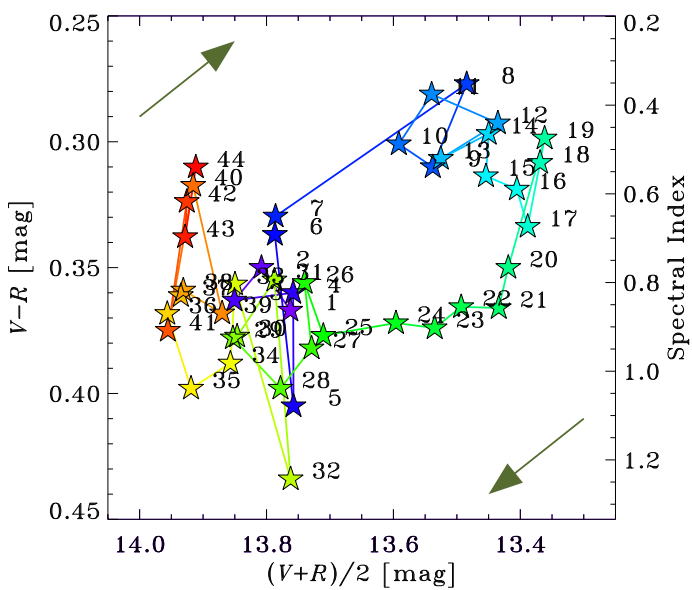}
\caption{Colour-magnitude diagram for the 2019 flare. The corresponding spectral index: $\alpha = [(V-R)-(A_V -A_R)-0.186]/0.07$, is also indicated. Here $A_V$ and $A_R$ stand for the Galactic extinction in the $VR$-bands, respectively (see Sect.\,\ref{sect:res:ltv:cmd}).
The clockwise spectral hysteresis can be clearly seen~-- the arrows are used to guide the eye about the hysteresis loop direction. The colours and numbers are sequential in time: violet ($\#1$)\;$\rightarrow$ red ($\#44$).}
\label{fig:loop}
\end{figure}

\begin{figure*}[t]
\centering
\begin{minipage}[t]{5.3cm}
\includegraphics[width=5.3cm,clip=true]{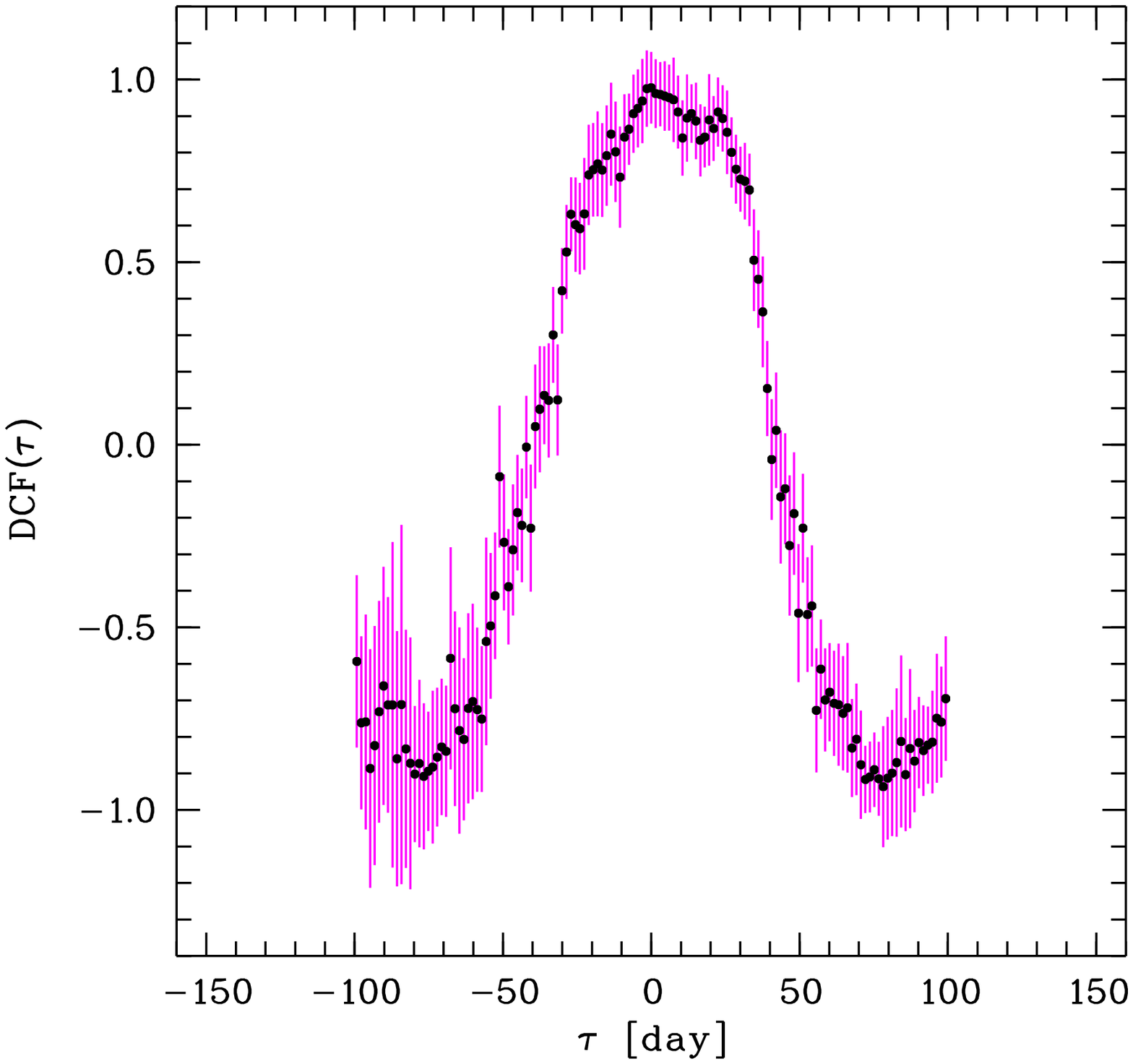}
\caption{Discrete correlation function. The fitted Gaussian is not shown for the sake of clarity.}
\label{fig:dcf}
\end{minipage}
  \hfill
\begin{minipage}[t]{5.3cm}
\includegraphics[width=5.3cm,clip=true]{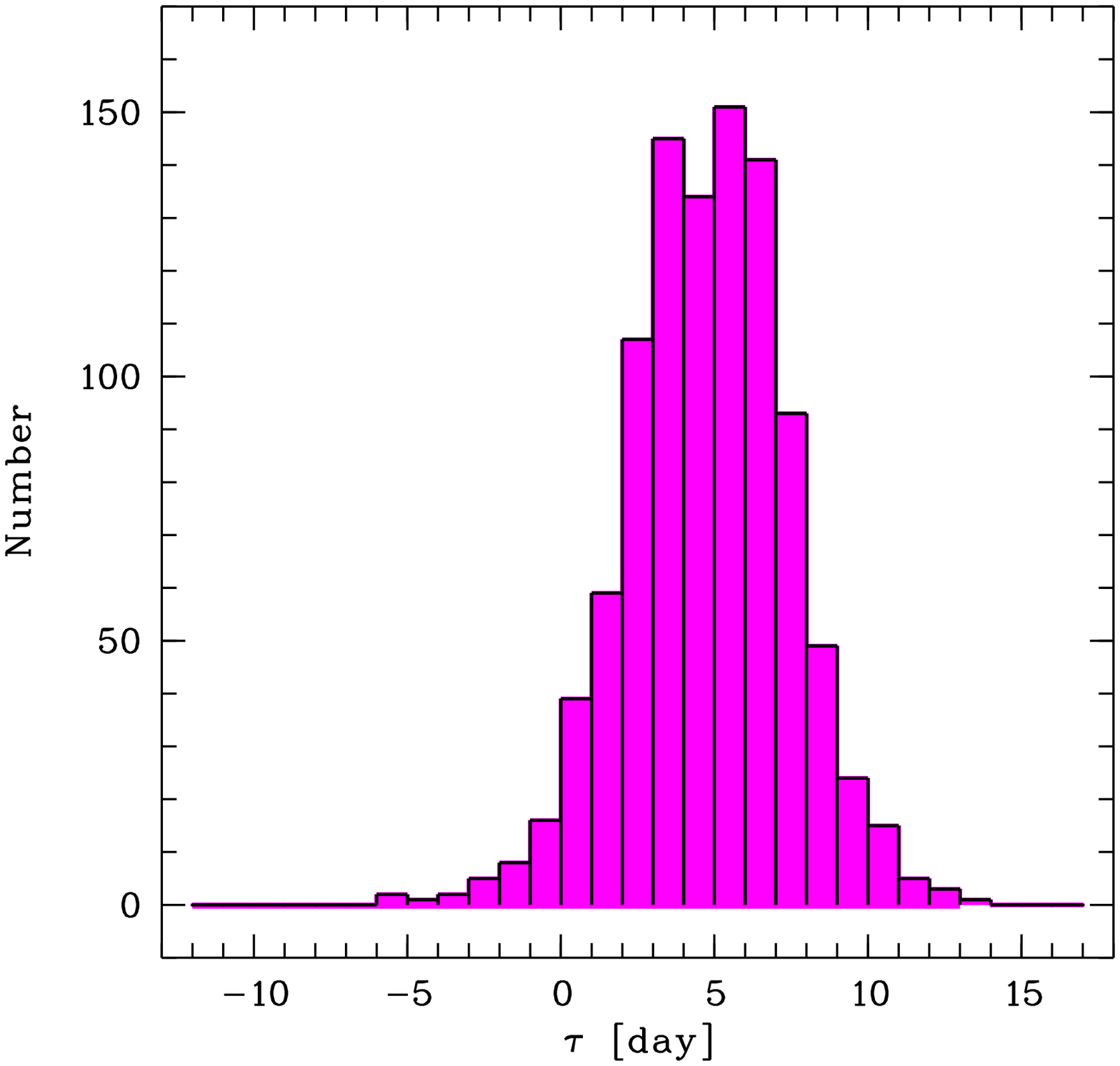}
\caption{Cross-correlation peak distribution for DCF.}
\label{fig:ccpd}
\end{minipage}
  \hfill
\begin{minipage}[t]{5.3cm}
\includegraphics[width=5.3cm,clip=true]{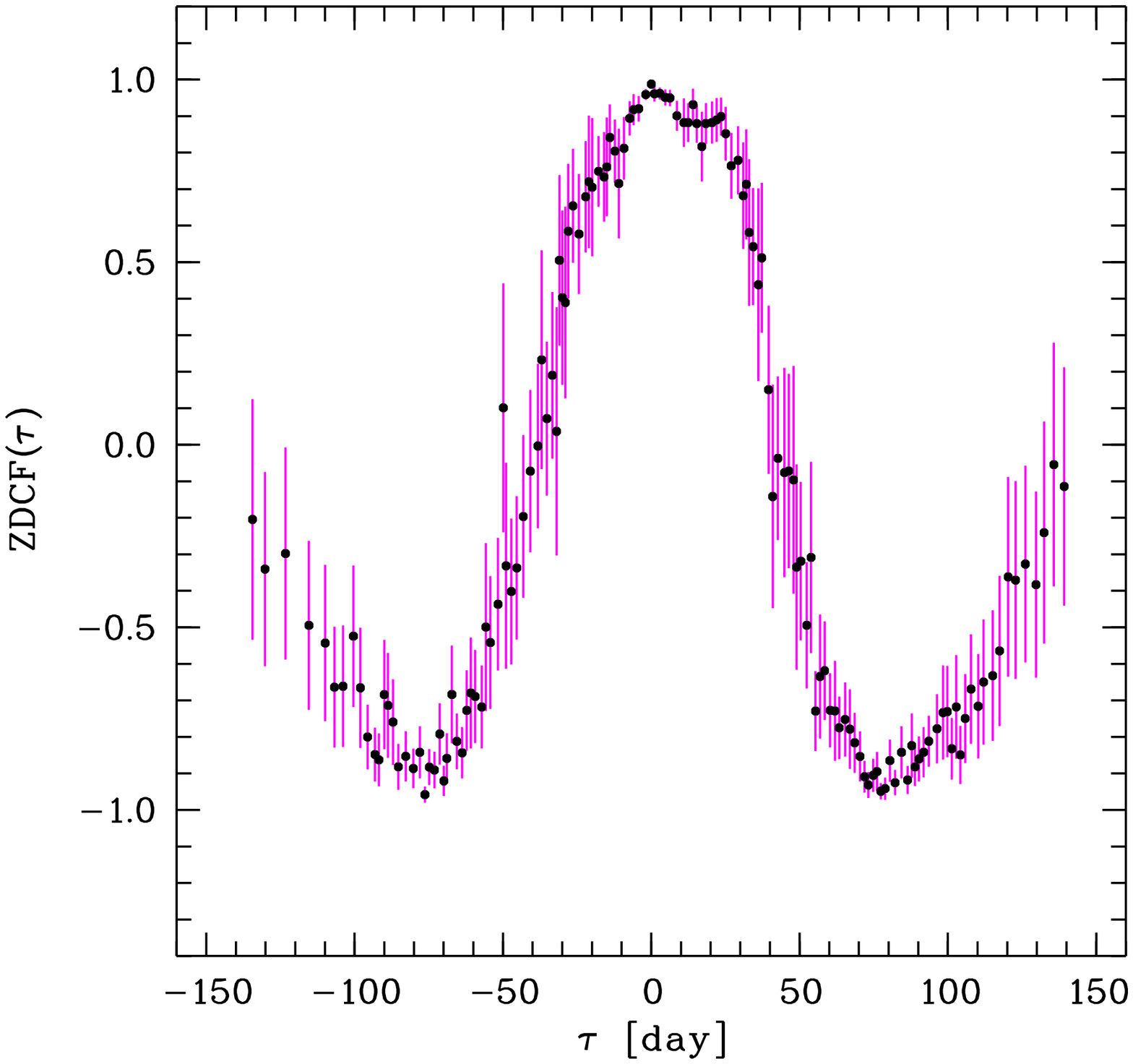}
\caption{Same as in Fig.\,\ref{fig:dcf}, but for ZDCF.}
\label{fig:zdcf}
\end{minipage}
\end{figure*}

The next assumption, based solely on the achromaticity, is that the base level variability is a result of the Doppler factor change. We recall that the Doppler factor changes if the viewing angle varies. To account for this relativistic boosting we divided each flux data point by the ratio, $C(t)$, between the spline value at that time and the spline global minimum value, $C(t) = F^{\rm (spl)}(t)/F^{\rm (spl)}_{\rm min}$.

The resulting `corrected' LCs are presented in Fig.\,\ref{fig:base_corr}. The variation amplitudes are comparable over the entire period covered by the historical data set. This should mean that we see the variations of the short-term component alone. 
The corresponding CMD, however, still does not show any clear chromatism (Fig.\,\ref{fig:vr_vr_part_corr_fx}). The characteristics of the linear fit presented in Fig.\,\ref{fig:vr_vr_part_corr_fx} are: a slope of $0.120 \pm 0.007$; $\chi^2_{\rm df}=5.3$; a linear Pearson correlation coefficient $r_{\rm CMD}=0.22$ with corresponding $p$-value of 0.003; a standard deviation about the fitted line of $0.131$\,mJy.

The scatter about the fitted line now is larger compared to the pre-corrected CMD scatter of 0.022\,mJy.
This could be attributed mainly to the unknown true base level and shape. Regarding the shape, we implicitly assumed that it could be represented by the cubic spline fitted to the binned LCs. 
Thus, by correcting as explained above, we introduce uncertainties, firstly, by the base level subtraction itself and, secondly, by the Doppler boosting correction based on this level.
The other source of uncertainties is the quality of the cubic spline interpolation, which depends on the LC sampling~-- both in terms of the total number of data points and the degree of their clustering. We, however, can consider this source of less importance regarding the CMD scatter.
Despite the large scatter, now the CMD shows no significant sub-structure compared to the pre-corrected CMD (see Fig.\,\ref{fig:vr_vr}). 

Regarding the short time span segments, we divided the historical LCs into, the situation does not change. All segments but the pre-flare and flare ones show either achromatism or non-significant chromatism. Now, the pre-flare shows non-significant BWB chromatism (after removing 3 deviant data points) and the flare BWB trend gets more significant~-- $r_{\rm CMD}=0.63$ and $p<10^{-5}$.

The above analysis shows that on both long and short timescales, the variability of \pg\ has no significant chromatism (except the 2019 flare).
Another notable result is that the corrected LCs show signs for quasi-periodicity (see Sect.\,\ref{period}).

Given the achromaticity alone, we cannot distinguish whether it reflects variations caused by geometric effects or by processes intrinsic to the jet. However, the observed periodicities, discussed in Sect.\,\ref{period}, could be considered in support of the geometric origin of the \pg\ variability \citep[e.g.][]{2017MNRAS.465..161S}.

\subsection{The 2019 flare}
\label{flare}

\subsubsection{Colour hysteresis and time lag}
\label{flare:lag}

During April 2019, \pg\ reached its historical brightness maximum showing a prominent flare. The flare shows some sub-structure~-- there is a flux increase (sub-flares?) around MJD 8550 and MJD 8610 (the latter is best traced in the $R$-band). The LCs sampling, however, is not good enough to derive further conclusions about these sub-structures.
The detailed inspection of the flux and colour (or spectral) index behaviour during the flare revealed a lag of the flux changes behind the spectral ones (see Fig.\,\ref{fig:vr_time}). 
The spectrum got its hardest value (data point $\#8$ in Fig.\,\ref{fig:loop}) before the flux maximum was reached (data point $\#19$) and then continuously softened till the end of the flare. 
The corresponding CMD shows a clockwise spectral hysteresis loop~-- the spectrum hardens (i.e. gets flatter $\equiv$ shows BWB chromatism) as the flux rises and softens (i.e. gets steeper $\equiv$ shows RWB chromatism) as the flux declines (Fig.\,\ref{fig:loop}). 
This kind of hysteresis is typically observed at X-rays in HBLs \citep[e.g.][]{1996ApJ...470L..89T} and signals to the existence of a time lag, ${\mathcal T}$: the flux variations at a high frequency ($V$-band) lead those at a low frequency ($R$-band; this is the so-called `soft' lag\footnote{This notation comes from X-ray passbands and means the hard X-rays lead the soft X-rays. The opposite is known as a `hard' lag.}).
The detected hysteresis loop could also be traced on the CMD built using only our and Steward data sets as well as on the CMD built using the historical $VR$ data sets after their correction for the long-term variations. So, we shall consider this hysteresis as real rather than an artefact from the heterogeneous data used.

To search for the expected time lag we utilized both the discrete correlation function \citep[DCF,][]{1988ApJ...333..646E} and the $z$-transformed DCF \citep[ZDCF,][]{1997ASSL..218..163A,2014ascl.soft04002A}. We cross-correlated the $V$- and $R$-band LCs transformed into fluxes.

The DCF is commonly used to search for correlation in two unevenly sampled time series. We did not take into account the measurement uncertainties in the construction of the DCF following \citet{1994PASP..106..879W}.
In our particular case the discrete correlations were binned using a time lag bin of width $\Delta{\mathcal T}=1.5$\,days and a Gaussian weighting scheme was applied, so that the higher importance is assigned to the unbinned values closer to the bin centre.
In our notation the positive lag means that the variations in the $V$-band lead those in the $R$-band. The time lag was determined using both the Gaussian fitting and the centroiding methods; we took into account only those lags, for which the correlation coefficient is larger than 0.5. 
We found ${\mathcal T}_{\rm gauss}=(5.76\pm1.48)$\,days using the Gaussian fit and ${\mathcal T}_{\rm cent}=3.71$\,days using the weighted centroid. The DCF is shown in Fig.\,\ref{fig:dcf}. 
For more reliable estimation of the time lag and its uncertainty, we applied the flux randomization/random subset selection method \citep[FR/RSS,][]{1998PASP..110..660P} based on Monte Carlo simulations. 
The data points counted more than once during the RSS process were rejected. For each FR/RSS realization the time lag was found by means of the Gaussian fit. We ran a total of 1000 cycles and the resulted time lag values were used to build the cross-correlation peaks distribution \citep[CCPD,][]{1989MNRAS.236...21M}. 
The CCPD is shown in Fig.\,\ref{fig:ccpd} and the derived time lag is ${\mathcal T}_{\rm CCPD}=4.86^{+2.50}_{-2.62}$\,days. The quoted time lag value is the 50$^{\rm th}$ percentile (the median) of the CCPD, while the 16$^{\rm th}$ and 85$^{\rm th}$ percentiles serve as the $1\sigma$ time lag uncertainties. 

The application of ZDCF gave similar results (Fig.\,\ref{fig:zdcf}). The corresponding estimates of the time lag are ${\mathcal T}_{\rm gauss}=(3.71\pm0.88)$\,days and ${\mathcal T}_{\rm cent}=3.85$\,days. It is worth noting the better agreement between these estimates compared to the DCF ones.
So, we detected a significant positive time lag in agreement with the clockwise spectral hysteresis. 

\subsubsection{Flare fit}
\label{flare:fit}

\begin{figure}[t]
\centering
\includegraphics[width=\columnwidth,clip=true]{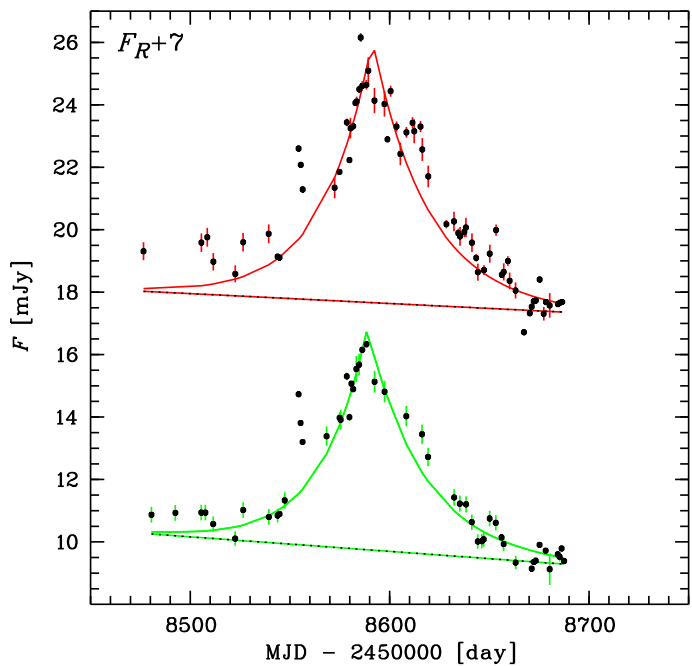}
\caption{Approximation of the 2019 flare with an exponential law plus a local linear base level. The possible sub-flares could be discerned around $\rm MJD=8550$ and $\rm MJD=8610$.}
\label{fig:fit}
\end{figure}

In order to quantify the flare, we fitted it with an exponential law \citep{1999ApJS..120...95V} plus a linear base level:
\[
F(t) =
  \begin{cases}
    a+b\,t+\exp[+(t-t_{\rm max})/{\mathcal T}_{\rm ris}] & \quad \text{if } t \le t_{\rm max} \\
    a+b\,t+\exp[-(t-t_{\rm max})/{\mathcal T}_{\rm dec}] & \quad \text{if } t \ge t_{\rm max}
  \end{cases},
\]
where $(a,b)$ are the parameters of the local linear base level, $t_{\rm max}$ the epoch of the flare maximum and $({\mathcal T}_{\rm ris},{\mathcal T}_{\rm dec})$ the flare rise and decay $e$-folding timescales, respectively.
The data points belonging to the possible sub-flare at MJD $\sim$8550 and the data point at the $R$-band maximum were excluded from the fit. The fitted maximum position along with the rise and decay $e$-folding timescales are listed in Table\,\ref{tab:fit}. The resulting fits are shown in Fig.\,\ref{fig:fit}.

The fits confirmed the time lag found by the cross-correlation techniques~-- the flare $V$-band LC leads the $R$-band one by ${\mathcal T}_{\rm fit}=(2.6 \pm 0.2)$\,days. This lag equals to the cross-correlation one to within the uncertainties quoted.

\begin{table}
\caption{Flare best fit parameters~-- observer-frame.}
\label{tab:fit}
\centering
\begin{tabular}{cccc}
\hline\hline\noalign{\smallskip}
Passband & $t_{\rm max}$ & ${\mathcal T}_{\rm ris}$ & ${\mathcal T}_{\rm dec}$ \\
 & [MJD] & [day] & [day] \\
\hline\noalign{\smallskip}
$V$ & $8588.7\pm0.2$ & $23.5\pm0.5$ & $27.5\pm0.4$ \\
$R$ & $8591.3\pm0.1$ & $25.0\pm0.1$ & $27.7\pm0.3$ \\
\hline
\end{tabular}
\end{table}

\begin{figure}[t]
\centering
\includegraphics[width=\columnwidth,clip=true]{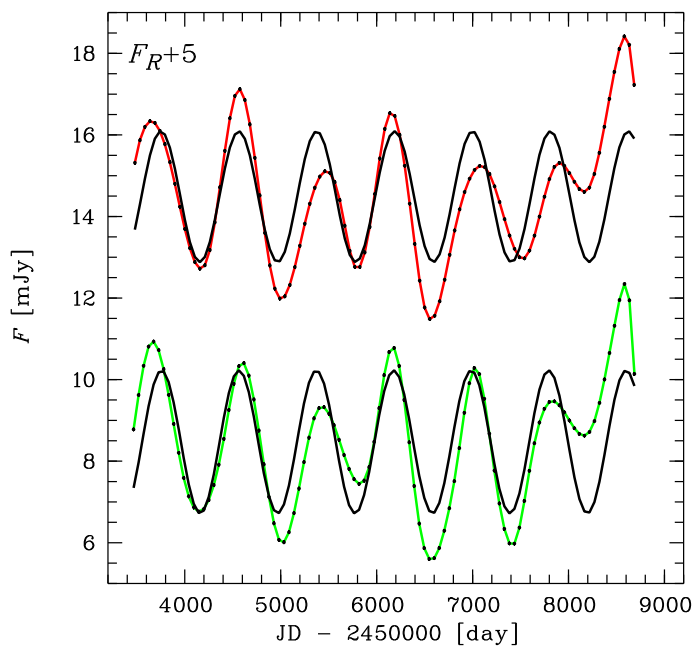}
\caption{Sine function fits to the $VR$ splines: $V$-band~-- lower green curve with black dots overplotted, $R$-band~-- upper red curve with black dots overplotted, sine fit~-- black curves of constant amplitude.}
\label{fig:sin}
\end{figure}

\begin{figure}[t]
\centering
\includegraphics[width=\columnwidth,clip=true]{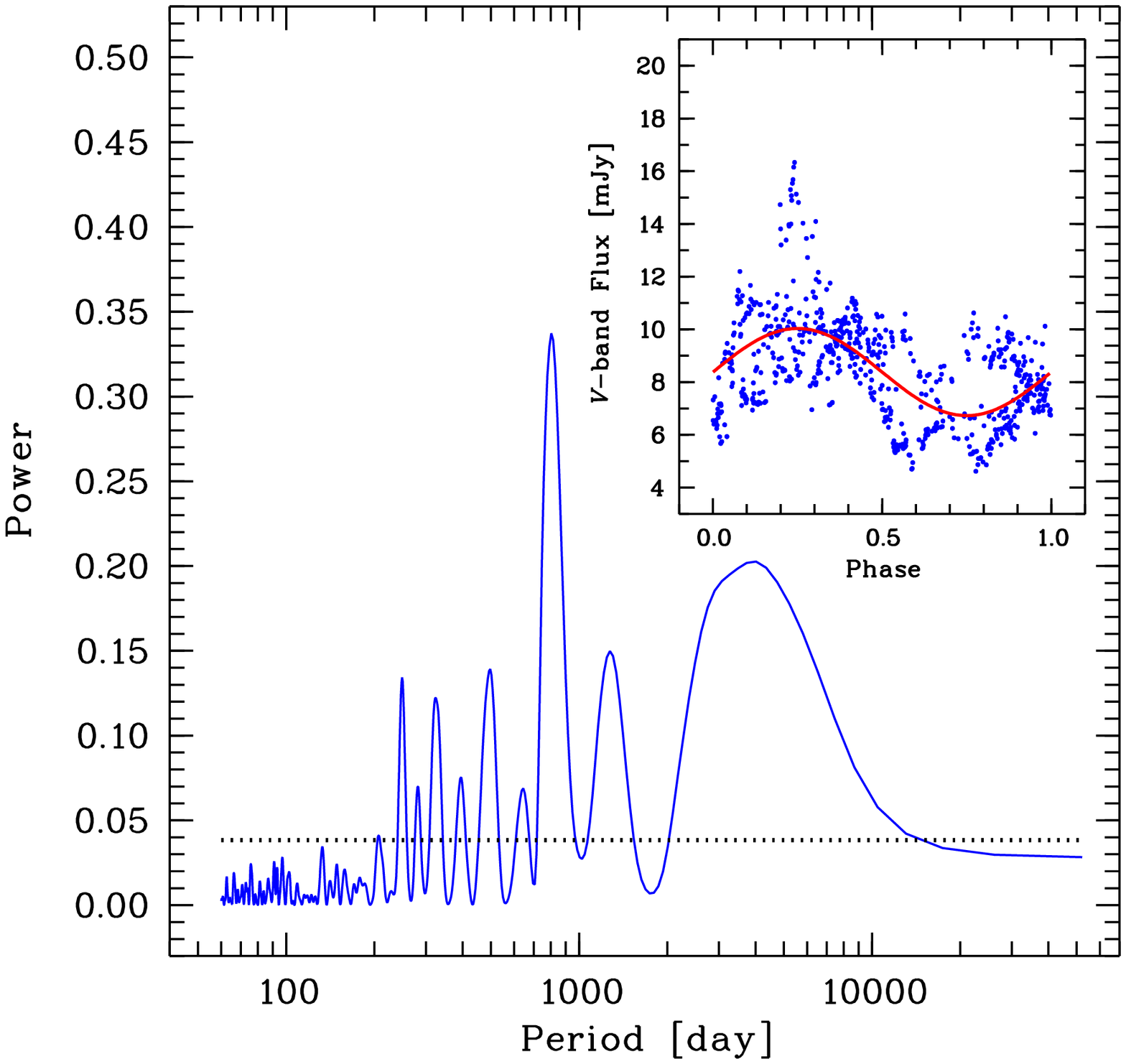}
\caption{Lomb-Scargle periodogram for the historical $V$-band LC. The dotted line marks the level corresponding to a maximum peak false alarm probability of 1\% under the assumption of white noise. The insert shows the LC folded with the most significant period; the red line denotes the fitted sine wave.}
\label{fig:ls_v}
\vspace{0.5cm}
\centering
\includegraphics[width=\columnwidth,clip=true]{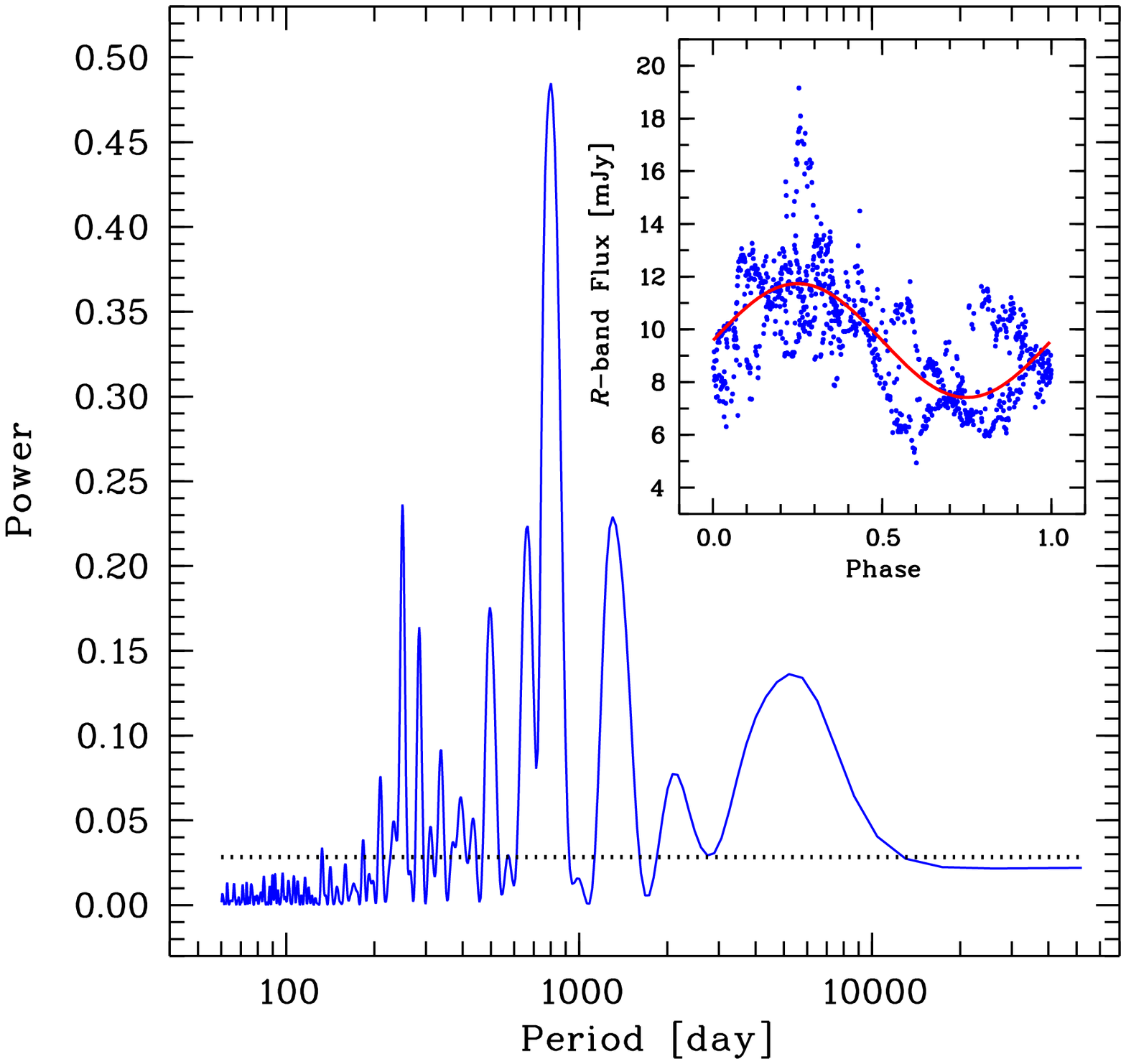}
\caption{Same as in Fig\,\ref{fig:ls_v}, but for the $R$-band.}
\label{fig:ls_r}
\end{figure}

\begin{figure}[t]
\centering
\includegraphics[width=\columnwidth,clip=true]{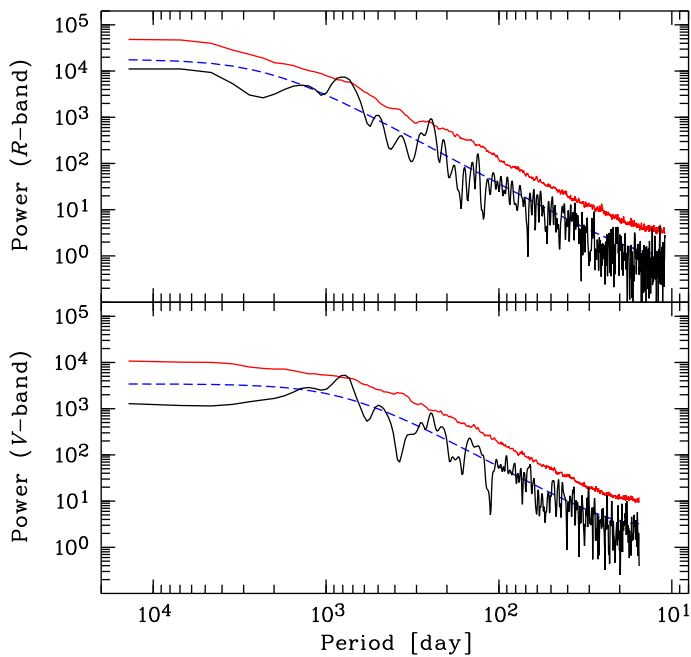}
\caption{Output from the {\tt REDFIT} programme. The black line shows the bias-corrected power spectrum, the blue dashed line marks the theoretical red noise spectrum and the red line marks the 99\% local significance level derived by means of Monte Carlo simulations.}
\label{fig:redfit}
\end{figure}

\subsection{Search for periodicity}
\label{period}

\begin{figure}
\centering
\includegraphics[width=\columnwidth,clip=true]{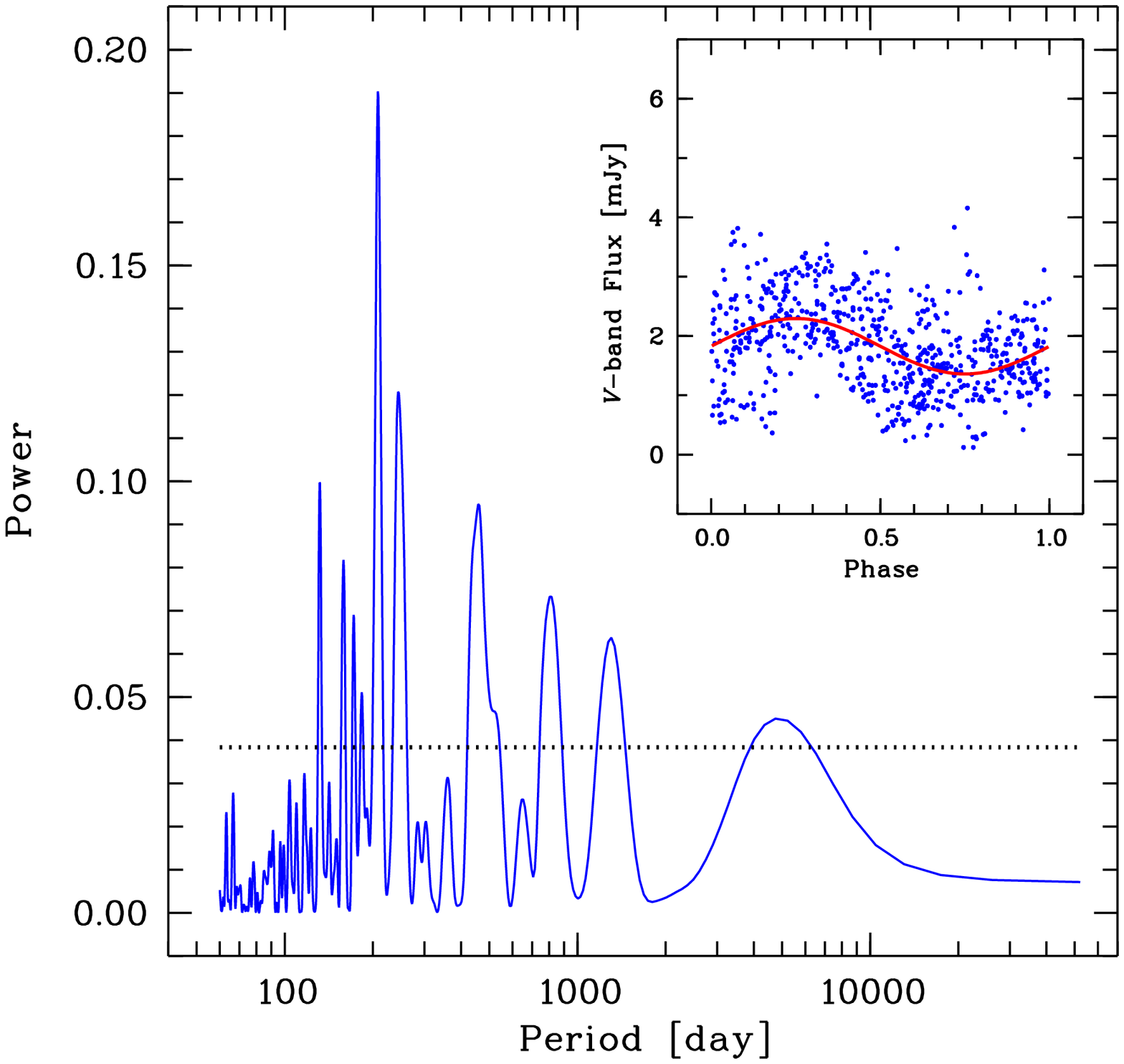}
\caption{Same as in Fig.\,\ref{fig:ls_v}, but for the historical $V$-band LC corrected for the long-term variations.}
\label{fig:ls_v:c}
\vspace{0.5cm}
\centering
\includegraphics[width=\columnwidth,clip=true]{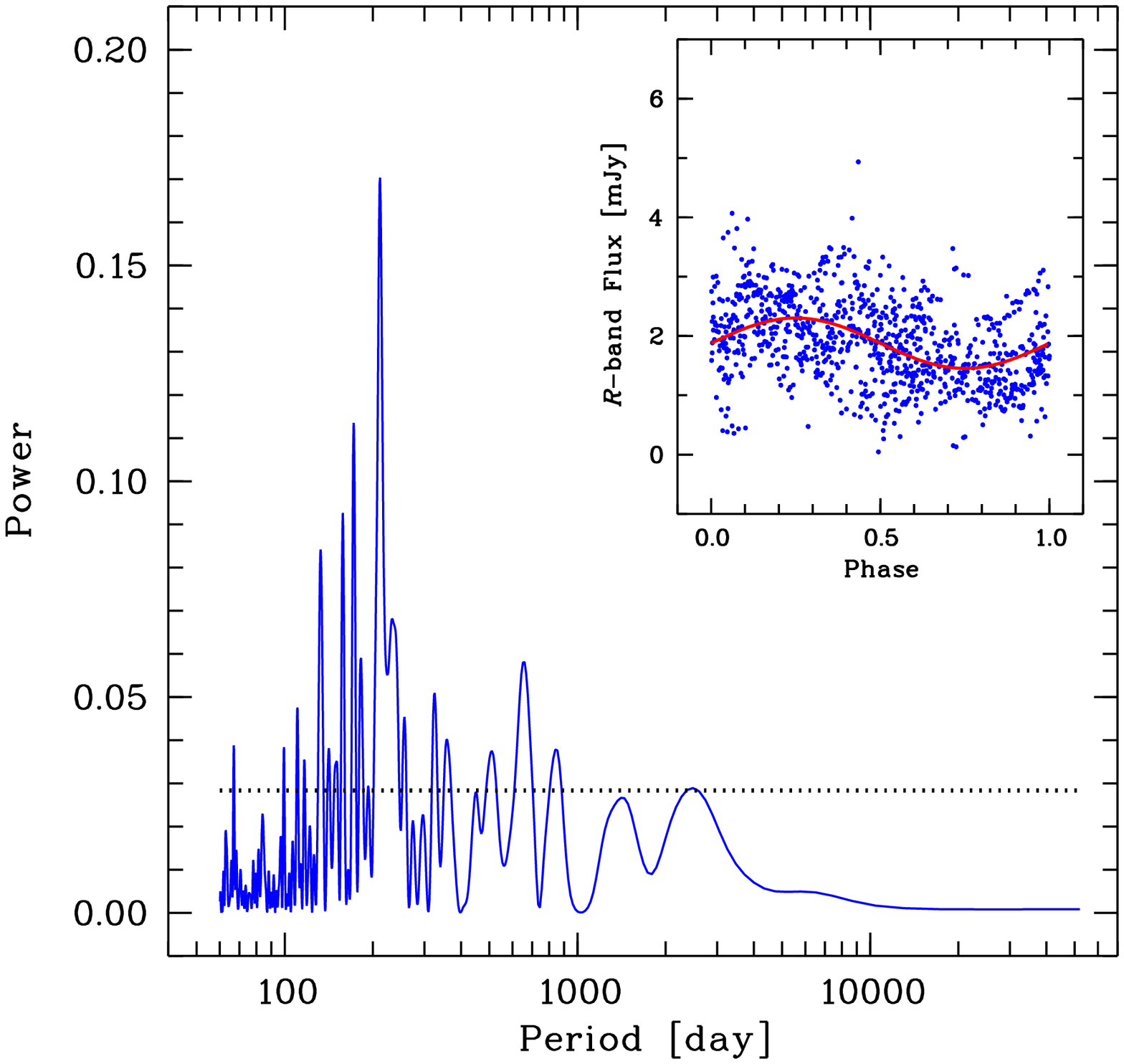}
\caption{Same as in Fig\,\ref{fig:ls_v:c}, but for the $R$-band.}
\label{fig:ls_r:c}
\end{figure}

\begin{figure}[t]
\centering
\includegraphics[width=\columnwidth,clip=true]{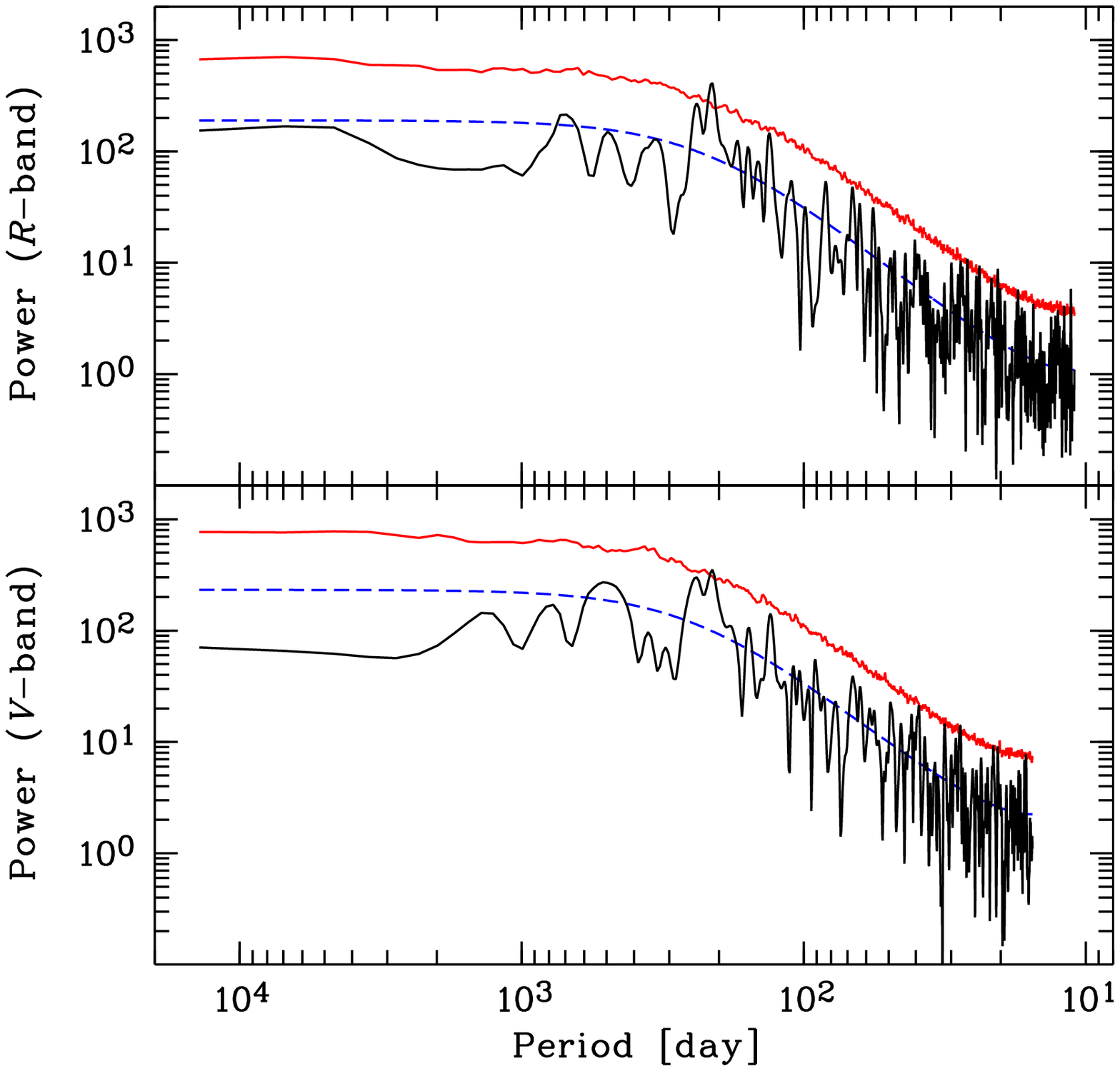}
\caption{Same as in Fig.\,\ref{fig:redfit}, but for the historical $VR$-band LCs corrected for the long-term variations.}
\label{fig:redfit:c}
\end{figure}

We addressed the \pg\ periodicity issue using our historical $VR$-band LCs and applying three different methods.
The periodicity presence is obvious if we consider the spline fits to the historical LCs (see Fig.\,\ref{fig:base}); we recall that the spline fits represent the long-term variability component.
To quantify the periodicity, we fitted a sine function of the form $F(t)\propto \sin(2\pi t/P+\varPhi)$ to the corresponding spline; here $P$ and $\varPhi$ stand for the period and phase, respectively. The resulting fits are shown in Fig.\,\ref{fig:sin}. 
We found periods of $(807\pm20)$\,days and $(812\pm30)$\,days considering the $V$- and $R$-band LCs, respectively. The corresponding mean differences between the maxima of the splines and of the fitted sine waves are $(9\pm20)$\,days and $(17\pm30)$\,days. The quoted standard errors of the mean were used as the uncertainties of the corresponding periods.

Next, we applied the classical Lomb-Scargle periodogram (LSP) analysis, which was proposed by
\citet{1976Ap&SS..39..447L}
and extended by \citet{1982ApJ...263..835S}. The LSP has been generalized for more practical
use by \citet{1989ApJ...338..277P}. More details are outlined in \citet{2018ApJS..236...16V}
and references therein. The statistical significance of the periods, found by means of LSP and {\tt REDFIT} (see below) methods, is evaluated locally \citep[e.g.][]{2020A&A...634A.120A}.

The periodograms are shown in Figs.\,\ref{fig:ls_v} and \ref{fig:ls_r}. The maxima of the most significant peaks correspond to periods of $(803 \pm 70)$\,days and $(801 \pm 62)$\,days, for the $V$- and $R$-band LCs, respectively. 
The quoted uncertainties represent the half-width at the half-power maximum derived by means of Gaussian fit; we ignore the possible asymmetry of the fitted peaks. These periods agree to within the uncertainties with the ones obtained by applying the sine fits to the splines. Because of this, we shall consider these periods as real. 
There are, however, a number of additional, less significant peaks~-- their existence is most probably caused either by the departure of the shape of the folded LCs from a sinusoid (see the inserts of Figs.\,\ref{fig:ls_v} and \ref{fig:ls_r}) or by aliasing due to the finite monitoring time span and irregular sampling. So, we shall not consider these peaks further.

Inspecting Fig.\,\ref{fig:ls_v} one can see that the amplitude of the power spectrum between the major peaks, that is, the noise-induced power spectrum decreases with decreasing period~-- this is a signature of the presence of red noise; generally, the red noise arises from stochastic processes and is frequency-dependent.
To test if the power spectrum peaks are significant against the red noise background, we utilized the improved translation of the {\tt REDFIT} programme \citep{2002CG.....28..421S} run within the {\tt R} environment\footnote{https://rdrr.io/cran/dplR/}. 
In this programme an auto-regressive process of first-order (AR1) is used to approximate the red noise.
Before actual computations, we checked whatever the AR1 model is appropriate to characterize
our historical $VR$-band LCs; the non-parametric runs test indicates that both the $VR$ spectra are consistent with the AR1 model.

We ran {\tt REDFIT} using the oversampling factor of 4, two overlapping by 50\% segments, and Welch spectral window. The results are shown in Fig.\,\ref{fig:redfit}.
The peaks exceeding the 99\% significance level correspond to periods of $(772 \pm 85)$\,days and $(817 \pm 107)$\,days, for the $V$- and $R$-band, respectively. The periods agree to within the uncertainties with the above estimates; the uncertainties are larger owing to the broader peaks.

There is another peak above the 99\% significance level corresponding to a period of $(248 \pm 12)$\,days (Fig.\,\ref{fig:redfit}, $R$-band\footnote{The peak could be traced onto the $V$-band {\tt REDFIT} periodogram as well, but at lower significance.}). 
A hint for quasi-periodicity on such timescales could be found inspecting Fig.\,\ref{fig:base_corr}, especially the $R$-band LC. 
A similar period was found by \citet{2018A&A...615A.118S}: $(250 \pm 60)$\,days. In addition, the most significant rest-frame period, reported by \citet{2018A&A...620A.185N}~-- 174\,days~-- is in agreement with the above periods, transformed to the rest-frame.

To check further the existence of a possible period in the range 200--250\,days, we apply the LSP method and the {\tt REDFIT} programme to the LCs corrected for the long-term variations. The results from the LSP application are presented in Figs.\,\ref{fig:ls_v:c} and \ref{fig:ls_r:c}. 
The most significant peaks correspond to periods of $(208 \pm 6)$\,days and $(212 \pm 6)$\,days, for the $VR$-band LCs, respectively. The {\tt REDFIT} found the following periods: $(211 \pm 10)$\,days and $(210 \pm 8)$\,days, for the $VR$-band LCs, respectively (Fig.\,\ref{fig:redfit:c}). 
The non-parametric runs test, however, indicated that the spectrum of the corrected $V$-band LC is inconsistent with the AR1 model, that is, non-AR1 components are present; the {\tt REDFIT} assumes that the noise in a
time series can be approximated by an AR1 process.

In Table\,\ref{tab:period} we summarize the periods found.
The local significance of the periods~-- except those, obtained by means of the sine wave fit~-- is greater than 2.5$\sigma$. The similar local significance of the same year-long period was found by \citet{2018A&A...615A.118S} using the AR1 model.

\begin{table}
\caption{Summary of the periods found in the historical $VR$-band LCs of \pg. The methods used are also listed.}
\label{tab:period}
\centering
\begin{tabular}{llr@{\,$\pm$\,}l}
\hline\hline\noalign{\smallskip}
Passband & Method & \multicolumn{2}{l}{Period} \\
 & & \multicolumn{2}{l}{[day]} \\
\hline\noalign{\smallskip}
$V$-band & Sine wave fit & $807$ & $20$ \\
 & LSP  & $803$ & $70$ \\
 & {\tt REDFIT} & $772$ & $85$ \\
 & LSP  & $208$ & $6$\tablefootmark{\,a} \\
 & {\tt REDFIT}\tablefootmark{\,b} & $211$ & $10$\tablefootmark{\,a} \\
\hline\noalign{\smallskip}
$R$-band & Sine wave fit & $812$ & $30$ \\
 & LSP  & $801$ & $62$ \\
 & {\tt REDFIT} & $817$ & $107$ \\
 & {\tt REDFIT} & $248$ & $12$ \\
 & LSP  & $212$ & $6$\tablefootmark{\,a} \\
 & {\tt REDFIT} & $210$ & $8$\tablefootmark{\,a} \\
\hline
\end{tabular}
\tablefoot{
\tablefoottext{a}{These periods refer to the LCs corrected for the long-term variations.} \tablefoottext{b}{$V$ power spectrum is inconsistent with the AR1 model.}
}
\end{table}

\subsection{Spectral energy distribution}
\label{sect:sed}

\begin{figure}[t]
\centering
\includegraphics[width=\columnwidth,clip=true]{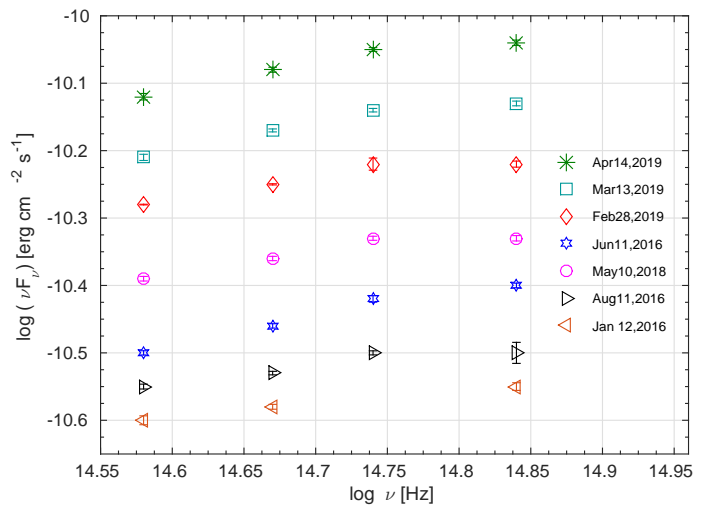}
\caption{Optical SEDs of \pg\ for selected brightness levels (see text).} 
\label{fig:sed2}
\end{figure}

To further understand spectral variations in the source, we studied its multi-band optical SEDs.
In Fig.\,\ref{fig:sed2}, we have generated optical SEDs of \pg\ considering all dates when we have quasi-simultaneous
observations in at least three optical passbands.
To display SED variations during our observing campaign, we selected different states of the source as follows: an outburst state (March 13 and April 4, 2019), intermediate state (June 11, 2016, May 10, 2018, and February 28, 2019) and low
state (January 12 and August 11, 2016). These seven SEDs display an increasing trend, which
could hint towards the fact that the synchrotron peak is located above 10$^{15}$\,Hz. This is in
accordance with the HBL classification. Therefore, based upon our optical study, we can conclude that the source behaves more like an HBL rather than an LBL. However, these optical SEDs cover a
narrow frequency range, therefore further multi-wavelength observations will offer a more powerful diagnosis of the true class of the source.

The SEDs do not show any significant shape change as the mean flux level varies.
This lack of variations in the SED shapes allows us to build the mean SED over the time interval covered by our monitoring campaign using the weighted mean magnitudes listed in Table\,\ref{tab:ltv}. To get the mean spectral index, we fitted a straight line of the form $\log(F_{\nu}) = -\alpha\,\log(\nu) + const$ to the mean SED. We got $\alpha = 0.89 \pm 0.06$ with $\chi^2_{\rm df}=1.1$, a result consistent with those of \citet{1994ApJS...93..125F} and \citet{2019ApJ...871..192P}.

\section{Discussion and conclusions}
\label{sect:disc}

In this paper, we have presented the results from the optical monitoring of the blazar \pg\ from 2016 to 2019 on intra-night and long-term timescales. We have also constructed and analysed the historical 2005--2019 $VR$-band LCs of the blazar.

In the course of our monitoring campaign, we recorded in April 2019 the brightest state of \pg\ for the period from 2005 to 2019, covered by the historical LC (see Figs.\,\ref{fig:ourLC} and \ref{fig:fullLC}): $R \simeq 13.2\,\rm mag$. 
In general, the variations of the \pg\ flux on both long and short timescales are achromatic, except the 2019 flare. The spectral index that characterizes the mean SED of the blazar over the period from 2005 to 2019 is $\alpha = 0.89 \pm 0.06$.
We shall discuss below some constraints on the jet parameters imposed by our observing results.

\subsection{Intra-night variability}

From the analysis of our and literature INM data during a total of 74 nights, we found a low value of the \pg\ DC: 10.8\% if the probably variable cases
are considered non-variable and 17.6\%~-- otherwise. This result agrees with the low DC reported for HBLs \citep[e.g.][]{2014RMxAC..44...95A}. The low value of the DC imposes some constraints on the physical jet parameters.

Let us assume that the INV in blazars is associated with interactions of shocks with small scale structures~-- like density inhomogeneities or eddies~-- in the otherwise steady jet flow. Then the low DC value of HBLs could be explained with the strong magnetic field present in this sub-class of blazars \citep[e.g.][]{1996ApJ...463..444S}.
\citet{1995Ap&SS.234...49R} found that the axial magnetic field halts the build-up of Kelvin-Helmholtz instabilities in the sub-parsec to parsec scale jets when the field value exceeds the critical value defined as:
\begin{equation}
{\mathcal B}_{\rm c} = \left[4 \pi n_{\rm e} m_{\rm e} c^2 \left(\Gamma^{\,2} - 1\right)\right]^{1/2}\Gamma^{\,-1},
\end{equation}
where $n_{\rm e}$ is the local electron density, $m_{\rm e}$ the rest mass of the electron, and $\Gamma$ the bulk Lorentz factor of the flow. 
Therefore, we have that, in case of HBLs, there will be no instabilities when ${\mathcal B} > {\mathcal B}_{\rm c}$, which in turn reduces the incidence of rapid variability in their optical LCs.

\subsection{The 2019 flare}

The 2019 flare marks the brightest state of \pg\ for the period from 2005 to 2019. The analysis of the flare $VR$-band LCs revealed a clockwise hysteresis loop and a soft lag, which means that the synchrotron cooling of accelerated electrons is the dominant emission mechanism of the flare \citep{1998A&A...333..452K}. 

The flare has a slight asymmetry, with the rising part being steeper. This asymmetry, however, is questionable because of the putative sub-flare at MJD $\sim$8610, which is not well resolved in order to be accounted for in the proper way. 
So, for further analysis, we shall assume that the 2019 flare is symmetric with the decay timescale equal to the rise one. This means that the cooling time is smaller than or comparable to the light crossing time, $t_{\rm cool}\la {\mathcal R}/c$.

\subsubsection{Radius of the emitting region}

Under the assumption that the injection time is less than the light crossing time, the rising part of the flare LC could be used to constrain the radius of the emitting region:
$
{\mathcal R} \la c {\mathcal T}_{\rm ris} \delta / (1+z),
$
where $\delta$ is the Doppler factor and $z$ the redshift.
Using a rise timescale of 24.9\,days, which is the weighted mean over the $V$- and $R$-band rise timescales, we got a radius of ${\mathcal R} \la 4.5\times10^{16} \delta\,\rm cm$. 

Regarding the Doppler factor of \pg, there exist two sets of values. The first set is based on the multi-wavelength SED modelling~-- $\delta$ varies from 20 \citep{2010MNRAS.401.1570T} to 40 \citep{2015MNRAS.450.4399A}. The second, single-value set is based on the Very Long Baseline Array observations~-- $\delta$ is equal to 1.4 \citep{2020A&A...634A..87L}. This discrepancy is typical for TeV HBLs and reflects the so called `Doppler crisis' \citep{2006tmgm.meet..512T}. 
The `Doppler crisis' can be resolved if one assumes the presence of the velocity gradient either in the axial or in the transverse direction \citep[for details see, e.g.][and references therein]{2018mgm..conf.3074P}. This assumption could reduce the Doppler factor that characterizes the one-zone synchrotron self-Compton models. 

The typical radius of the emitting region, used in the SED modelling, is ${\mathcal R}\la 10^{17}\,\rm cm$ \citep[e.g.][]{2015MNRAS.450.4399A,2019MNRAS.487..845B}. {Therefore,} the Doppler factor must be $\delta\la 3$ given our estimate of the weighted mean rise timescale. 
The Doppler factor of 20--40, as derived from the SED modelling, would result in a too large radius, ${\mathcal R}\simeq (9\!-\!18)\times10^{17}\,\rm cm$.

\subsubsection{Magnetic field strength and electron energies in the emitting region}

Assuming $t_{\rm cool}\la {\mathcal R}/c$, we could estimate the lower limit of the magnetic field strength in the emitting region using the rise timescales listed in Table\,\ref{tab:fit} \citep{1997A&A...327...61G}:
\begin{equation}
{\mathcal B}\delta^{1/3}\ga 0.67\,\nu_{15}^{-1/3}\,{\mathcal T}_{\rm ris}^{\,-2/3}\,(1+U_{\rm S}/U_{\rm B})^{\,-2/3}\,(1+z)^{1/3},
\end{equation}
where ${\mathcal B}$ is in units of Gauss, ${\mathcal T}_{\rm ris}$ is in units of days, $\nu_{15}$ is the observed frequency (in units of $10^{15}\,\rm Hz$), $U_{\rm S}$ the radiation energy density of the synchrotron emission, and $U_{\rm B}$ the magnetic energy density; the energy densities are measured in the comoving frame and
we assume that $U_{\rm S}=U_{\rm B}$.
We got ${\mathcal B}\delta^{1/3}\ga 0.071\,\rm G$ for the $V$-band and ${\mathcal B}\delta^{1/3}\ga 0.072\,\rm G$ for the $R$-band. So, if we use Doppler factors of 3 and 20, then the respective lower limits are 0.05\,G and 0.03\,G and they are the same for both $VR$-bands.

An independent way to estimate the magnetic field strength is to use the soft time lag and the following expression \citep[e.g.][]{2003A&A...397..565P}:
\begin{equation}
{\mathcal B}\delta^{1/3}\simeq 300\,\left(\frac{1+z}{\nu_R}\right)^{1/3}\,\left[\frac{1-(\nu_R/\nu_V)^{1/2}}{{\mathcal T}}\right]^{2/3},
\end{equation}
where ${\mathcal B}$ is in units of Gauss, $\nu_V$ and $\nu_R$ are the frequencies of the $VR$-bands (in units of $10^{17}$\,Hz) and ${\mathcal T}$ the soft time lag (in units of seconds). Here it is assumed that the lag resulted from synchrotron cooling of high-energy electrons to lower energies \citep{1999ApJ...521..552C}.
We got ${\mathcal B}\delta^{1/3}\simeq 0.066\,\rm G$ if ${\mathcal T}={\mathcal T}_{\rm CCPD}$ and ${\mathcal B}\delta^{1/3}\simeq 0.100\,\rm G$ if ${\mathcal T}={\mathcal T}_{\rm fit}$. If we use Doppler factors of 3 and 20 then the respective values are 0.05\,G and 0.02\,G using ${\mathcal T}_{\rm CCPD}$ and 0.07\,G and 0.04\,G using ${\mathcal T}_{\rm fit}$.
The magnetic field strengths estimated in both ways agree well to each other as well as with the values obtained by \citet{2010MNRAS.401.1570T} for TeV blazars by means of SED modelling.

Finally, we obtained the lower limit of the energy, $\gamma_{\rm obs}$ (in units of $m_{\rm e}c^2$), of the electrons emitting at the observed frequencies corresponding to $VR$-bands \citep{1997A&A...327...61G}:
\begin{equation}
\gamma_{\rm obs}\delta^{1/3}\la 2\times10^4\,\nu_{15}^{2/3}\,[{\mathcal T}_{\rm ris}\,(1+U_{\rm S}/U_{\rm B})\,(1+z)]^{1/3}.
\end{equation}
We got $\gamma_{\rm obs}\delta^{1/3}\la 5.5\times10^4$ for the $V$-band and $\gamma_{\rm obs}\delta^{1/3}\la 5.0\times10^4$ for the $R$-band. So, if we use Doppler factors of 3 and 20 then the respective upper limits are $3.8\times10^4$ and $2.0\times10^4$ for the $V$-band and $3.5\times10^4$ and $1.8\times10^4$ for the $R$-band.

\subsection{Periodicity analysis}

Our periodicity analysis of the historical $VR$-band LCs using three different methods gives a median (over the methods and $VR$-bands) period of $[2.21 \pm 0.04\,\rm(s.d.)]$\,years, which confirms the previous findings. 
In addition, our results increase the number of the wavelengths at which the year-long period is detected. This fact, coupled with the findings of \citet{2018A&A...615A.118S} and \citet{2020ApJ...895..122C} strongly suggest that this period is real feature in the \pg\ variability pattern.

The year-long periodicity in the \pg\ emission could be explained 
by the jet precession in a SMBH binary system \citep{2017MNRAS.465..161S,2017ApJ...851L..39C,2018ApJ...854...11T}. The precession will change the jet viewing angle and, hence, the corresponding Doppler factor. In the case of a continuous jet, the observed flux is related to the intrinsic one via $F^{(\rm obs)} = \delta^{\,(\alpha+2)} F$. 
If the viewing angle changes because of the jet precession then $\delta=\delta(t)$ and, hence, $F=F(t)$. We already assumed that the mean spline, derived in Sect.\,\ref{sect:res:ltv:cmd}, represents the long-term variability component of geometric origin (jet precession). 
So, to account for the flux changes of the mean spline the Doppler factor has to be changed up to 1.3 times relative to its value corresponding to the minimum spline flux.

We also detected a possible secondary period of $\sim$210\,days using the LCs corrected for the long-term
variations. As long as the flux variations of the corrected LCs are achromatic, we should rely on geometric causes of this periodicity, for example, the emitting region follows a spiral path within the jet or the jet itself has a helical geometry. The jet wobbling in \pg, recently found by \citet{2020A&A...634A..87L} could contribute to some extent to the periodic STV.
It is worth noting that \citet{2016ApJ...832...47B} found a secondary period of $\sim$400\,days for the blazar OJ\,287 in addition to the main one of $\sim$12 years.

The update and further analysis of the historical LCs are of substantial importance in order to confirm or reject the putative secondary period in \pg. 
The observed periodicity (if we assume it is real) coupled with the observed achromatic flux variation support the geometric origin of the STV and LTV of \pg.

\begin{acknowledgements}
We thank the anonymous referee whose comments improved the paper.
BM and LSM are supported by the Bulgarian National Science Fund of the Ministry of Education and
Science under grant DN 18/13-2017.
SZ acknowledges NCN grant No. 1028/29/B/ST9/01793.
We thank TUBITAK National Observatory for partial support in using T60 and T100 telescopes with project number  1505 and 1486, respectively. AO was supported by the Scientific Research Project Coordination Unit of Ataturk University, Project ID 8418.
AR acknowledges the Research
Associate Fellowship with Order No. 03(1428)/18/EMR-II
under the Council of Scientific and Industrial Research (CSIR).
Data from the Steward Observatory spectropolarimetric monitoring project were used. This program is supported by Fermi Guest Investigator grants NNX08AW56G, NNX09AU10G, NNX12AO93G, and NNX15AU81G.
Based on observations obtained with the Samuel Oschin 48-inch Telescope at the Palomar Observatory as part of the Zwicky Transient Facility project. ZTF is supported by the National Science Foundation under Grant No. AST-1440341 and collaboration including Caltech, IPAC, the Weizmann Institute for Science, the Oskar Klein Center at Stockholm University, the University of Maryland, the University of Washington, Deutsches Elektronen-Synchrotron and Humboldt University, Los Alamos National Laboratories, the TANGO Consortium of Taiwan, the University of Wisconsin at Milwaukee, and Lawrence Berkeley National Laboratories. Operations are conducted by COO, IPAC, and UW. 
The iPTF project is a scientific collaboration between Caltech; Los Alamos National Laboratory; the University of Wisconsin, Milwaukee; the Oskar Klein Centre in Sweden; the Weizmann Institute of Science in Israel; the TANGO Program of the University System of Taiwan; and the Kavli Institute for the Physics and Mathematics of the Universe in Japan.
The CSS survey is funded by the National Aeronautics and Space
Administration under Grant No. NNG05GF22G issued through the Science
Mission Directorate Near-Earth Objects Observations Program.  The CRTS
survey is supported by the U.S.~National Science Foundation under
grants AST-0909182.
Based on data acquired at Complejo Astron{\'o}mico El
Leoncito, operated under agreement between the Consejo Nacional de
Investigaciones Cient{\'i}ficas y T{\'e}cnicas de la Rep{\'u}blica Argentina and the National Universities of La Plata, C{\'o}rdoba and San Juan (proposals JS-2019A-10, JS-2019A-16, HSH-2018B-03, and Staff time).
This research has made use of the NASA/IPAC Extragalactic Database (NED), which is funded by the National Aeronautics and Space Administration and operated by the California Institute of Technology.
\end{acknowledgements}

\bibliographystyle{aa}
\bibliography{PG1553}

\begin{thebibliography}{105}
\expandafter\ifx\csname natexlab\endcsname\relax\def\natexlab#1{#1}\fi

\bibitem[{{Abdo} {et~al.}(2010){Abdo}, {Ackermann}, {Agudo}, {Ajello}, {Aller},
  {Aller}, {Angelakis}, {Arkharov}, {Axelsson}, {Bach}, {Baldini}, {Ballet},
  {Barbiellini}, {Bastieri}, {Baughman}, {Bechtol}, {Bellazzini}, {Benitez},
  {Berdyugin}, {Berenji}, {Bland ford}, {Bloom}, {Boettcher}, {Bonamente},
  {Borgland}, {Bregeon}, {Brez}, {Brigida}, {Bruel}, {Burnett}, {Burrows},
  {Buson}, {Caliandro}, {Calzoletti}, {Cameron}, {Capalbi}, {Caraveo},
  {Carosati}, {Casand jian}, {Cavazzuti}, {Cecchi}, {{\c{C}}elik}, {Charles},
  {Chaty}, {Chekhtman}, {Chen}, {Chiang}, {Chincarini}, {Ciprini}, {Claus},
  {Cohen-Tanugi}, {Colafrancesco}, {Cominsky}, {Conrad}, {Costamante},
  {Cutini}, {D'ammando}, {Deitrick}, {D'Elia}, {Dermer}, {de Angelis}, {de
  Palma}, {Digel}, {Donnarumma}, {Silva}, {Drell}, {Dubois}, {Dultzin},
  {Dumora}, {Falcone}, {Farnier}, {Favuzzi}, {Fegan}, {Focke}, {Forn{\'e}},
  {Fortin}, {Frailis}, {Fuhrmann}, {Fukazawa}, {Funk}, {Fusco}, {G{\'o}mez},
  {Gargano}, {Gasparrini}, {Gehrels}, {Germani}, {Giebels}, {Giglietto},
  {Giommi}, {Giordano}, {Giuliani}, {Glanzman}, {Godfrey}, {Grenier},
  {Gronwall}, {Grove}, {Guillemot}, {Guiriec}, {Gurwell}, {Hadasch},
  {Hanabata}, {Harding}, {Hayashida}, {Hays}, {Healey}, {Heidt}, {Hiriart},
  {Horan}, {Hoversten}, {Hughes}, {Itoh}, {Jackson}, {J{\'o}hannesson},
  {Johnson}, {Johnson}, {Jorstad}, {Kadler}, {Kamae}, {Katagiri}, {Kataoka},
  {Kawai}, {Kennea}, {Kerr}, {Kimeridze}, {Kn{\"o}dlseder}, {Kocian},
  {Kopatskaya}, {Koptelova}, {Konstantinova}, {Kovalev}, {Kovalev},
  {Kurtanidze}, {Kuss}, {Lande}, {Larionov}, {Latronico}, {Leto}, {Lindfors},
  {Longo}, {Loparco}, {Lott}, {Lovellette}, {Lubrano}, {Madejski}, {Makeev},
  {Marchegiani}, {Marscher}, {Marshall}, {Max-Moerbeck}, {Mazziotta},
  {McConville}, {McEnery}, {Meurer}, {Michelson}, {Mitthumsiri}, {Mizuno},
  {Moiseev}, {Monte}, {Monzani}, {Morselli}, {Moskalenko}, {Murgia},
  {Nestoras}, {Nilsson}, {Nizhelsky}, {Nolan}, {Norris}, {Nuss}, {Ohsugi},
  {Ojha}, {Omodei}, {Orlando}, {Ormes}, {Osborne}, {Ozaki}, {Pacciani},
  {Padovani}, {Pagani}, {Page}, {Paneque}, {Panetta}, {Parent}, {Pasanen},
  {Pavlidou}, {Pelassa}, {Pepe}, {Perri}, {Pesce-Rollins}, {Piranomonte},
  {Piron}, {Pittori}, {Porter}, {Puccetti}, {Rahoui}, {Rain{\`o}}, {Raiteri},
  {Rando}, {Razzano}, {Reimer}, {Reimer}, {Reposeur}, {Richards}, {Ritz},
  {Rochester}, {Rodriguez}, {Romani}, {Ros}, {Roth}, {Roustazadeh}, {Ryde},
  {Sadrozinski}, {Sadun}, {Sanchez}, {Sander}, {Saz Parkinson}, {Scargle},
  {Sellerholm}, {Sgr{\`o}}, {Shaw}, {Sigua}, {Siskind}, {Smith}, {Smith},
  {Spandre}, {Spinelli}, {Starck}, {Stevenson}, {Stratta}, {Strickman},
  {Suson}, {Tajima}, {Takahashi}, {Takahashi}, {Takalo}, {Tanaka}, {Thayer},
  {Thayer}, {Thompson}, {Tibaldo}, {Torres}, {Tosti}, {Tramacere}, {Uchiyama},
  {Usher}, {Vasileiou}, {Verrecchia}, {Vilchez}, {Villata}, {Vitale}, {Waite},
  {Wang}, {Winer}, {Wood}, {Ylinen}, {Zensus}, {Zhekanis}, \&
  {Ziegler}}]{2010ApJ...716...30A}
{Abdo}, A.~A., {Ackermann}, M., {Agudo}, I., {et~al.} 2010, \apj, 716, 30

\bibitem[{{Ackermann} {et~al.}(2015){Ackermann}, {Ajello}, {Albert}, {Atwood},
  {Baldini}, {Ballet}, {Barbiellini}, {Bastieri}, {Becerra Gonzalez},
  {Bellazzini}, {Bissaldi}, {Blandford}, {Bloom}, {Bonino}, {Bottacini},
  {Bregeon}, {Bruel}, {Buehler}, {Buson}, {Caliandro}, {Cameron}, {Caputo},
  {Caragiulo}, {Caraveo}, {Cavazzuti}, {Cecchi}, {Chekhtman}, {Chiang},
  {Chiaro}, {Ciprini}, {Cohen-Tanugi}, {Conrad}, {Cutini}, {D'Ammand o}, {de
  Angelis}, {de Palma}, {Desiante}, {Di Venere}, {Dom{\'\i}nguez}, {Drell},
  {Favuzzi}, {Fegan}, {Ferrara}, {Focke}, {Fuhrmann}, {Fukazawa}, {Fusco},
  {Gargano}, {Gasparrini}, {Giglietto}, {Giommi}, {Giordano}, {Giroletti},
  {Godfrey}, {Green}, {Grenier}, {Grove}, {Guiriec}, {Harding}, {Hays},
  {Hewitt}, {Hill}, {Horan}, {Jogler}, {J{\'o}hannesson}, {Johnson}, {Kamae},
  {Kuss}, {Larsson}, {Latronico}, {Li}, {Li}, {Longo}, {Loparco}, {Lott},
  {Lovellette}, {Lubrano}, {Magill}, {Maldera}, {Manfreda}, {Max-Moerbeck},
  {Mayer}, {Mazziotta}, {McEnery}, {Michelson}, {Mizuno}, {Monzani},
  {Morselli}, {Moskalenko}, {Murgia}, {Nuss}, {Ohno}, {Ohsugi}, {Ojha},
  {Omodei}, {Orlando}, {Ormes}, {Paneque}, {Pearson}, {Perkins}, {Perri},
  {Pesce-Rollins}, {Petrosian}, {Piron}, {Pivato}, {Porter}, {Rain{\`o}},
  {Rando}, {Razzano}, {Readhead}, {Reimer}, {Reimer}, {Schulz}, {Sgr{\`o}},
  {Siskind}, {Spada}, {Spandre}, {Spinelli}, {Suson}, {Takahashi}, {Thayer},
  {Thompson}, {Tibaldo}, {Torres}, {Tosti}, {Troja}, {Uchiyama}, {Vianello},
  {Wood}, {Wood}, {Zimmer}, {Berdyugin}, {Corbet}, {Hovatta}, {Lindfors},
  {Nilsson}, {Reinthal}, {Sillanp{\"a}{\"a}}, {Stamerra}, {Takalo}, \&
  {Valtonen}}]{2015ApJ...813L..41A}
{Ackermann}, M., {Ajello}, M., {Albert}, A., {et~al.} 2015, \apjl, 813, L41

\bibitem[{{Agarwal} {et~al.}(2019){Agarwal}, {Cellone}, {Andruchow}, {Mammana},
  {Singh}, {Anupama}, {Mihov}, {Raj}, {Slavcheva-Mihova}, {{\"O}zd{\"o}nmez},
  \& {Ege}}]{2019MNRAS.488.4093A}
{Agarwal}, A., {Cellone}, S.~A., {Andruchow}, I., {et~al.} 2019, \mnras, 488,
  4093

\bibitem[{{Agarwal} \& {Gupta}(2015)}]{2015MNRAS.450..541A}
{Agarwal}, A. \& {Gupta}, A.~C. 2015, \mnras, 450, 541

\bibitem[{{Agarwal} {et~al.}(2017){Agarwal}, {Mohan}, {Gupta}, {Mangalam},
  {Volvach}, {Aller}, {Aller}, {Gu}, {L{\"a}hteenm{\"a}ki}, {Tornikoski}, \&
  {Volvach}}]{2017MNRAS.469..813A}
{Agarwal}, A., {Mohan}, P., {Gupta}, A.~C., {et~al.} 2017, \mnras, 469, 813

\bibitem[{{Aharonian} {et~al.}(2006){Aharonian}, {Akhperjanian}, {Bazer-Bachi},
  {Beilicke}, {Benbow}, {Berge}, {Bernl{\"o}hr}, {Boisson}, {Bolz}, {Borrel},
  {Braun}, {Breitling}, {Brown}, {B{\"u}hler}, {Carrigan}, {Chadwick},
  {Chounet}, {Cornils}, {Costamante}, {Degrange}, {Dickinson},
  {Djannati-Ata{\"\i}}, {O'C. Drury}, {Dubus}, {Egberts}, {Emmanoulopoulos},
  {Espigat}, {Feinstein}, {Fontaine}, {Funk}, {Gallant}, {Giebels},
  {Glicenstein}, {Goret}, {Hadjichristidis}, {Hauser}, {Hauser}, {Heinzelmann},
  {Henri}, {Hermann}, {Hinton}, {Hofmann}, {Holleran}, {Horns}, {Jacholkowska},
  {de Jager}, {Kh{\'e}lifi}, {Komin}, {Konopelko}, {Latham}, {Le Gallou},
  {Lemi{\`e}re}, {Lemoine-Goumard}, {Lohse}, {Martin}, {Martineau-Huynh},
  {Marcowith}, {Masterson}, {McComb}, {de Naurois}, {Nedbal}, {Nolan},
  {Noutsos}, {Orford}, {Osborne}, {Ouchrif}, {Panter}, {Pelletier}, {Pita},
  {P{\"u}hlhofer}, {Punch}, {Raubenheimer}, {Raue}, {Rayner}, {Reimer},
  {Reimer}, {Ripken}, {Rob}, {Rolland}, {Rowell}, {Sahakian}, {Saug{\'e}},
  {Schlenker}, {Schlickeiser}, {Schuster}, {Schwanke}, {Siewert}, {Sol},
  {Spangler}, {Steenkamp}, {Stegmann}, {Superina}, {Tavernet}, {Terrier},
  {Th{\'e}oret}, {Tluczykont}, {van Eldik}, {Vasileiadis}, {Venter}, {Vincent},
  {V{\"o}lk}, {Wagner}, \& {Ward}}]{2006A&A...448L..19A}
{Aharonian}, F., {Akhperjanian}, A.~G., {Bazer-Bachi}, A.~R., {et~al.} 2006,
  \aap, 448, L19

\bibitem[{{Ait Benkhali} {et~al.}(2020){Ait Benkhali}, {Hofmann}, {Rieger}, \&
  {Chakraborty}}]{2020A&A...634A.120A}
{Ait Benkhali}, F., {Hofmann}, W., {Rieger}, F.~M., \& {Chakraborty}, N. 2020,
  \aap, 634, A120

\bibitem[{{Aleksi{\'c}} {et~al.}(2015){Aleksi{\'c}}, {Ansoldi}, {Antonelli},
  {Antoranz}, {Babic}, {Bangale}, {Barrio}, {Becerra Gonz{\'a}lez}, {Bednarek},
  {Bernardini}, {Biasuzzi}, {Biland}, {Blanch}, {Bonnefoy}, {Bonnoli},
  {Borracci}, {Bretz}, {Carmona}, {Carosi}, {Colin}, {Colombo}, {Contreras},
  {Cortina}, {Covino}, {da Vela}, {Dazzi}, {de Angelis}, {de Caneva}, {de
  Lotto}, {de O{\~n}a Wilhelmi}, {Delgado Mendez}, {Dominis Prester}, {Dorner},
  {Doro}, {Einecke}, {Eisenacher}, {Elsaesser}, {Fidalgo}, {Fonseca}, {Font},
  {Frantzen}, {Fruck}, {Galindo}, {Garc{\'\i}a L{\'o}pez}, {Garczarczyk},
  {Garrido Terrats}, {Gaug}, {Godinovi{\'c}}, {Gonz{\'a}lez Mu{\~n}oz},
  {Gozzini}, {Hadasch}, {Hanabata}, {Hayashida}, {Herrera}, {Hildebrand},
  {Hose}, {Hrupec}, {Idec}, {Kadenius}, {Kellermann}, {Knoetig}, {Kodani},
  {Konno}, {Krause}, {Kubo}, {Kushida}, {La Barbera}, {Lelas}, {Lewandowska},
  {Lindfors}, {Lombardi}, {Longo}, {L{\'o}pez}, {L{\'o}pez-Coto},
  {L{\'o}pez-Oramas}, {Lorenz}, {Lozano}, {Makariev}, {Mallot}, {Maneva},
  {Mannheim}, {Maraschi}, {Marcote}, {Mariotti}, {Mart{\'\i}nez}, {Mazin},
  {Menzel}, {Mirand a}, {Mirzoyan}, {Moralejo}, {Munar-Adrover}, {Nakajima},
  {Neustroev}, {Niedzwiecki}, {Nilsson}, {Nishijima}, {Noda}, {Orito},
  {Overkemping}, {Paiano}, {Palatiello}, {Paneque}, {Paoletti}, {Paredes},
  {Paredes-Fortuny}, {Persic}, {Poutanen}, {Prada Moroni}, {Prandini},
  {Puljak}, {Reinthal}, {Rhode}, {Rib{\'o}}, {Rico}, {Rodriguez Garcia},
  {R{\"u}gamer}, {Saito}, {Saito}, {Satalecka}, {Scalzotto}, {Scapin},
  {Schultz}, {Schweizer}, {Sillanp{\"a}{\"a}}, {Sitarek}, {Snidaric},
  {Sobczynska}, {Spanier}, {Stamerra}, {Steinbring}, {Storz}, {Strzys},
  {Takalo}, {Takami}, {Tavecchio}, {Temnikov}, {Terzi{\'c}}, {Tescaro},
  {Teshima}, {Thaele}, {Tibolla}, {Torres}, {Toyama}, {Treves}, {Vogler},
  {Will}, {Zanin}, {MAGIC Collaboration}, {D'Ammando}, {Buson},
  {L{\"a}hteenm{\"a}ki}, {Tornikoski}, {Hovatta}, {Readhead}, {Max-Moerbeck},
  \& {Richards}}]{2015MNRAS.450.4399A}
{Aleksi{\'c}}, J., {Ansoldi}, S., {Antonelli}, L.~A., {et~al.} 2015, \mnras,
  450, 4399

\bibitem[{{Alexander}(1997)}]{1997ASSL..218..163A}
{Alexander}, T. 1997, Astrophysics and Space Science Library, Vol. 218, {Is AGN
  Variability Correlated with Other AGN Properties? ZDCF Analysis of Small
  Samples of Sparse Light Curves}, ed. D.~{Maoz}, A.~{Sternberg}, \& E.~M.
  {Leibowitz}, 163

\bibitem[{{Alexander}(2014)}]{2014ascl.soft04002A}
{Alexander}, T. 2014, {ZDCF: Z-Transformed Discrete Correlation Function}

\bibitem[{{Andruchow} {et~al.}(2007){Andruchow}, {Cellone}, \&
  {Romero}}]{2007BAAA...50..299A}
{Andruchow}, I., {Cellone}, S.~A., \& {Romero}, G.~E. 2007, Bolet{\'i}n de la
  Asociaci{\'o}n Argentina de Astronom{\'i}a, 50, 299

\bibitem[{{Andruchow} {et~al.}(2014){Andruchow}, {Cellone}, \&
  {Romero}}]{2014RMxAC..44...95A}
{Andruchow}, I., {Cellone}, S.~A., \& {Romero}, G.~E. 2014, in Revista Mexicana
  de Astronom{\'i}a y Astrof{\'i}sica Conference Series, Vol.~44, 95

\bibitem[{{Andruchow} {et~al.}(2011){Andruchow}, {Combi},
  {Mu{\~n}oz-Arjonilla}, {Romero}, {Cellone}, \&
  {Mart{\'\i}}}]{2011A&A...531A..38A}
{Andruchow}, I., {Combi}, J.~A., {Mu{\~n}oz-Arjonilla}, A.~J., {et~al.} 2011,
  \aap, 531, A38

\bibitem[{{Bachev}(2018)}]{2018BlgAJ..28...22B}
{Bachev}, R. 2018, Bulgarian Astronomical Journal, 28, 22

\bibitem[{{Banerjee} {et~al.}(2019){Banerjee}, {Joshi}, {Majumdar},
  {Williamson}, {Jorstad}, \& {Marscher}}]{2019MNRAS.487..845B}
{Banerjee}, B., {Joshi}, M., {Majumdar}, P., {et~al.} 2019, \mnras, 487, 845

\bibitem[{{Bessell}(1990)}]{1990PASP..102.1181B}
{Bessell}, M.~S. 1990, \pasp, 102, 1181

\bibitem[{{Bessell} {et~al.}(1998){Bessell}, {Castelli}, \&
  {Plez}}]{1998A&A...333..231B}
{Bessell}, M.~S., {Castelli}, F., \& {Plez}, B. 1998, \aap, 333, 231

\bibitem[{{Bhatta} {et~al.}(2016){Bhatta}, {Zola}, {Stawarz}, {Ostrowski},
  {Winiarski}, {Og{\l}oza}, {Dr{\'o}{\.z}d{\.z}}, {Siwak}, {Liakos},
  {Kozie{\l}-Wierzbowska}, {Gazeas}, {Debski}, {Kundera}, {Stachowski}, \&
  {Paliya}}]{2016ApJ...832...47B}
{Bhatta}, G., {Zola}, S., {Stawarz}, {\L}., {et~al.} 2016, \apj, 832, 47

\bibitem[{{Bonev}(2011)}]{2011gfun.conf...89B}
{Bonev}, T. 2011, in Gaia follow-up network for the solar system objects : Gaia
  FUN-SSO workshop proceedings, 89--92

\bibitem[{{Caproni} {et~al.}(2017){Caproni}, {Abraham}, {Motter}, \&
  {Monteiro}}]{2017ApJ...851L..39C}
{Caproni}, A., {Abraham}, Z., {Motter}, J.~C., \& {Monteiro}, H. 2017, \apjl,
  851, L39

\bibitem[{{Cellone} {et~al.}(2007){Cellone}, {Romero}, \&
  {Araudo}}]{2007MNRAS.374..357C}
{Cellone}, S.~A., {Romero}, G.~E., \& {Araudo}, A.~T. 2007, \mnras, 374, 357

\bibitem[{{Chiappetti} {et~al.}(1999){Chiappetti}, {Maraschi}, {Tavecchio},
  {Celotti}, {Fossati}, {Ghisellini}, {Giommi}, {Pian}, {Tagliaferri},
  {Treves}, {Urry}, \& {Zhang}}]{1999ApJ...521..552C}
{Chiappetti}, L., {Maraschi}, L., {Tavecchio}, F., {et~al.} 1999, \apj, 521,
  552

\bibitem[{{Clements} {et~al.}(2003){Clements}, {Jenks}, \&
  {Torres}}]{2003AJ....126...37C}
{Clements}, S.~D., {Jenks}, A., \& {Torres}, Y. 2003, \aj, 126, 37

\bibitem[{{Covino} {et~al.}(2020){Covino}, {Landoni}, {Sandrinelli}, \&
  {Treves}}]{2020ApJ...895..122C}
{Covino}, S., {Landoni}, M., {Sandrinelli}, A., \& {Treves}, A. 2020, \apj,
  895, 122

\bibitem[{{Covino} {et~al.}(2019){Covino}, {Sandrinelli}, \&
  {Treves}}]{2019MNRAS.482.1270C}
{Covino}, S., {Sandrinelli}, A., \& {Treves}, A. 2019, \mnras, 482, 1270

\bibitem[{{Cutini} {et~al.}(2016){Cutini}, {Ciprini}, {Stamerra}, {Thompson},
  \& {Perri}}]{2016agnt.confE..58C}
{Cutini}, S., {Ciprini}, S., {Stamerra}, A., {Thompson}, D.~J., \& {Perri}, M.
  2016, in Active Galactic Nuclei 12: A Multi-Messenger Perspective (AGN12), 58

\bibitem[{{Drake} {et~al.}(2009){Drake}, {Djorgovski}, {Mahabal}, {Beshore},
  {Larson}, {Graham}, {Williams}, {Christensen}, {Catelan}, {Boattini},
  {Gibbs}, {Hill}, \& {Kowalski}}]{2009ApJ...696..870D}
{Drake}, A.~J., {Djorgovski}, S.~G., {Mahabal}, A., {et~al.} 2009, \apj, 696,
  870

\bibitem[{{Edelson} \& {Krolik}(1988)}]{1988ApJ...333..646E}
{Edelson}, R.~A. \& {Krolik}, J.~H. 1988, \apj, 333, 646

\bibitem[{{Falomo} {et~al.}(1994){Falomo}, {Scarpa}, \&
  {Bersanelli}}]{1994ApJS...93..125F}
{Falomo}, R., {Scarpa}, R., \& {Bersanelli}, M. 1994, \apjs, 93, 125

\bibitem[{{Falomo} \& {Treves}(1990)}]{1990PASP..102.1120F}
{Falomo}, R. \& {Treves}, A. 1990, \pasp, 102, 1120

\bibitem[{{Filippenko} {et~al.}(2001){Filippenko}, {Li}, {Treffers}, \&
  {Modjaz}}]{2001ASPC..246..121F}
{Filippenko}, A.~V., {Li}, W.~D., {Treffers}, R.~R., \& {Modjaz}, M. 2001,
  Astronomical Society of the Pacific Conference Series, Vol. 246, {The Lick
  Observatory Supernova Search with the Katzman Automatic Imaging Telescope},
  ed. B.~{Paczynski}, W.-P. {Chen}, \& C.~{Lemme}, 121

\bibitem[{{Gaur} {et~al.}(2012){Gaur}, {Gupta}, {Strigachev}, {Bachev},
  {Semkov}, {Wiita}, {Peneva}, {Boeva}, {Slavcheva-Mihova}, {Mihov}, {Latev},
  \& {Pandey}}]{2012MNRAS.425.3002G}
{Gaur}, H., {Gupta}, A.~C., {Strigachev}, A., {et~al.} 2012, \mnras, 425, 3002

\bibitem[{{Ghisellini} {et~al.}(1997){Ghisellini}, {Villata}, {Raiteri},
  {Bosio}, {de Francesco}, {Latini}, {Maesano}, {Massaro}, {Montagni}, {Nesci},
  {Tosti}, {Fiorucci}, {Pian}, {Maraschi}, {Treves}, {Comastri}, \&
  {Mignoli}}]{1997A&A...327...61G}
{Ghisellini}, G., {Villata}, M., {Raiteri}, C.~M., {et~al.} 1997, \aap, 327, 61

\bibitem[{{Gopal-Krishna} {et~al.}(2011){Gopal-Krishna}, {Goyal}, {Joshi},
  {Karthick}, {Sagar}, {Wiita}, {Anupama}, \& {Sahu}}]{2011MNRAS.416..101G}
{Gopal-Krishna}, {Goyal}, A., {Joshi}, S., {et~al.} 2011, \mnras, 416, 101

\bibitem[{{Green} {et~al.}(1986){Green}, {Schmidt}, \&
  {Liebert}}]{1986ApJS...61..305G}
{Green}, R.~F., {Schmidt}, M., \& {Liebert}, J. 1986, \apjs, 61, 305

\bibitem[{{Gupta} {et~al.}(2016){Gupta}, {Agarwal}, {Bhagwan}, {Strigachev},
  {Bachev}, {Semkov}, {Gaur}, {Damljanovic}, {Vince}, \&
  {Wiita}}]{2016MNRAS.458.1127G}
{Gupta}, A.~C., {Agarwal}, A., {Bhagwan}, J., {et~al.} 2016, \mnras, 458, 1127

\bibitem[{{Gupta} {et~al.}(2008){Gupta}, {Cha}, {Lee}, {Jin}, {Pak}, {Cho},
  {Moon}, {Park}, {Yuk}, {Nam}, \& {Kyeong}}]{2008AJ....136.2359G}
{Gupta}, A.~C., {Cha}, S.-M., {Lee}, S., {et~al.} 2008, \aj, 136, 2359

\bibitem[{{Heeschen} {et~al.}(1987){Heeschen}, {Krichbaum}, {Schalinski}, \&
  {Witzel}}]{1987AJ.....94.1493H}
{Heeschen}, D.~S., {Krichbaum}, T., {Schalinski}, C.~J., \& {Witzel}, A. 1987,
  \aj, 94, 1493

\bibitem[{{Howell} {et~al.}(1988){Howell}, {Mitchell}, \&
  {Warnock}}]{1988AJ.....95..247H}
{Howell}, S.~B., {Mitchell}, K.~J., \& {Warnock}, A., I. 1988, \aj, 95, 247

\bibitem[{{Ikejiri} {et~al.}(2011){Ikejiri}, {Uemura}, {Sasada}, {Ito},
  {Yamanaka}, {Sakimoto}, {Arai}, {Fukazawa}, {Ohsugi}, {Kawabata}, {Yoshida},
  {Sato}, \& {Kino}}]{2011PASJ...63..639I}
{Ikejiri}, Y., {Uemura}, M., {Sasada}, M., {et~al.} 2011, \pasj, 63, 639

\bibitem[{{Itoh} {et~al.}(2016){Itoh}, {Nalewajko}, {Fukazawa}, {Uemura},
  {Tanaka}, {Kawabata}, {Madejski}, {Schinzel}, {Kanda}, {Shiki}, {Akitaya},
  {Kawabata}, {Moritani}, {Nakaoka}, {Ohsugi}, {Sasada}, {Takaki}, {Takata},
  {Ui}, {Yamanaka}, \& {Yoshida}}]{2016ApJ...833...77I}
{Itoh}, R., {Nalewajko}, K., {Fukazawa}, Y., {et~al.} 2016, \apj, 833, 77

\bibitem[{{Jankowsky} \& {Wagner}(2019)}]{2019ATel12631....1J}
{Jankowsky}, F. \& {Wagner}, S. 2019, The Astronomer's Telegram, 12631, 1

\bibitem[{{Jayasinghe} {et~al.}(2019){Jayasinghe}, {Stanek}, {Kochanek},
  {Shappee}, {Holoien}, {Thompson}, {Prieto}, {Dong}, {Pawlak}, {Pejcha},
  {Shields}, {Pojmanski}, {Otero}, {Hurst}, {Britt}, \&
  {Will}}]{2019MNRAS.485..961J}
{Jayasinghe}, T., {Stanek}, K.~Z., {Kochanek}, C.~S., {et~al.} 2019, \mnras,
  485, 961

\bibitem[{{Jockers} {et~al.}(2000){Jockers}, {Credner}, {Bonev}, {Kisele},
  {Korsun}, {Kulyk}, {Rosenbush}, {Andrienko}, {Karpov}, {Sergeev}, \&
  {Tarady}}]{2000KFNTS...3...13J}
{Jockers}, K., {Credner}, T., {Bonev}, T., {et~al.} 2000, Kinematika i Fizika
  Nebesnykh Tel Supplement, 3, 13

\bibitem[{{Johnson} {et~al.}(2019){Johnson}, {Mulchaey}, {Chen}, {Wijers},
  {Connor}, {Muzahid}, {Schaye}, {Cen}, {Carlsten}, {Charlton}, {Drout},
  {Goulding}, {Hansen}, \& {Walth}}]{JMC2019}
{Johnson}, S.~D., {Mulchaey}, J.~S., {Chen}, H.-W., {et~al.} 2019, \apjl, 884,
  L31

\bibitem[{{Kinman}(1975)}]{1975IAUS...67..573K}
{Kinman}, T.~D. 1975, in IAU Symposium, Vol.~67, Variable Stars and Stellar
  Evolution, ed. V.~E. {Sherwood} \& L.~{Plaut}, 573

\bibitem[{{Kirk} {et~al.}(1998){Kirk}, {Rieger}, \&
  {Mastichiadis}}]{1998A&A...333..452K}
{Kirk}, J.~G., {Rieger}, F.~M., \& {Mastichiadis}, A. 1998, \aap, 333, 452

\bibitem[{{Kochanek} {et~al.}(2017){Kochanek}, {Shappee}, {Stanek}, {Holoien},
  {Thompson}, {Prieto}, {Dong}, {Shields}, {Will}, {Britt}, {Perzanowski}, \&
  {Pojma{\'n}ski}}]{2017PASP..129j4502K}
{Kochanek}, C.~S., {Shappee}, B.~J., {Stanek}, K.~Z., {et~al.} 2017, \pasp,
  129, 104502

\bibitem[{{K{\"o}nigl}(1981)}]{1981ApJ...243..700K}
{K{\"o}nigl}, A. 1981, \apj, 243, 700

\bibitem[{{Law} {et~al.}(2009){Law}, {Kulkarni}, {Dekany}, {Ofek}, {Quimby},
  {Nugent}, {Surace}, {Grillmair}, {Bloom}, {Kasliwal}, {Bildsten}, {Brown},
  {Cenko}, {Ciardi}, {Croner}, {Djorgovski}, {van Eyken}, {Filippenko}, {Fox},
  {Gal-Yam}, {Hale}, {Hamam}, {Helou}, {Henning}, {Howell}, {Jacobsen},
  {Laher}, {Mattingly}, {McKenna}, {Pickles}, {Poznanski}, {Rahmer}, {Rau},
  {Rosing}, {Shara}, {Smith}, {Starr}, {Sullivan}, {Velur}, {Walters}, \&
  {Zolkower}}]{2009PASP..121.1395L}
{Law}, N.~M., {Kulkarni}, S.~R., {Dekany}, R.~G., {et~al.} 2009, \pasp, 121,
  1395

\bibitem[{{Li} {et~al.}(2003){Li}, {Filippenko}, {Chornock}, \&
  {Jha}}]{2003PASP..115..844L}
{Li}, W., {Filippenko}, A.~V., {Chornock}, R., \& {Jha}, S. 2003, \pasp, 115,
  844

\bibitem[{{Lico} {et~al.}(2020){Lico}, {Liu}, {Giroletti}, {Orienti},
  {G{\'o}mez}, {Piner}, {MacDonald}, {D'Ammand o}, \&
  {Fuentes}}]{2020A&A...634A..87L}
{Lico}, R., {Liu}, J., {Giroletti}, M., {et~al.} 2020, \aap, 634, A87

\bibitem[{{Lomb}(1976)}]{1976Ap&SS..39..447L}
{Lomb}, N.~R. 1976, \apss, 39, 447

\bibitem[{{Maoz} \& {Netzer}(1989)}]{1989MNRAS.236...21M}
{Maoz}, D. \& {Netzer}, H. 1989, \mnras, 236, 21

\bibitem[{{Marscher} \& {Gear}(1985)}]{1985ApJ...298..114M}
{Marscher}, A.~P. \& {Gear}, W.~K. 1985, \apj, 298, 114

\bibitem[{{Masci} {et~al.}(2019){Masci}, {Laher}, {Rusholme}, {Shupe}, {Groom},
  {Surace}, {Jackson}, {Monkewitz}, {Beck}, {Flynn}, {Terek}, {Landry},
  {Hacopians}, {Desai}, {Howell}, {Brooke}, {Imel}, {Wachter}, {Ye}, {Lin},
  {Cenko}, {Cunningham}, {Rebbapragada}, {Bue}, {Miller}, {Mahabal}, {Bellm},
  {Patterson}, {Juri{\'c}}, {Golkhou}, {Ofek}, {Walters}, {Graham}, {Kasliwal},
  {Dekany}, {Kupfer}, {Burdge}, {Cannella}, {Barlow}, {Van Sistine}, {Giomi},
  {Fremling}, {Blagorodnova}, {Levitan}, {Riddle}, {Smith}, {Helou}, {Prince},
  \& {Kulkarni}}]{2019PASP..131a8003M}
{Masci}, F.~J., {Laher}, R.~R., {Rusholme}, B., {et~al.} 2019, \pasp, 131,
  018003

\bibitem[{{Massaro} \& {Tr\`evese}(1996)}]{1996A&A...312..810M}
{Massaro}, E. \& {Tr\`evese}, D. 1996, \aap, 312, 810

\bibitem[{{Meng} {et~al.}(2018){Meng}, {Zhang}, {Wu}, {Ma}, \&
  {Zhou}}]{2018ApJS..237...30M}
{Meng}, N., {Zhang}, X., {Wu}, J., {Ma}, J., \& {Zhou}, X. 2018, \apjs, 237, 30

\bibitem[{{Miller} {et~al.}(1988){Miller}, {Carini}, {Gaston}, \&
  {Hutter}}]{1988ESASP.281b.303M}
{Miller}, H.~R., {Carini}, M.~T., {Gaston}, B.~J., \& {Hutter}, D.~J. 1988, in
  ESA Special Publication, Vol.~2, ESA Special Publication, 303--304

\bibitem[{{Miller} \& {Green}(1983)}]{1983BAAS...15..957M}
{Miller}, H.~R. \& {Green}, R.~F. 1983, in \baas, Vol.~15, 957

\bibitem[{{Monet} {et~al.}(2003){Monet}, {Levine}, {Canzian}, {Ables}, {Bird},
  {Dahn}, {Guetter}, {Harris}, {Henden}, {Leggett}, {Levison}, {Luginbuhl},
  {Martini}, {Monet}, {Munn}, {Pier}, {Rhodes}, {Riepe}, {Sell}, {Stone},
  {Vrba}, {Walker}, {Westerhout}, {Brucato}, {Reid}, {Schoening}, {Hartley},
  {Read}, \& {Tritton}}]{2003AJ....125..984M}
{Monet}, D.~G., {Levine}, S.~E., {Canzian}, B., {et~al.} 2003, \aj, 125, 984

\bibitem[{{Nilsson} {et~al.}(2018){Nilsson}, {Lindfors}, {Takalo}, {Reinthal},
  {Berdyugin}, {Sillanp{\"a}{\"a}}, {Ciprini}, {Halkola}, {Hein{\"a}m{\"a}ki},
  {Hovatta}, {Kadenius}, {Nurmi}, {Ostorero}, {Pasanen}, {Rekola}, {Saarinen},
  {Sainio}, {Tuominen}, {Villforth}, {Vornanen}, \&
  {Zaprudin}}]{2018A&A...620A.185N}
{Nilsson}, K., {Lindfors}, E., {Takalo}, L.~O., {et~al.} 2018, \aap, 620, A185

\bibitem[{{Osterman} {et~al.}(2006){Osterman}, {Miller}, {Campbell},
  {Marshall}, {McFarland}, {Aller}, {Aller}, {Fried}, {Kurtanidze},
  {Nikolashvili}, {Tornikoski}, \& {Valtaoja}}]{2006AJ....132..873O}
{Osterman}, M.~A., {Miller}, H.~R., {Campbell}, A.~M., {et~al.} 2006, \aj, 132,
  873

\bibitem[{{Padovani} \& {Giommi}(1995)}]{1995ApJ...444..567P}
{Padovani}, P. \& {Giommi}, P. 1995, \apj, 444, 567

\bibitem[{{Pandey} {et~al.}(2019){Pandey}, {Gupta}, {Wiita}, \&
  {Tiwari}}]{2019ApJ...871..192P}
{Pandey}, A., {Gupta}, A.~C., {Wiita}, P.~J., \& {Tiwari}, S.~N. 2019, \apj,
  871, 192

\bibitem[{{Papadakis} {et~al.}(2003){Papadakis}, {Boumis}, {Samaritakis}, \&
  {Papamastorakis}}]{2003A&A...397..565P}
{Papadakis}, I.~E., {Boumis}, P., {Samaritakis}, V., \& {Papamastorakis}, J.
  2003, \aap, 397, 565

\bibitem[{{Papadakis} {et~al.}(2007){Papadakis}, {Villata}, \&
  {Raiteri}}]{2007A&A...470..857P}
{Papadakis}, I.~E., {Villata}, M., \& {Raiteri}, C.~M. 2007, \aap, 470, 857

\bibitem[{{Pasierb} {et~al.}(2020){Pasierb}, {Goyal}, {Ostrowski}, {Stawarz},
  {Wiita}, {Gopal-Krishna}, {Larionov}, {Morozova}, {Itoh}, {Alicavus},
  {Erdem}, {Joshi}, {Zola}, {Borman}, {Grishina}, {Kopatskaya}, {Larionova},
  {Savchenko}, {Nikiforova}, {Troitskaya}, {Troitsky}, {Akitaya}, {Kawabata},
  \& {Nakaoka}}]{2020MNRAS.492.1295P}
{Pasierb}, M., {Goyal}, A., {Ostrowski}, M., {et~al.} 2020, \mnras, 492, 1295

\bibitem[{{Pe{\~n}il} {et~al.}(2020){Pe{\~n}il}, {Dom{\'\i}nguez}, {Buson},
  {Ajello}, {Otero-Santos}, {Barrio}, {Nemmen}, {Cutini}, {Rani},
  {Franckowiak}, \& {Cavazzuti}}]{2020ApJ...896..134P}
{Pe{\~n}il}, P., {Dom{\'\i}nguez}, A., {Buson}, S., {et~al.} 2020, \apj, 896,
  134

\bibitem[{{Peterson} {et~al.}(1998){Peterson}, {Wanders}, {Horne}, {Collier},
  {Alexander}, {Kaspi}, \& {Maoz}}]{1998PASP..110..660P}
{Peterson}, B.~M., {Wanders}, I., {Horne}, K., {et~al.} 1998, \pasp, 110, 660

\bibitem[{{Piner} \& {Edwards}(2018)}]{2018mgm..conf.3074P}
{Piner}, B.~G. \& {Edwards}, P.~G. 2018, in Fourteenth Marcel Grossmann Meeting
  - MG14, ed. M.~{Bianchi}, R.~T. {Jansen}, \& R.~{Ruffini}, 3074--3079

\bibitem[{{Press} \& {Rybicki}(1989)}]{1989ApJ...338..277P}
{Press}, W.~H. \& {Rybicki}, G.~B. 1989, \apj, 338, 277

\bibitem[{{Prokhorov} \& {Moraghan}(2017)}]{2017MNRAS.471.3036P}
{Prokhorov}, D.~A. \& {Moraghan}, A. 2017, \mnras, 471, 3036

\bibitem[{{Raiteri} {et~al.}(2015){Raiteri}, {Stamerra}, {Villata}, {Larionov},
  {Acosta-Pulido}, {Ar{\'e}valo}, {Arkharov}, {Bachev}, {Ben{\'\i}tez},
  {Bozhilov}, {Borman}, {Buemi}, {Calcidese}, {Carnerero}, {Carosati},
  {Chigladze}, {Damljanovic}, {Di Paola}, {Doroshenko}, {Efimova},
  {Ehgamberdiev}, {Giroletti}, {Gonz{\'a}lez-Morales}, {Grinon-Marin},
  {Grishina}, {Hiriart}, {Ibryamov}, {Klimanov}, {Kopatskaya}, {Kurtanidze},
  {Kurtanidze}, {Kurtenkov}, {Larionova}, {Larionova}, {L{\'a}zaro},
  {L{\"a}hteenm{\"a}ki}, {Leto}, {Markovic}, {Mirzaqulov}, {Mokrushina},
  {Morozova}, {M{\'u}jica}, {Nazarov}, {Nikolashvili}, {Ohlert}, {Ovcharov},
  {Paiano}, {Pastor Yabar}, {Prandini}, {Ramakrishnan}, {Sadun}, {Semkov},
  {Sigua}, {Strigachev}, {Tammi}, {Tornikoski}, {Trigilio}, {Troitskaya},
  {Troitsky}, {Umana}, {Velasco}, \& {Vince}}]{2015MNRAS.454..353R}
{Raiteri}, C.~M., {Stamerra}, A., {Villata}, M., {et~al.} 2015, \mnras, 454,
  353

\bibitem[{{Rau} {et~al.}(2009){Rau}, {Kulkarni}, {Law}, {Bloom}, {Ciardi},
  {Djorgovski}, {Fox}, {Gal-Yam}, {Grillmair}, {Kasliwal}, {Nugent}, {Ofek},
  {Quimby}, {Reach}, {Shara}, {Bildsten}, {Cenko}, {Drake}, {Filippenko},
  {Helfand}, {Helou}, {Howell}, {Poznanski}, \&
  {Sullivan}}]{2009PASP..121.1334R}
{Rau}, A., {Kulkarni}, S.~R., {Law}, N.~M., {et~al.} 2009, \pasp, 121, 1334

\bibitem[{{Rector} \& {Perlman}(2003)}]{2003AJ....126...47R}
{Rector}, T.~A. \& {Perlman}, E.~S. 2003, \aj, 126, 47

\bibitem[{{Romero}(1995)}]{1995Ap&SS.234...49R}
{Romero}, G.~E. 1995, \apss, 234, 49

\bibitem[{{Romero} {et~al.}(1999){Romero}, {Cellone}, \&
  {Combi}}]{1999A&AS..135..477R}
{Romero}, G.~E., {Cellone}, S.~A., \& {Combi}, J.~A. 1999, \aaps, 135, 477

\bibitem[{{Sambruna} {et~al.}(1996){Sambruna}, {Maraschi}, \&
  {Urry}}]{1996ApJ...463..444S}
{Sambruna}, R.~M., {Maraschi}, L., \& {Urry}, C.~M. 1996, \apj, 463, 444

\bibitem[{{Sandrinelli} {et~al.}(2018){Sandrinelli}, {Covino}, {Treves},
  {Holgado}, {Sesana}, {Lindfors}, \& {Ramazani}}]{2018A&A...615A.118S}
{Sandrinelli}, A., {Covino}, S., {Treves}, A., {et~al.} 2018, \aap, 615, A118

\bibitem[{{Scargle}(1982)}]{1982ApJ...263..835S}
{Scargle}, J.~D. 1982, \apj, 263, 835

\bibitem[{{Schulz} \& {Mudelsee}(2002)}]{2002CG.....28..421S}
{Schulz}, M. \& {Mudelsee}, M. 2002, Computers and Geosciences, 28, 421

\bibitem[{{Shappee} {et~al.}(2014){Shappee}, {Prieto}, {Grupe}, {Kochanek},
  {Stanek}, {De Rosa}, {Mathur}, {Zu}, {Peterson}, {Pogge}, {Komossa}, {Im},
  {Jencson}, {Holoien}, {Basu}, {Beacom}, {Szczygie{\l}}, {Brimacombe},
  {Adams}, {Campillay}, {Choi}, {Contreras}, {Dietrich}, {Dubberley},
  {Elphick}, {Foale}, {Giustini}, {Gonzalez}, {Hawkins}, {Howell}, {Hsiao},
  {Koss}, {Leighly}, {Morrell}, {Mudd}, {Mullins}, {Nugent}, {Parrent},
  {Phillips}, {Pojmanski}, {Rosing}, {Ross}, {Sand}, {Terndrup}, {Valenti},
  {Walker}, \& {Yoon}}]{2014ApJ...788...48S}
{Shappee}, B.~J., {Prieto}, J.~L., {Grupe}, D., {et~al.} 2014, \apj, 788, 48

\bibitem[{{Smith} {et~al.}(2009){Smith}, {Montiel}, {Rightley}, {Turner},
  {Schmidt}, \& {Jannuzi}}]{2009arXiv0912.3621S}
{Smith}, P.~S., {Montiel}, E., {Rightley}, S., {et~al.} 2009, in 2009 Fermi
  Symposium, eConf Proceedings C0911022, arXiv:0912.3621

\bibitem[{{Sobacchi} {et~al.}(2017){Sobacchi}, {Sormani}, \&
  {Stamerra}}]{2017MNRAS.465..161S}
{Sobacchi}, E., {Sormani}, M.~C., \& {Stamerra}, A. 2017, \mnras, 465, 161

\bibitem[{{Stalin} {et~al.}(2005){Stalin}, {Gupta}, {Gopal-Krishna}, {Wiita},
  \& {Sagar}}]{2005MNRAS.356..607S}
{Stalin}, C.~S., {Gupta}, A.~C., {Gopal-Krishna}, {Wiita}, P.~J., \& {Sagar},
  R. 2005, \mnras, 356, 607

\bibitem[{{Stetson}(1987)}]{S1987PASP}
{Stetson}, P.~B. 1987, \pasp, 99, 191

\bibitem[{{Stetson}(1992)}]{S1992ASPC}
{Stetson}, P.~B. 1992, in Astronomical Society of the Pacific Conference
  Series, Vol.~25, Astronomical Data Analysis Software and Systems I, ed. D.~M.
  {Worrall}, C.~{Biemesderfer}, \& J.~{Barnes}, 297

\bibitem[{{Takahashi} {et~al.}(1996){Takahashi}, {Tashiro}, {Madejski}, {Kubo},
  {Kamae}, {Kataoka}, {Kii}, {Makino}, {Makishima}, \&
  {Yamasaki}}]{1996ApJ...470L..89T}
{Takahashi}, T., {Tashiro}, M., {Madejski}, G., {et~al.} 1996, \apjl, 470, L89

\bibitem[{{Takalo} {et~al.}(2008){Takalo}, {Nilsson}, {Lindfors},
  {Sillanp{\"a}{\"a}}, {Berdyugin}, \& {Pasanen}}]{2008AIPC.1085..705T}
{Takalo}, L.~O., {Nilsson}, K., {Lindfors}, E., {et~al.} 2008, in American
  Institute of Physics Conference Series, Vol. 1085, American Institute of
  Physics Conference Series, ed. F.~A. {Aharonian}, W.~{Hofmann}, \&
  F.~{Rieger}, 705--707

\bibitem[{{Tavani} {et~al.}(2018){Tavani}, {Cavaliere}, {Munar-Adrover}, \&
  {Argan}}]{2018ApJ...854...11T}
{Tavani}, M., {Cavaliere}, A., {Munar-Adrover}, P., \& {Argan}, A. 2018, \apj,
  854, 11

\bibitem[{{Tavecchio}(2006)}]{2006tmgm.meet..512T}
{Tavecchio}, F. 2006, in The Tenth Marcel Grossmann Meeting. On recent
  developments in theoretical and experimental general relativity, gravitation
  and relativistic field theories, 512

\bibitem[{{Tavecchio} {et~al.}(2010){Tavecchio}, {Ghisellini}, {Ghirlanda},
  {Foschini}, \& {Maraschi}}]{2010MNRAS.401.1570T}
{Tavecchio}, F., {Ghisellini}, G., {Ghirlanda}, G., {Foschini}, L., \&
  {Maraschi}, L. 2010, \mnras, 401, 1570

\bibitem[{{Torres Zafra} {et~al.}(2017){Torres Zafra}, {Cellone}, {Andruchow},
  {Buzzoni}, \& {Portilla Barbosa}}]{TZCA2017RMxAC}
{Torres Zafra}, J., {Cellone}, S.~A., {Andruchow}, I., {Buzzoni}, A., \&
  {Portilla Barbosa}, J.~G. 2017, in Revista Mexicana de Astronomia y
  Astrofisica Conference Series, Vol.~49, 138

\bibitem[{{Treves} {et~al.}(2007){Treves}, {Falomo}, \&
  {Uslenghi}}]{2007A&A...473L..17T}
{Treves}, A., {Falomo}, R., \& {Uslenghi}, M. 2007, \aap, 473, L17

\bibitem[{{Valtaoja} {et~al.}(1999){Valtaoja}, {L{\"a}hteenm{\"a}ki},
  {Ter{\"a}sranta}, \& {Lainela}}]{1999ApJS..120...95V}
{Valtaoja}, E., {L{\"a}hteenm{\"a}ki}, A., {Ter{\"a}sranta}, H., \& {Lainela},
  M. 1999, \apjs, 120, 95

\bibitem[{{VanderPlas}(2018)}]{2018ApJS..236...16V}
{VanderPlas}, J.~T. 2018, \apjs, 236, 16

\bibitem[{{Villata} {et~al.}(2004){Villata}, {Raiteri}, {Kurtanidze},
  {Nikolashvili}, {Ibrahimov}, {Papadakis}, {Tosti}, {Hroch}, {Takalo},
  {Sillanp{\"a}{\"a}}, {Hagen-Thorn}, {Larionov}, {Schwartz}, {Basler},
  {Brown}, {Balonek}, {Ben{\'\i}tez}, {Ram{\'\i}rez}, {Sadun}, {Boltwood},
  {Carini}, {Barnaby}, {Coloma}, {Ros}, {Dai}, {Xie}, {Mattox}, {Rodriguez},
  {Asfand iyarov}, {Atkerson}, {Beem}, {Bloom}, {Chanturiya}, {Ciprini},
  {Crapanzano}, {de Diego}, {Efimova}, {Gardiol}, {Guerra}, {Kahharov},
  {Kapanadze}, {Karttunen}, {Kato}, {Kimeridze}, {Kudryavtseva}, {Lainela},
  {Lanteri}, {Larionova}, {Maesano}, {Marchili}, {Massone}, {Monroe},
  {Montagni}, {Nesci}, {Nilsson}, {Noble}, {Nucciarelli}, {Ostorero},
  {Papamastorakis}, {Pasanen}, {Peters}, {Pursimo}, {Reig}, {Ryle}, {Sclavi},
  {Sigua}, {Uemura}, \& {Wills}}]{2004A&A...421..103V}
{Villata}, M., {Raiteri}, C.~M., {Kurtanidze}, O.~M., {et~al.} 2004, \aap, 421,
  103

\bibitem[{{Villata} {et~al.}(2002){Villata}, {Raiteri}, {Kurtanidze},
  {Nikolashvili}, {Ibrahimov}, {Papadakis}, {Tsinganos}, {Sadakane}, {Okada},
  {Takalo}, {Sillanp{\"a}{\"a}}, {Tosti}, {Ciprini}, {Frasca}, {Marilli},
  {Robb}, {Noble}, {Jorstad}, {Hagen-Thorn}, {Larionov}, {Nesci}, {Maesano},
  {Schwartz}, {Basler}, {Gorham}, {Iwamatsu}, {Kato}, {Pullen}, {Ben{\'\i}tez},
  {de Diego}, {Moilanen}, {Oksanen}, {Rodriguez}, {Sadun}, {Kelly}, {Carini},
  {Miller}, {Catalano}, {Dultzin-Hacyan}, {Fan}, {Ishioka}, {Karttunen},
  {Kein{\"a}nen}, {Kudryavtseva}, {Lainela}, {Lanteri}, {Larionova},
  {Matsumoto}, {Mattox}, {Montagni}, {Nucciarelli}, {Ostorero},
  {Papamastorakis}, {Pasanen}, {Sobrito}, \& {Uemura}}]{2002A&A...390..407V}
{Villata}, M., {Raiteri}, C.~M., {Kurtanidze}, O.~M., {et~al.} 2002, \aap, 390,
  407

\bibitem[{{Wagner} \& {Witzel}(1995)}]{1995ARA&A..33..163W}
{Wagner}, S.~J. \& {Witzel}, A. 1995, \araa, 33, 163

\bibitem[{{White} \& {Peterson}(1994)}]{1994PASP..106..879W}
{White}, R.~J. \& {Peterson}, B.~M. 1994, \pasp, 106, 879

\bibitem[{{Wierzcholska} {et~al.}(2015){Wierzcholska}, {Ostrowski}, {Stawarz},
  {Wagner}, \& {Hauser}}]{2015A&A...573A..69W}
{Wierzcholska}, A., {Ostrowski}, M., {Stawarz}, {\L}., {Wagner}, S., \&
  {Hauser}, M. 2015, \aap, 573, A69

\bibitem[{{Xie} {et~al.}(2004){Xie}, {Zhou}, {Li}, {Dai}, {Chen}, \&
  {Ma}}]{2004MNRAS.348..831X}
{Xie}, G.~Z., {Zhou}, S.~B., {Li}, K.~H., {et~al.} 2004, \mnras, 348, 831

\bibitem[{{Zibecchi} {et~al.}(2020){Zibecchi}, {Andruchow}, {Cellone}, \&
  {Carpintero}}]{2020MNRAS.498.3013Z}
{Zibecchi}, L., {Andruchow}, I., {Cellone}, S.~A., \& {Carpintero}, D.~D. 2020,
  \mnras, 498, 3013

\bibitem[{{Zibecchi} {et~al.}(2017){Zibecchi}, {Andruchow}, {Cellone},
  {Carpintero}, {Romero}, \& {Combi}}]{2017MNRAS.467..340Z}
{Zibecchi}, L., {Andruchow}, I., {Cellone}, S.~A., {et~al.} 2017, \mnras, 467,
  340

\end{thebibliography}

\begin{appendix}

\section{Results from INM}
\label{app:A}

We present the results from the INV tests in Table\,\ref{tab:var_res}. The individual intra-night LCs, derived during the 28 nights of INM, are shown in Fig.\,\ref{LC_BL1}.

\begin{table*}
\caption{Results of INV observations of \pg.} 
\label{tab:var_res}
\centering 
\resizebox{0.82\textwidth}{!}{ 
\begin{tabular}{lcccccccccccc} 
\hline\hline \noalign{\smallskip} 
 Date of observation& Passband & $N$ & $C$-test      & $F$-test     & $\Gamma$ & $\sigma_2$ & Variable? \\ \noalign{\smallskip} 
 (yyyy mm dd)       &          &     & $C_{1},C_{2}$ &$F_{1},F_{2}$ & $\Gamma_1$,$\Gamma_2$ &            &           \\ \noalign{\smallskip} 
 (1) & (2) & (3) & (4) & (5) & (6) & (7) & (8) \\
 \noalign{\smallskip} \hline\noalign{\smallskip} 
 
 2018 05 07   & $B$     & 17  & 1.03, 1.04 & 1.06, 1.08 & 1.0161, 0.9990 & 0.0109 & NV \\
              & $I$     & 16  & 0.71, 0.93 & 0.50, 0.86 & 1.0938, 1.0470 & 0.0087 & NV \\
              & $R$     & 17  & 0.63, 0.60 & 0.40, 0.36 & 1.1334, 1.1006 & 0.0130 & NV \\
              & $V$     & 17  & 1.06, 1.30 & 1.12, 1.69 & 1.1093, 1.0797 & 0.0086 & NV \\
 2018 05 10   & $B$     & 57  & 0.88, 0.70 & 0.77, 0.49 & 1.0315, 1.0218 & 0.0099 & NV \\
              & $I$     & 61  & 1.04, 1.23 & 1.08, 1.51 & 1.1112, 1.0708 & 0.0061 & NV \\
              & $R$     & 61  & 1.39, 1.42 & 1.93, 2.02 & 1.1265, 1.1013 & 0.0057 & NV \\
              & $V$     & 61  & 1.09, 1.23 & 1.19, 1.51 & 1.1053, 1.0921 & 0.0063 & NV \\
 2018 05 12   & $B$     & 59  & 1.38, 1.39 & 1.90, 1.93 & 1.0322, 1.0191 & 0.0072 & NV \\
              & $I$     & 67  & 0.86, 1.10 & 0.74, 1.21 & 1.1039, 1.0658 & 0.0068 & NV \\
              & $R$     & 67  & 0.85, 0.92 & 0.72, 0.85 & 1.1395, 1.1092 & 0.0076 & NV \\
              & $V$     & 67  & 1.10, 1.27 & 1.21, 1.61 & 1.1202, 1.0937 & 0.0085 & NV \\
 2018 05 13   & $R$     & 40  & 1.03, 1.04 & 1.06, 1.08 & 1.1367, 1.1040 & 0.0070 & NV \\
 2018 05 28   & $B$     & 51  & 0.97, 0.89 & 0.94, 0.79 & 0.9884, 0.9731 & 0.0159 & NV \\
              & $I$     & 52  & 0.74, 1.04 & 0.55, 1.08 & 1.0809, 1.0219 & 0.0200 & NV \\
              & $R$     & 53  & 0.99, 0.95 & 0.98, 0.90 & 1.1026, 1.0643 & 0.0094 & NV \\
              & $V$     & 54  & 0.79, 0.94 & 0.62, 0.88 & 1.0865, 1.0520 & 0.0117 & NV \\
 2018 05 29   & $B$     & 39  & 0.97, 0.85 & 0.94, 0.72 & 1.0029, 0.9873 & 0.0231 & NV \\
              & $I$     & 39  & 0.90, 1.13 & 0.81, 1.28 & 1.1006, 1.0444 & 0.0136 & NV \\
              & $R$     & 39  & 1.03, 1.04 & 1.06, 1.08 & 1.1416, 1.0977 & 0.0137 & NV \\
              & $V$     & 34  & 0.78, 1.03 & 0.61, 1.06 & 1.1127, 1.0804 & 0.0151 & NV \\
 2018 06 04   & $B$     & 32  & 0.79, 0.81 & 0.62, 0.66 & 1.0236, 1.0074 & 0.0141 & NV \\
              & $I$     & 34  & 1.13, 1.23 & 1.28, 1.51 & 1.0878, 1.0461 & 0.0079 & NV \\
              & $R$     & 34  & 0.96, 1.02 & 0.92, 1.04 & 1.1337, 1.0961 & 0.0086 & NV \\
              & $V$     & 34  & 0.99, 1.21 & 0.98, 1.46 & 1.1033, 1.0796 & 0.0076 & NV \\
  2018 06 05  & $B$     & 17  & 1.52, 1.04 & 2.31, 1.08 & 1.0263, 1.0083 & 0.1100 & NV \\
              & $I$     & 17  & 0.85, 1.29 & 0.72, 1.66 & 1.0965, 1.0492 & 0.0076 & NV \\
              & $R$     & 17  & 1.20, 0.84 & 1.44, 0.70 & 1.1397, 1.1024 & 0.0094 & NV \\
              & $V$     & 17  & 0.89, 0.79 & 0.79, 0.62 & 1.1154, 1.0887 & 0.0088 & NV \\
 2018 07 21   & $B$     & 22  & 1.48, 2.16 & 2.19, 4.66 & 0.8547, 0.8898 & 0.6287 & NV \\
              & $I$     & 25  & 1.60, 2.25 & 2.56, 5.06 & 0.9834, 1.0010 & 0.3396 & NV \\
              & $R$     & 24  & 1.08, 1.76 & 1.17, 3.10 & 1.0306, 1.0457 & 0.6810 & NV \\
              & $V$     & 24  & 1.45, 2.31 & 2.10, 5.34 & 1.0637, 1.0237 & 0.6322 & NV \\
 2019 03 01   & $R$     & 11  & 0.27, 1.18 & 0.07, 1.39 & 1.0258, 1.0086 & 0.0084 & NV \\
              & $V$     & 14  & 0.94, 1.07 & 0.88, 1.14 & 1.0111, 1.0073 & 0.0252 & NV \\
 2019 03 12   & $R$     & 81  & 0.98, 1.21 & 0.96, 1.46 & 0.9789, 0.9376 & 0.0064 & NV \\
 2019 03 13   & $R$     & 83  & 0.95, 1.03 & 0.90, 1.06 & 0.9893, 0.9541 & 0.0058 & NV \\
 2019 04 01   & $R$     & 112 & 0.41, 1.16 & 0.17, 1.35 & 0.9736, 0.9552 & 0.1210 & NV \\
 2019 04 04   & $B$     & 34  & 0.28, 1.17 & 0.08, 1.37 & 0.8726, 0.8893 & 0.1237 & NV \\
              & $I$     & 32  & 0.16, 1.05 & 0.03, 1.10 & 0.9381, 0.9034 & 0.0604 & NV \\
              & $R$     & 36  & 0.06, 1.08 & 0.004, 1.17 & 0.9467, 0.9329 & 0.1312 & NV \\
              & $V$     & 35  & 0.10, 1.11 & 0.01, 1.23 & 0.9064, 0.9274 & 0.1799 & NV \\
 2019 04 06   & $R$     & 57  & 0.75, 1.24 & 0.56, 1.54 & 0.9626, 0.9215 & 0.0111 & NV \\
 2019 04 07   & $R$     & 50  & 1.07, 1.04 & 1.14, 1.08 & 0.9581, 0.9184 & 0.0077 & NV \\
 2019 04 08   & $R$     & 62  & 0.41, 1.03 & 0.17, 1.06 & 0.9581, 0.9362 & 0.0096 & NV \\
 2019 04 09   & $R$     & 61  & 0.50, 1.45 & 0.25, 2.10 & 0.9467, 0.9285 & 0.0088 & NV \\
 2019 04 09   & $R$     & 57  & 0.95, 1.02 & 0.90, 1.04 & 0.9484, 0.9045 & 0.0077 & NV \\
 2019 04 10   & $R$     & 45  & 0.92, 0.89 & 0.85, 0.79 & 0.9375, 0.8963 & 0.0075 & NV \\
 2019 04 10   & $R$     & 37  & 0.37, 0.84 & 0.14, 0.70 & 0.9506, 0.9268 & 0.1280 & NV \\
 2019 04 11   & $R$     & 72  & 0.62, 1.51 & 0.38, 2.28 & 0.9414, 0.9230 & 0.0114 & NV \\
 2019 06 21   & $R$     & 343 & 1.52, 1.55 & 2.31, 2.40 & 1.0501, 1.0234 & 0.0040 & NV \\
 2019 07 10   & $R$     & 121 & 0.84, 1.14 & 0.71, 1.30 & 1.0550, 1.0266 & 0.0039 & NV \\
 2019 07 13   & $R$     & 111 & 0.92, 0.99 & 0.85, 0.98 & 1.0997, 1.0680 & 0.0076 & NV \\
 2019 07 19   & $R$     & 55  & 1.04, 1.35 & 1.08, 1.82 & 1.0837, 1.0577 & 0.0047 & NV \\
 2019 07 20   & $R$     & 65  & 1.25, 1.06 & 1.56, 1,12 & 1.0794, 1.0532 & 0.0082 & NV \\
 2019 07 22   & $R$     & 11  & 0.21, 1.15 & 0.04, 1.32 & 1.0771, 1.0484 & 0.0434 & NV \\
\hline
\end{tabular}}
\tablefoot{Table columns read: (2) Passband of observation. (3) Number of data points in the given passband. (4)--(5) Results for $C$- and $F$-test, respectively. (6) Corresponding scale factor. (7) Dispersion of the corresponding control-comparison star LC. (8) Variability status denoted as follows: Var = variable, NV = non-variable.}
\end{table*}

\begin{figure*}[t]

\epsfig{figure=  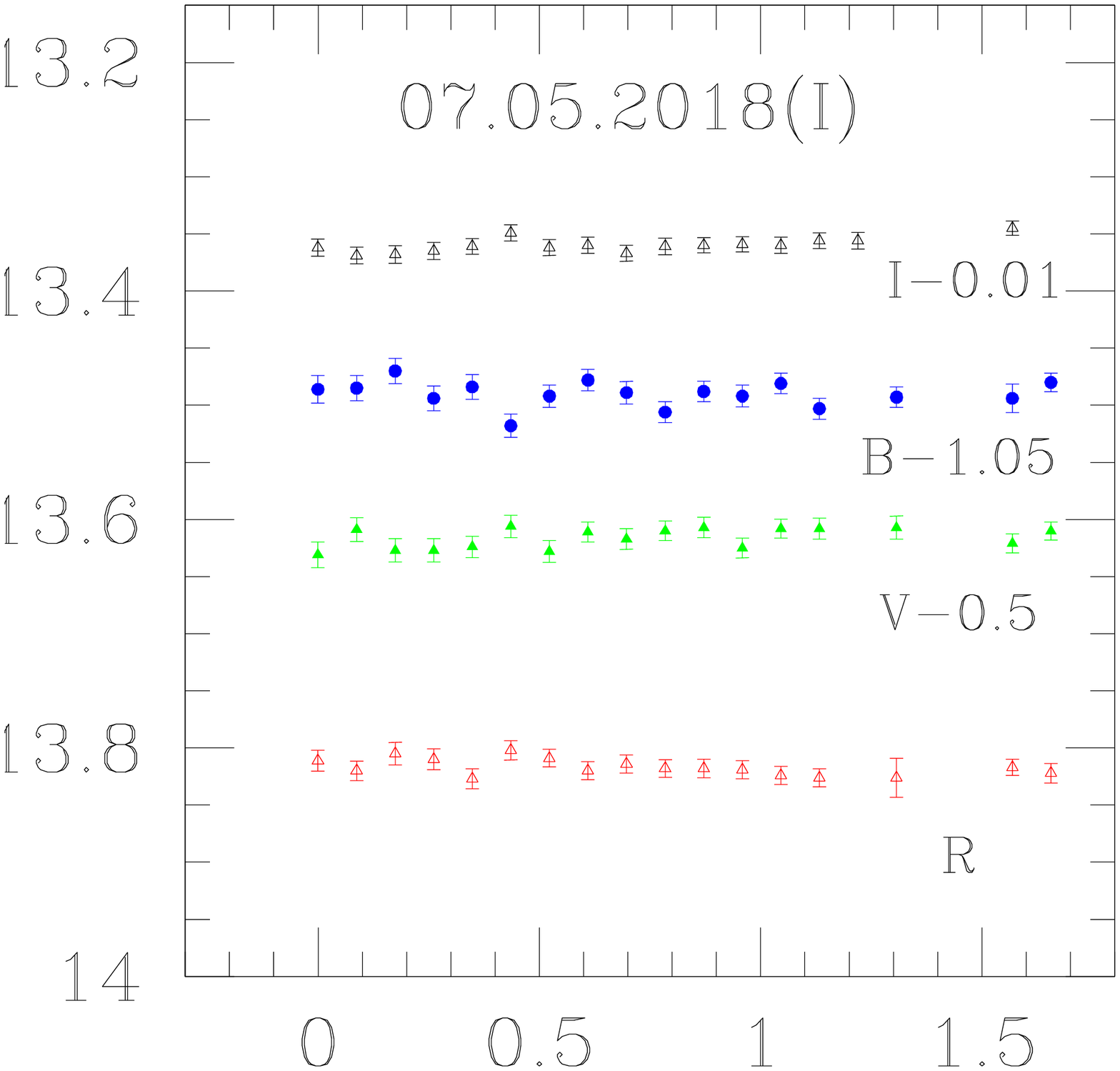,height=2.1in,width=2in,angle=0}
\epsfig{figure=  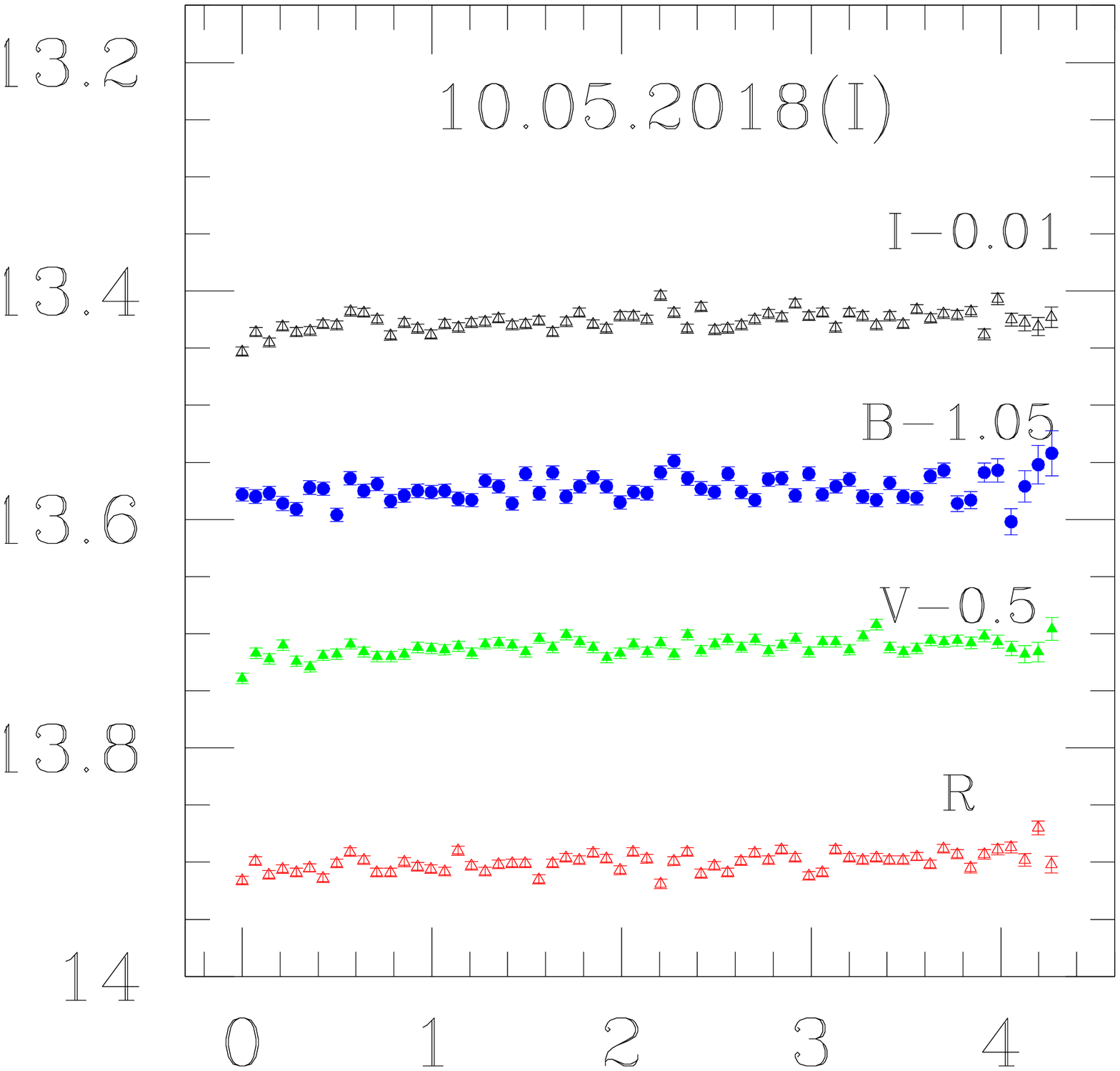,height=2.1in,width=2in,angle=0}
\epsfig{figure=  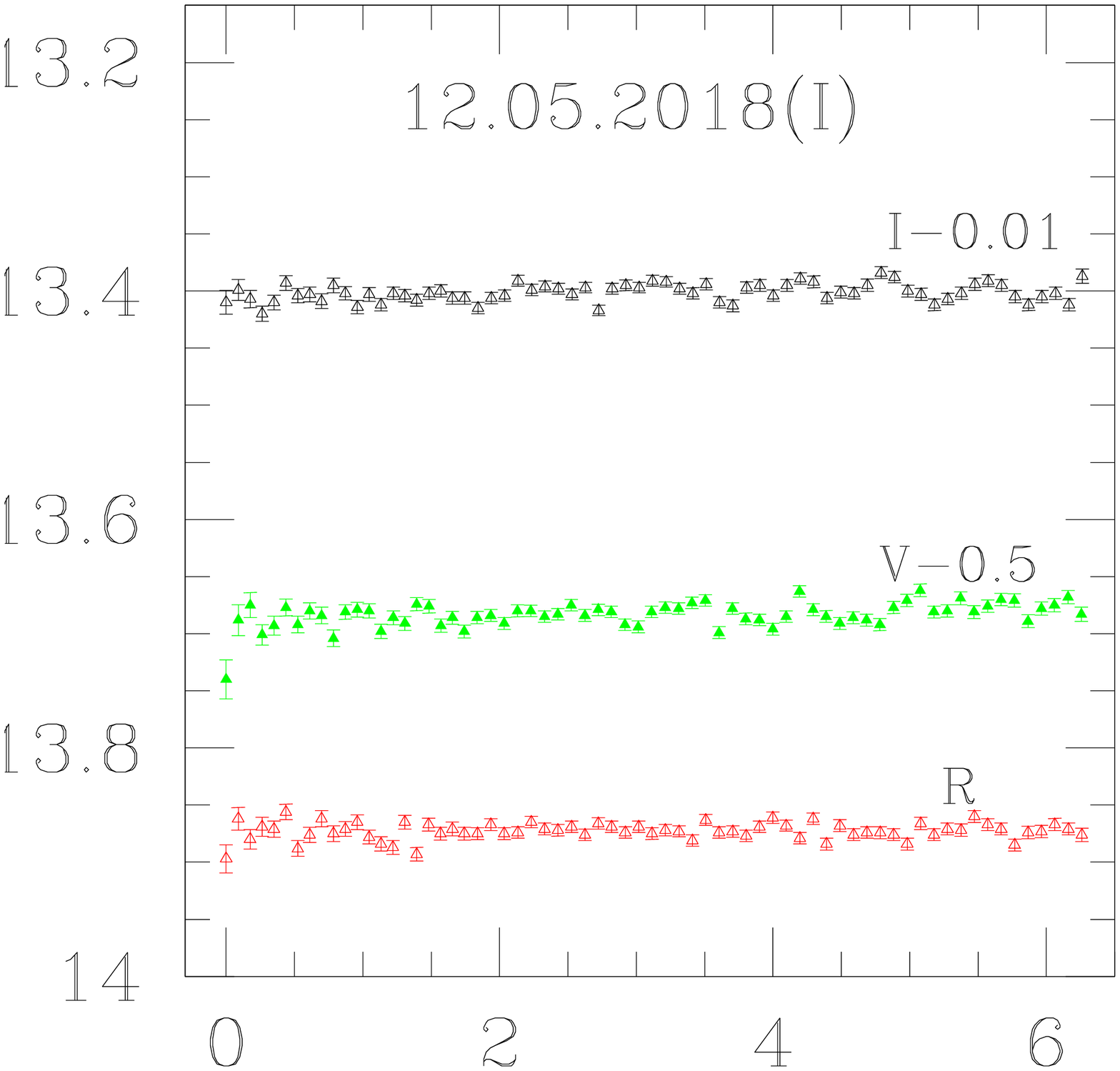,height=2.1in,width=2in,angle=0}
\epsfig{figure=  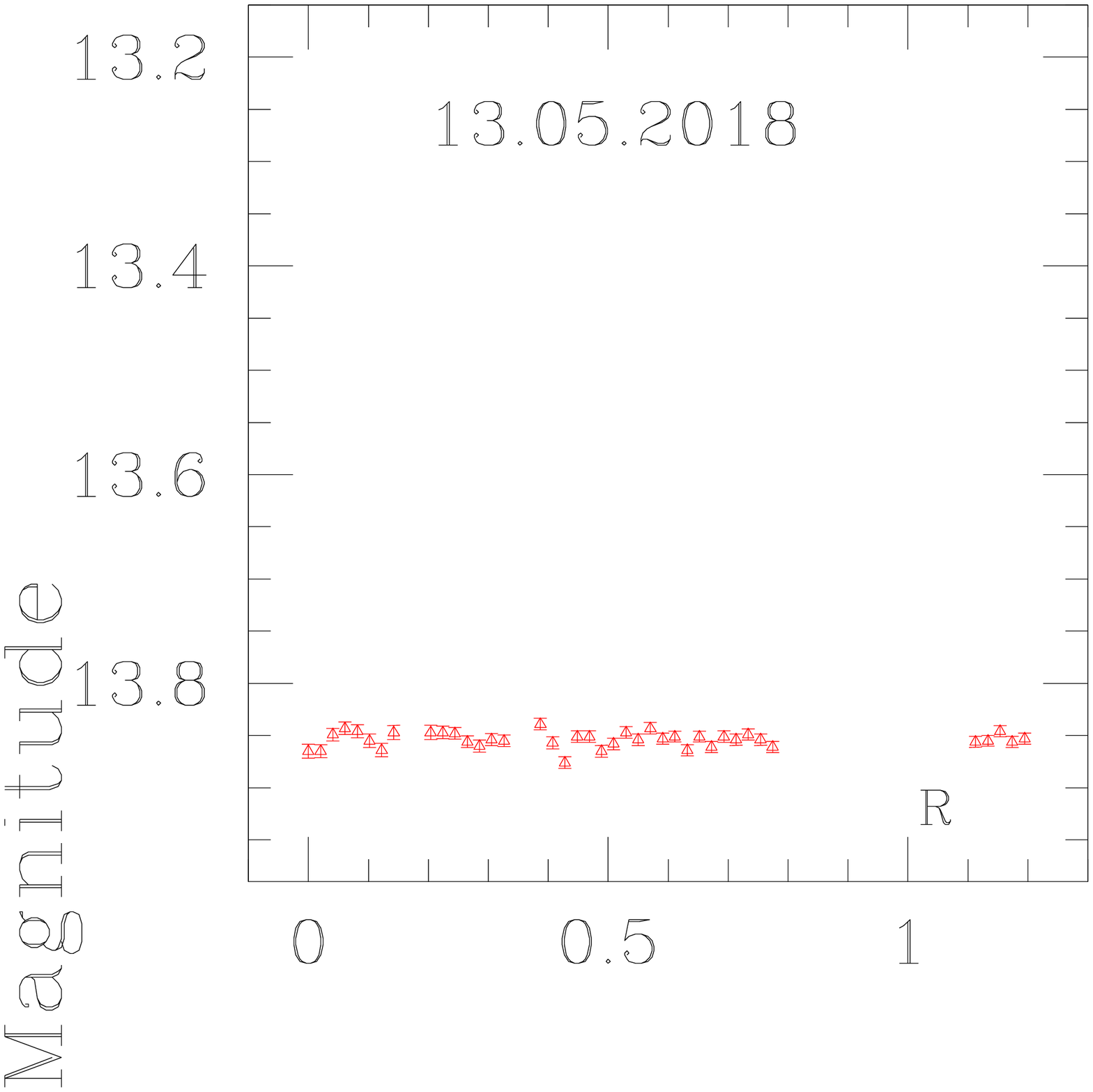,height=2.1in,width=2in,angle=0}
\epsfig{figure=  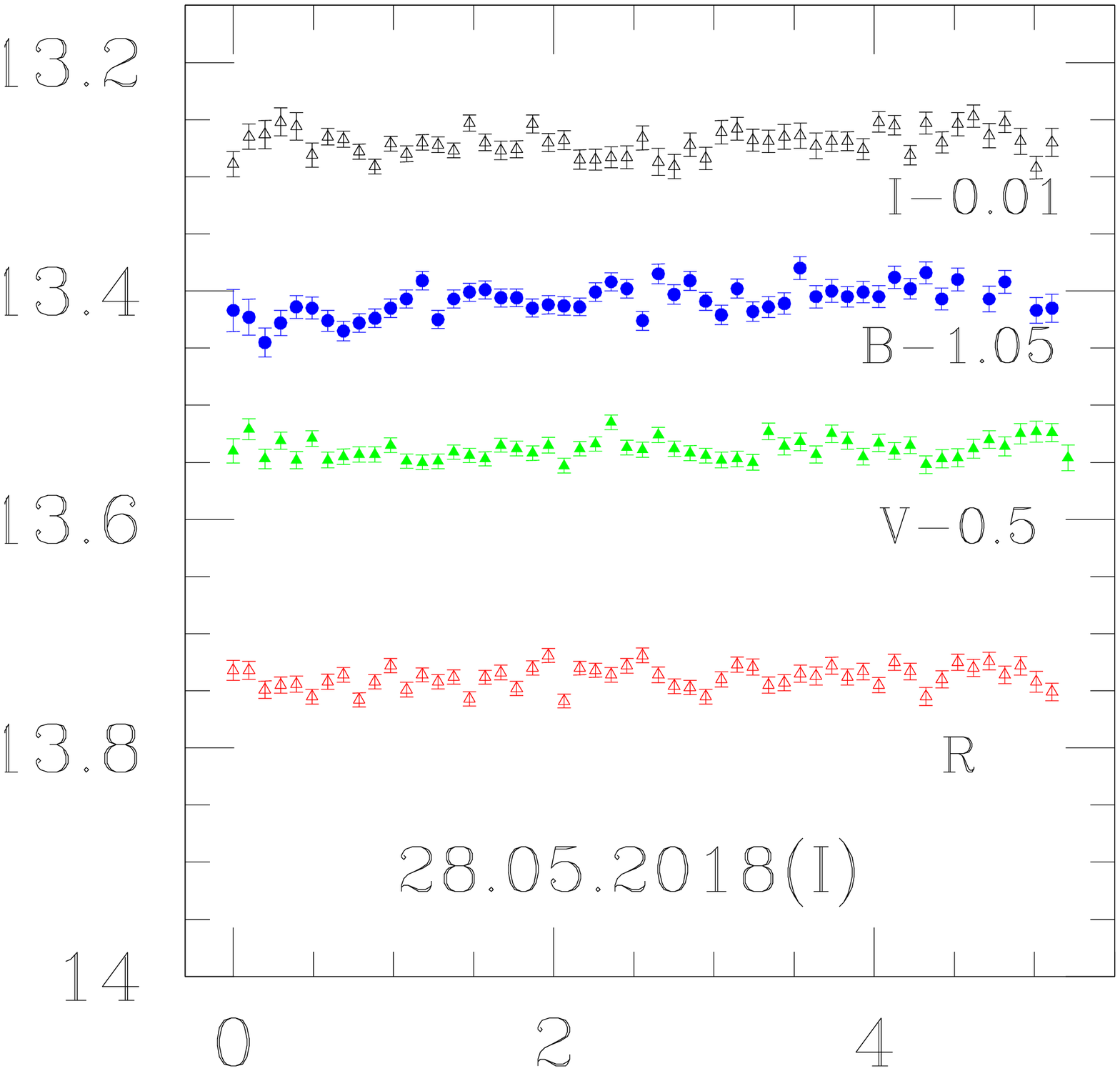,height=2.1in,width=2in,angle=0}
\epsfig{figure=  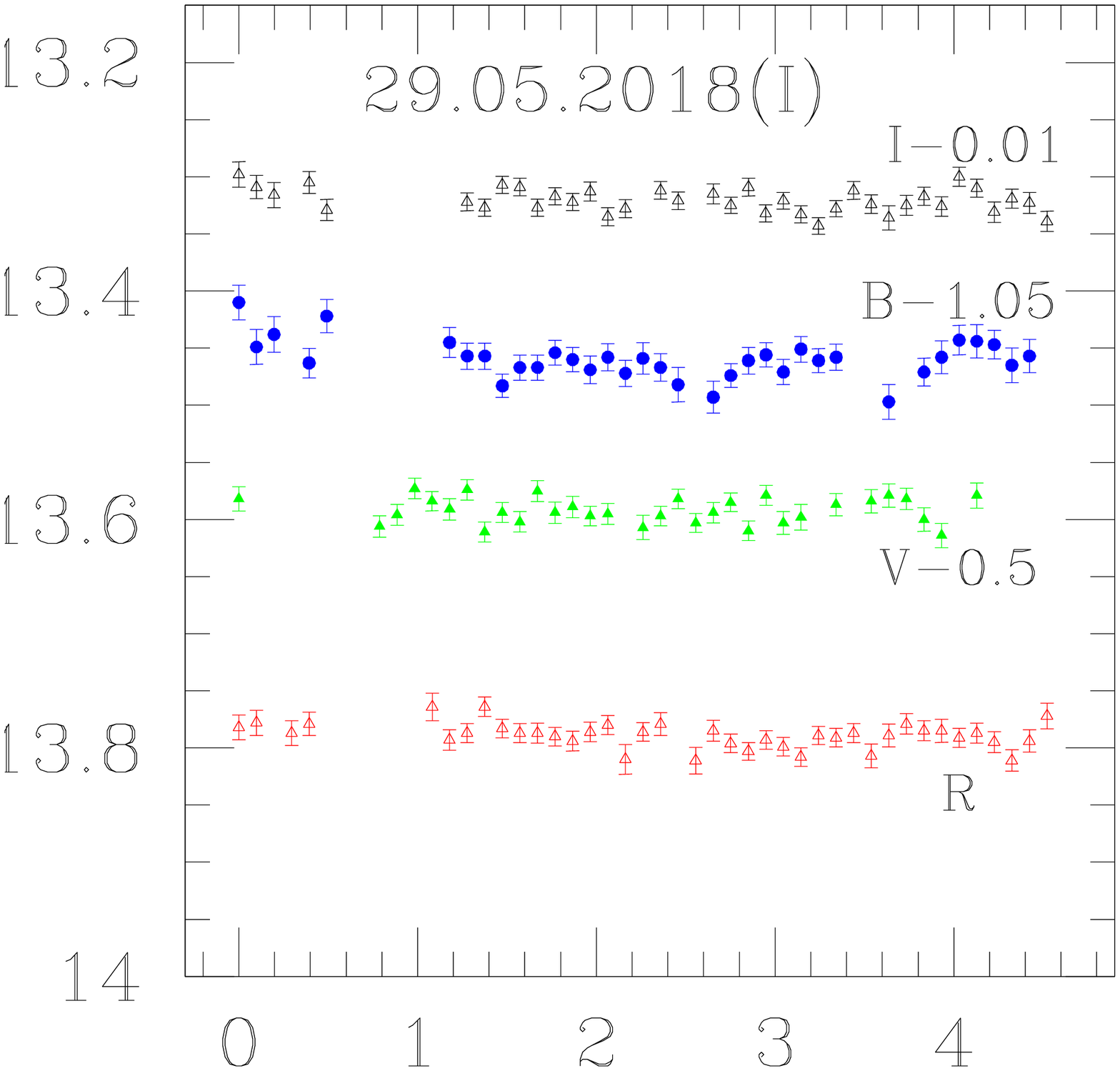,height=2.1in,width=2in,angle=0}
\epsfig{figure=  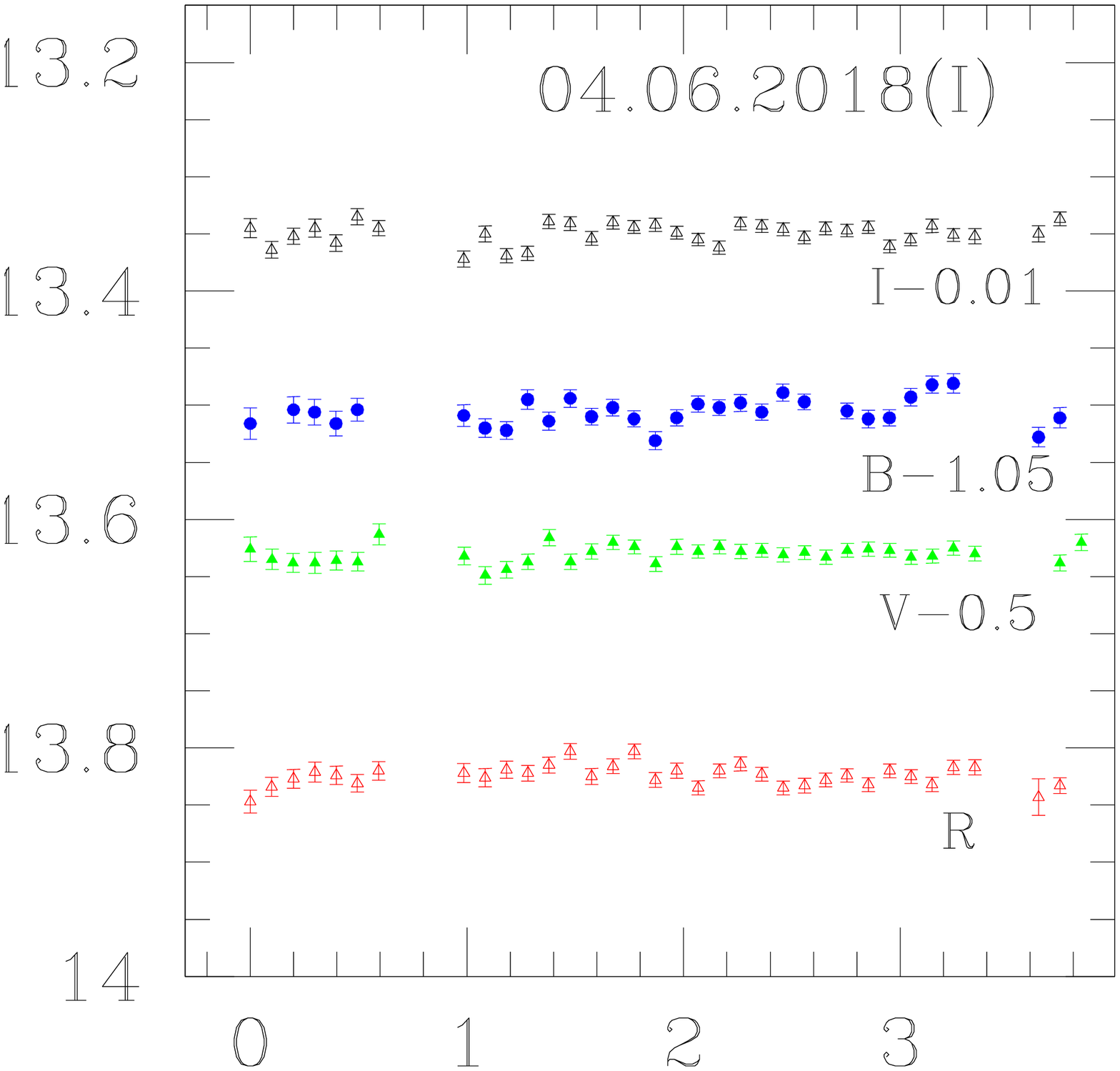,height=2.1in,width=2in,angle=0}
\epsfig{figure=  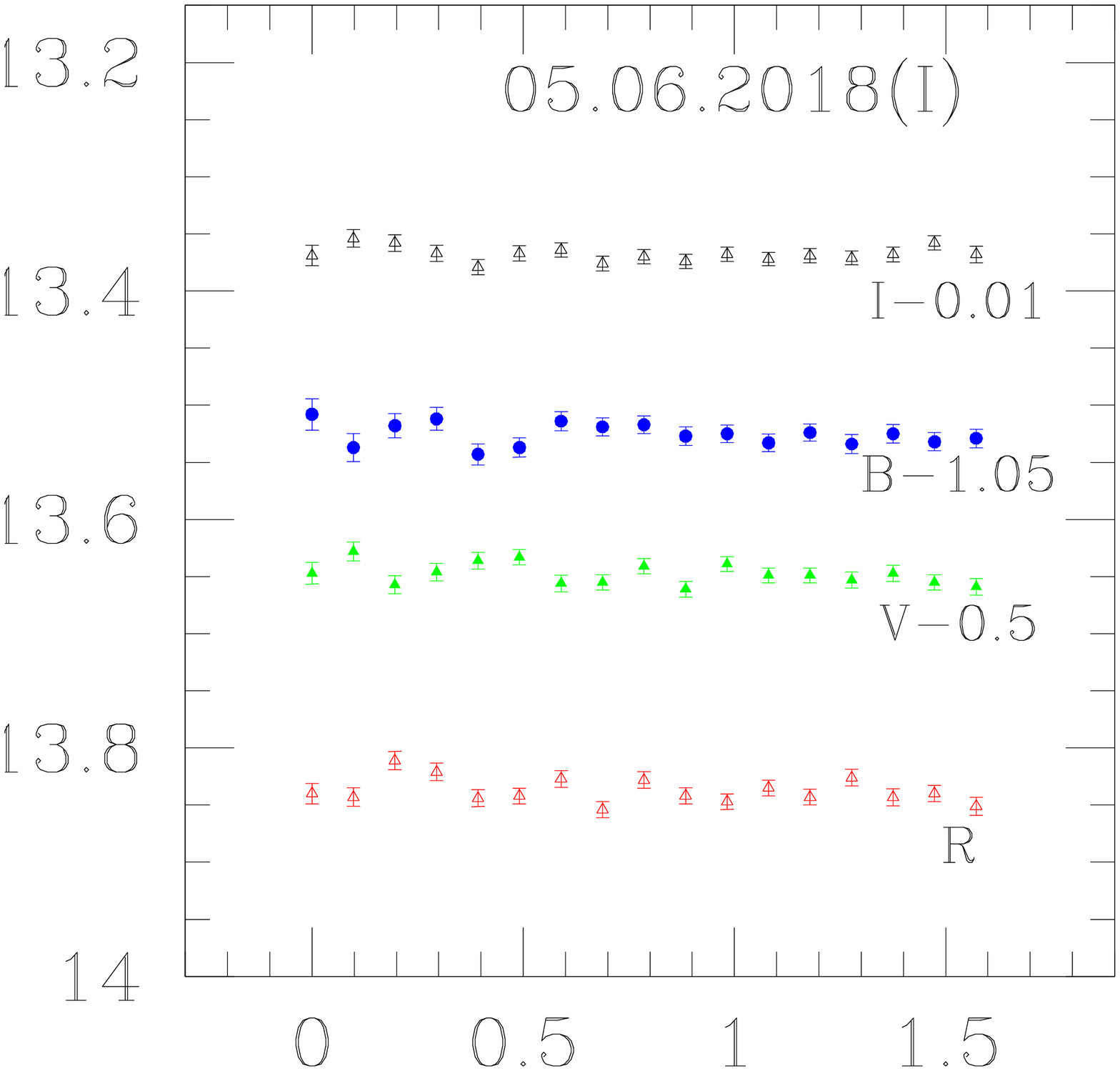,height=2.1in,width=2in,angle=0}
\epsfig{figure=  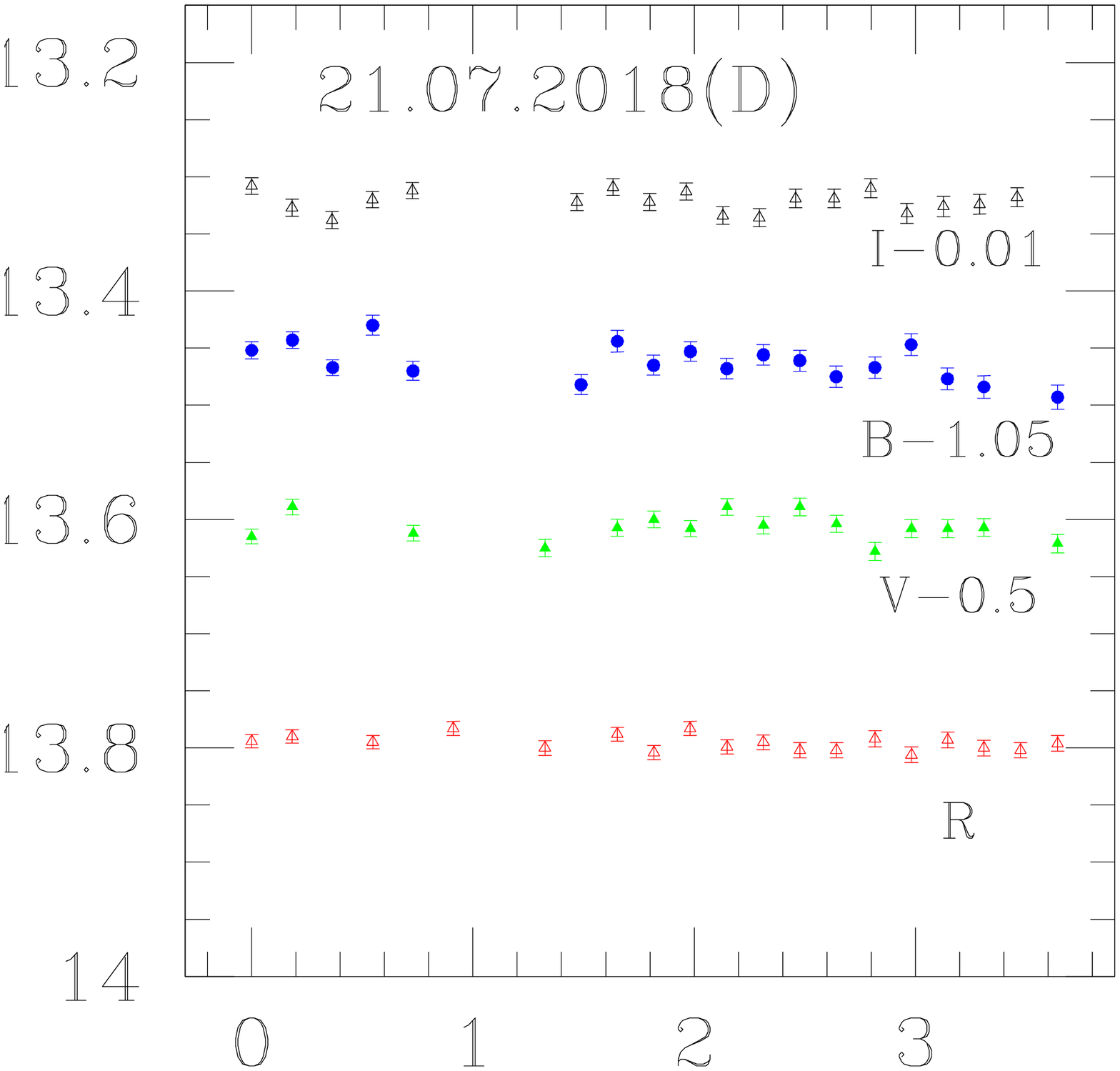,height=2.1in,width=2in,angle=0}
\epsfig{figure=  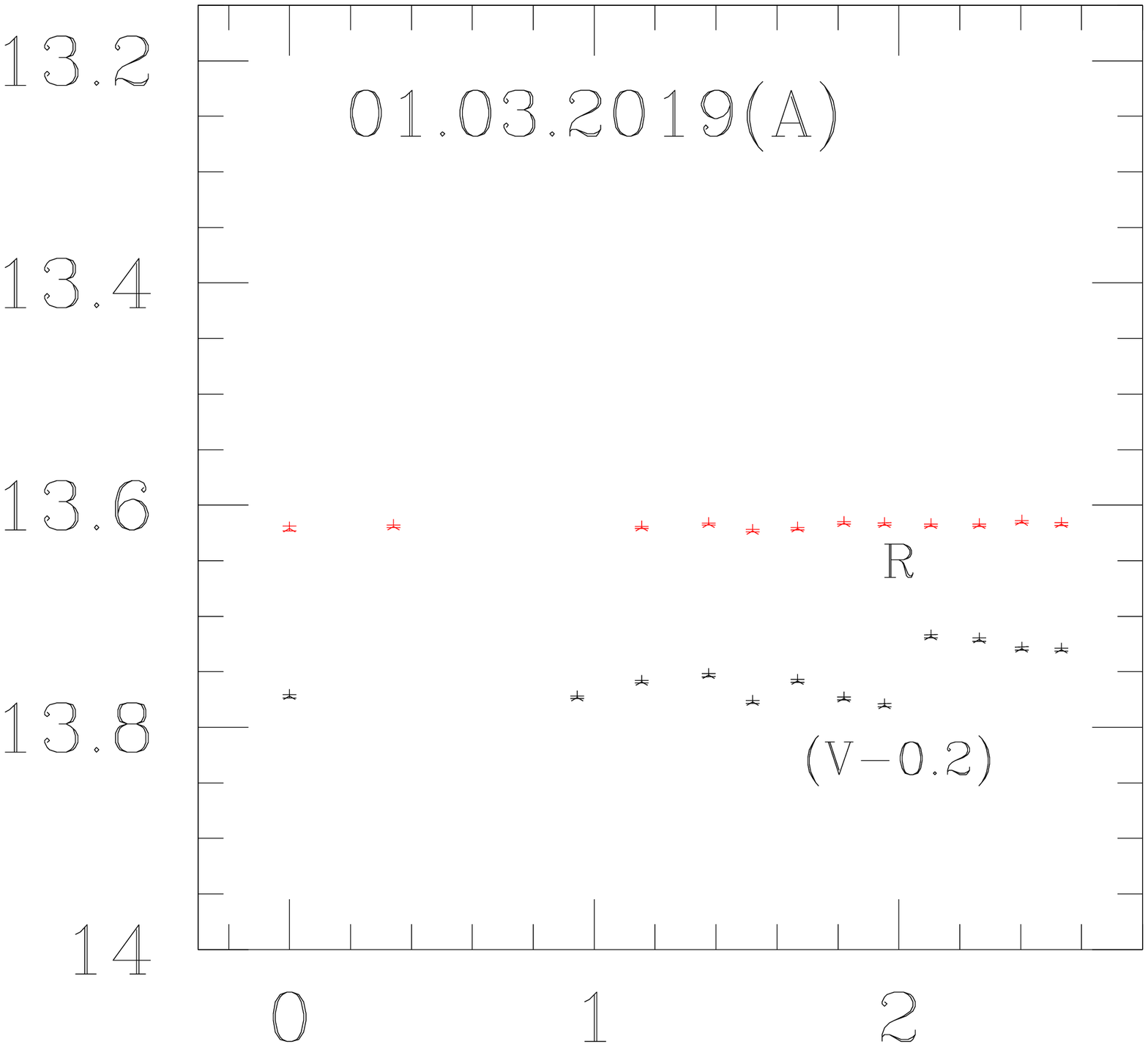,height=2.1in,width=2in,angle=0}
\hspace{0.55in}
\epsfig{figure=  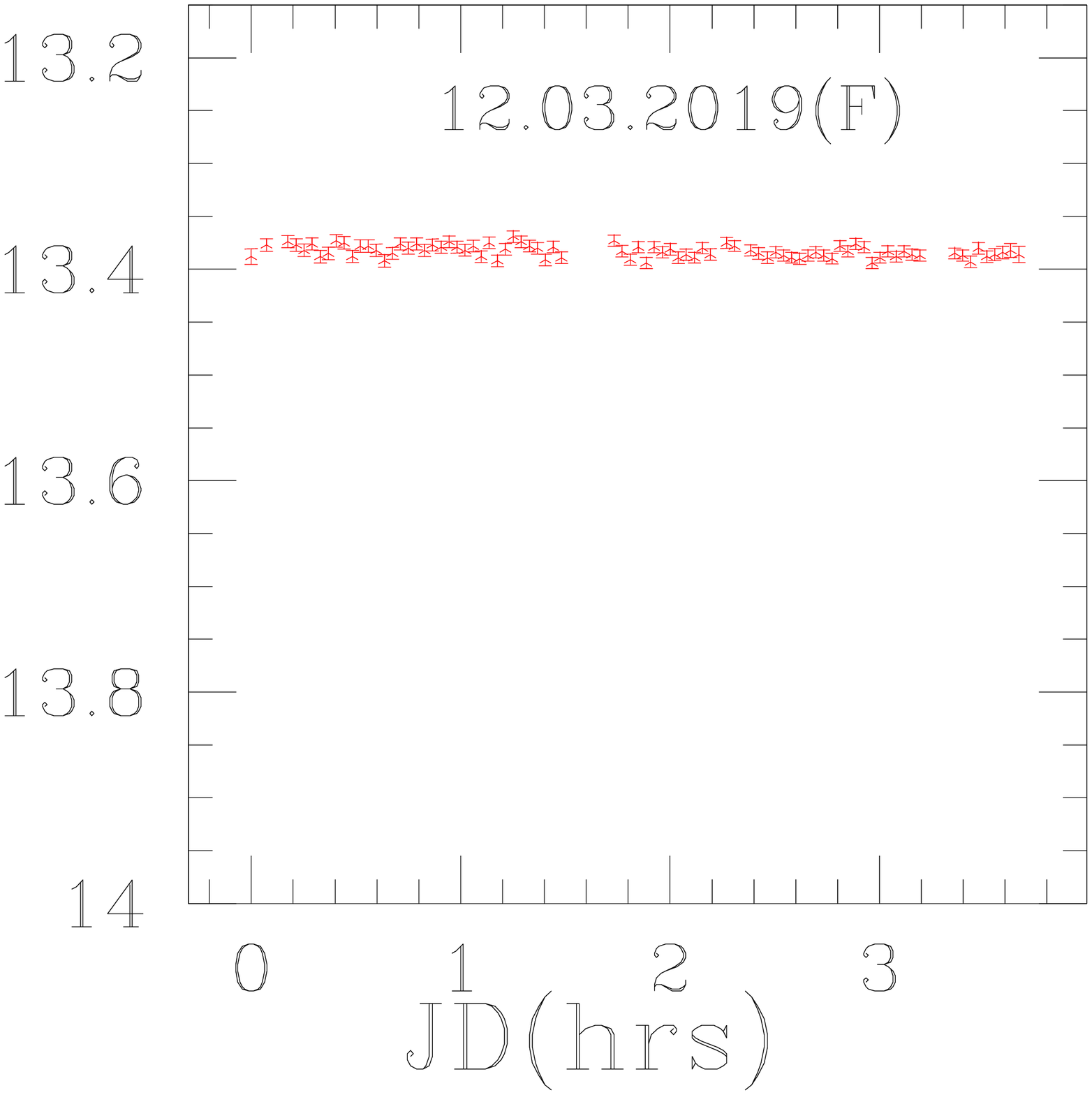,height=2.1in,width=2in,angle=0}
\hspace{0.55in}
\epsfig{figure=  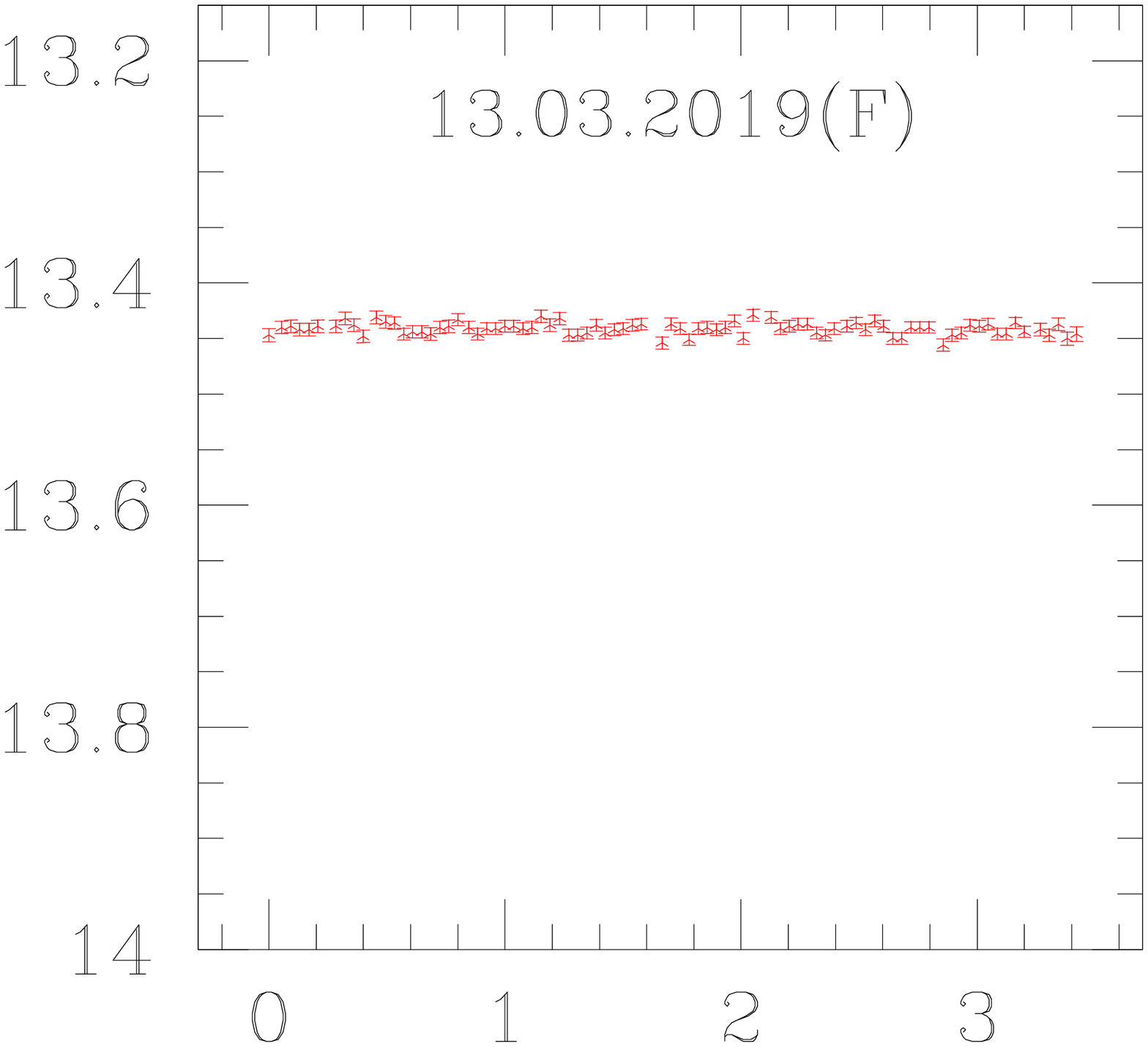,height=2.1in,width=2in,angle=0}
\caption{Intra-night LCs for \pg: blue denotes $B$-band, green~-- $V$-band, red~-- $R$-band, and black~-- $I$-band. In each plot the $x$-axis is the duration of the INM in hours.
Date of observation and the telescope used are indicated in each plot.}
\label{LC_BL1}
\end{figure*}

\addtocounter{figure}{-1}
\begin{figure*}[t]

\epsfig{figure=  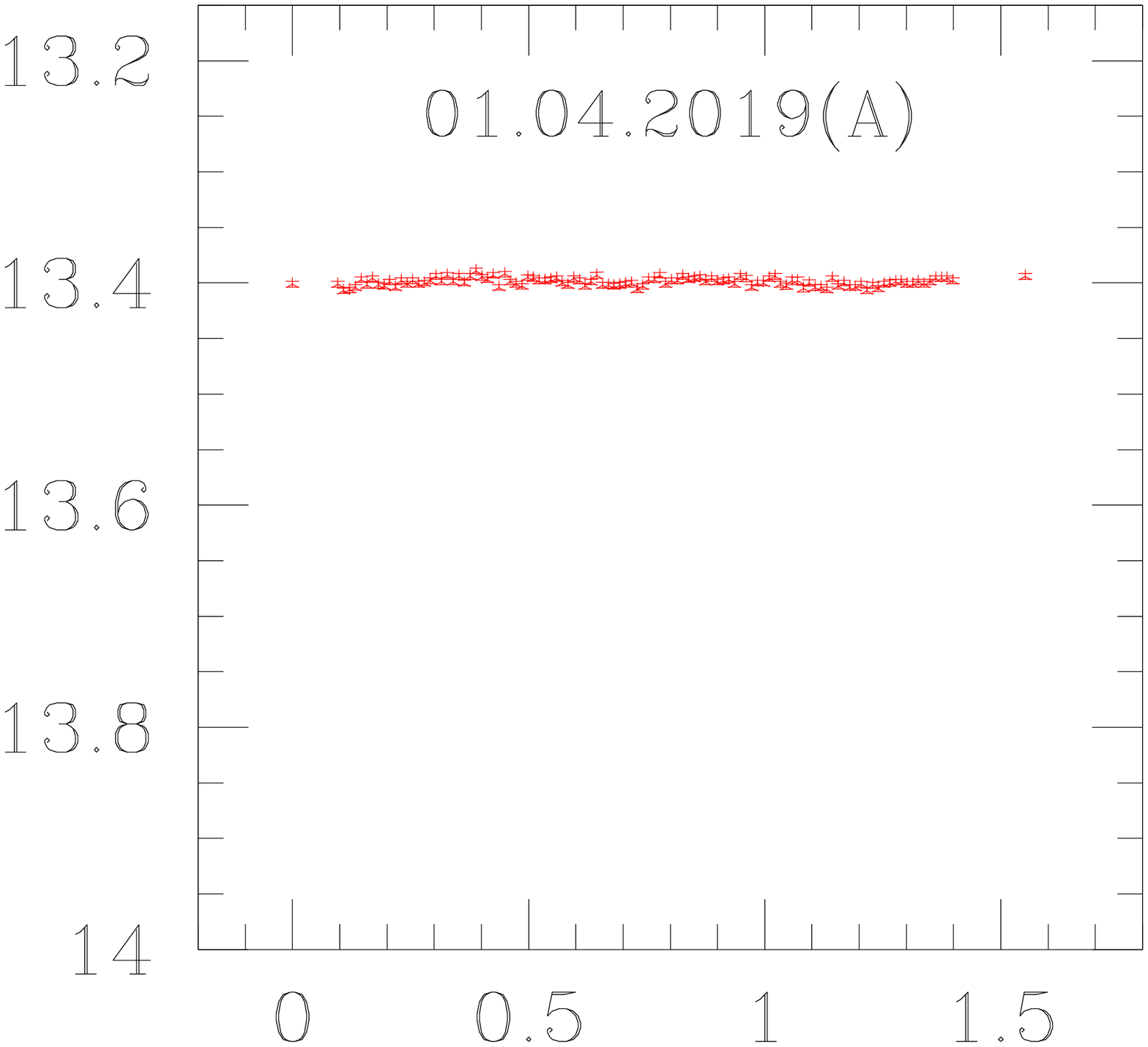,height=2.1in,width=2in,angle=0}
\epsfig{figure=  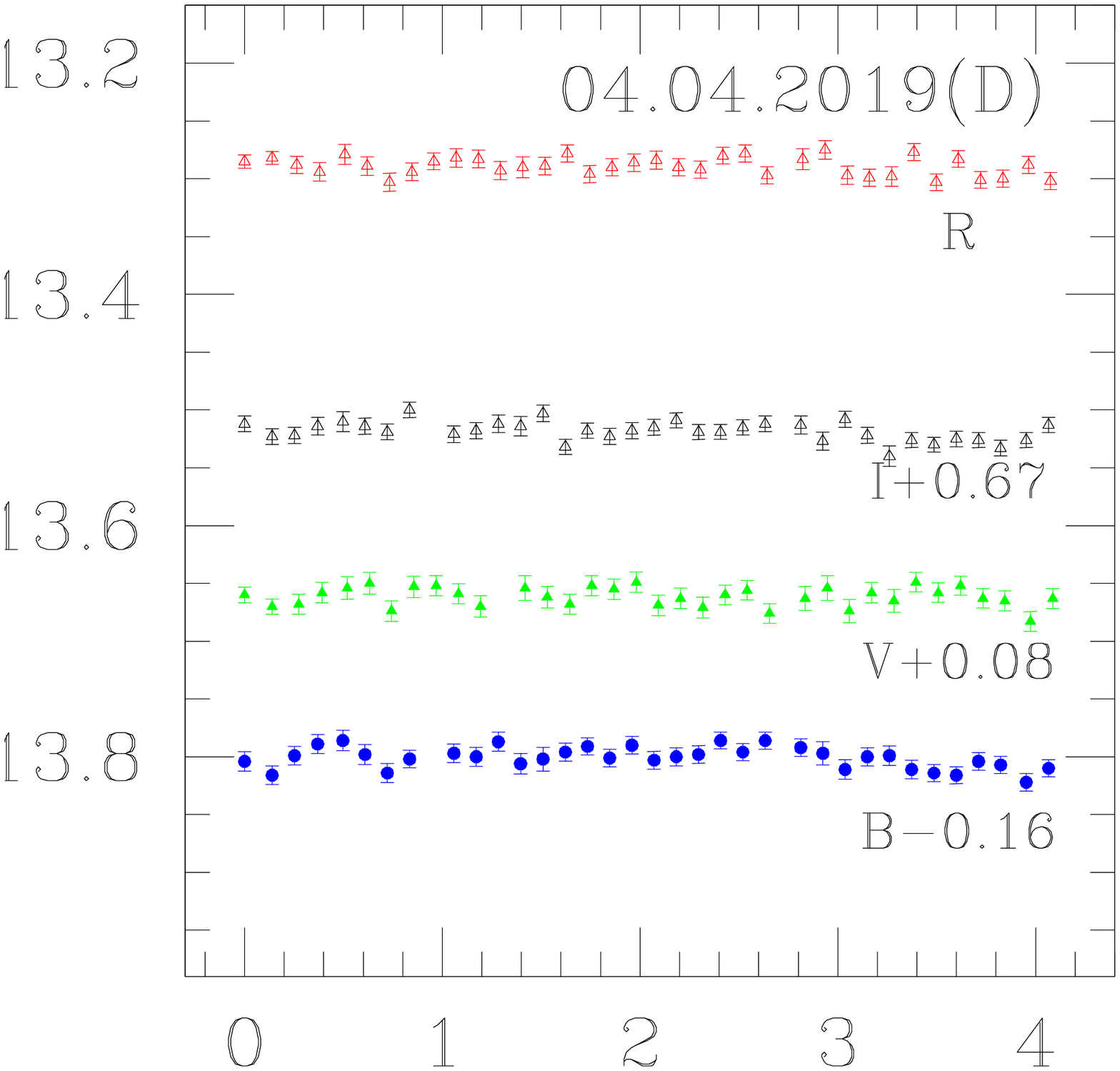,height=2.1in,width=2in,angle=0}
\epsfig{figure=  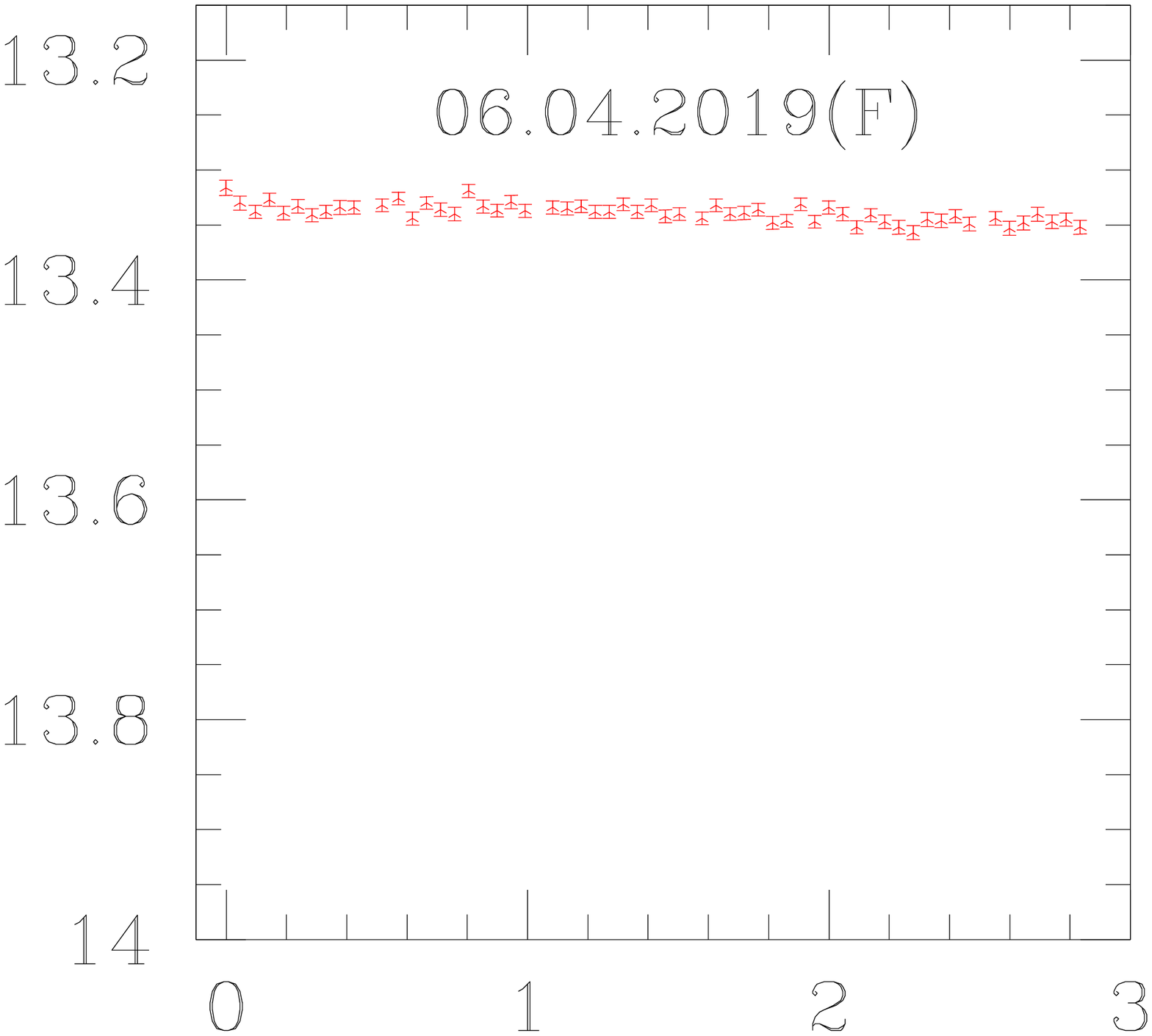,height=2.1in,width=2in,angle=0}
\epsfig{figure=  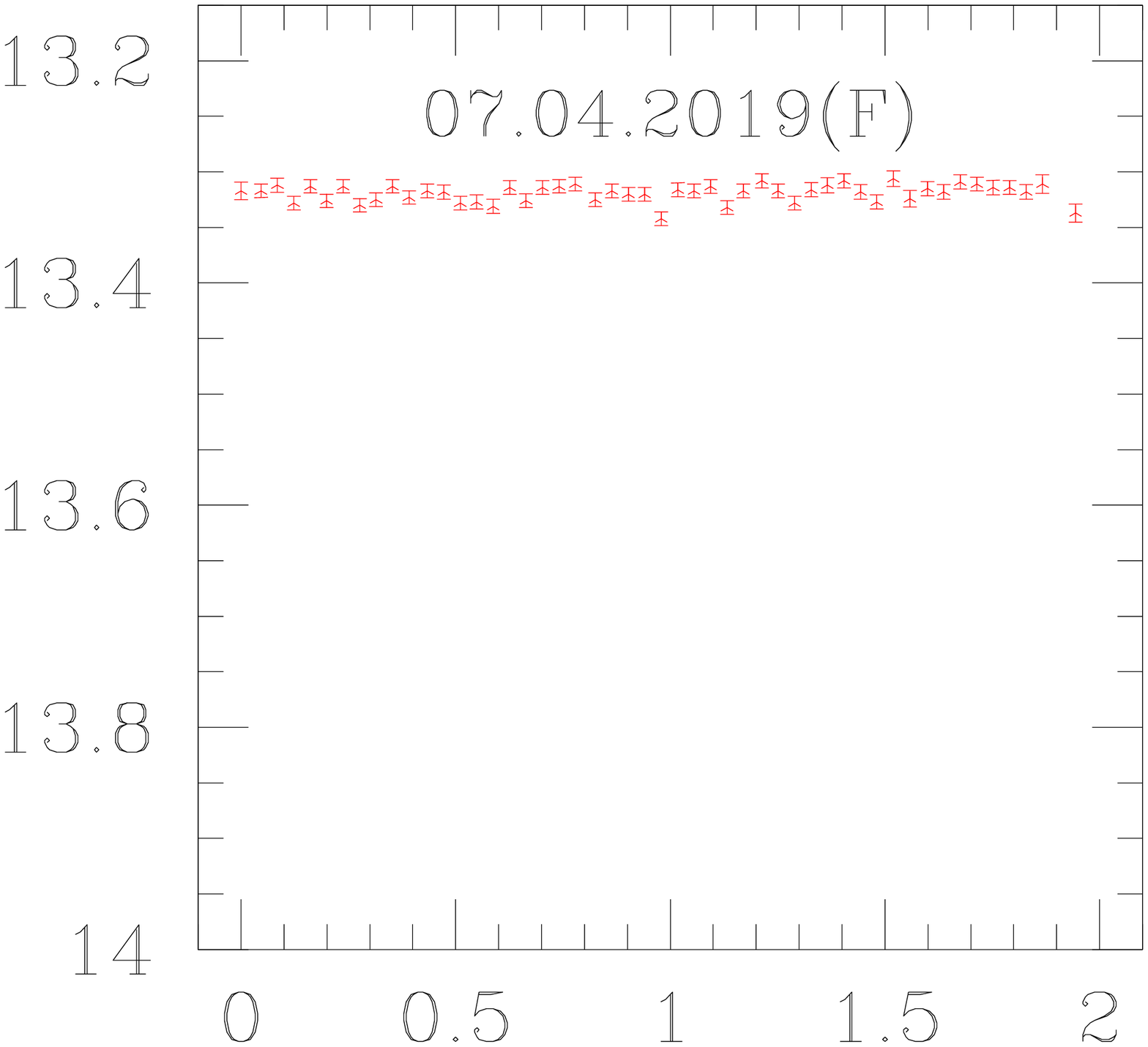,height=2.1in,width=2in,angle=0}
\epsfig{figure=  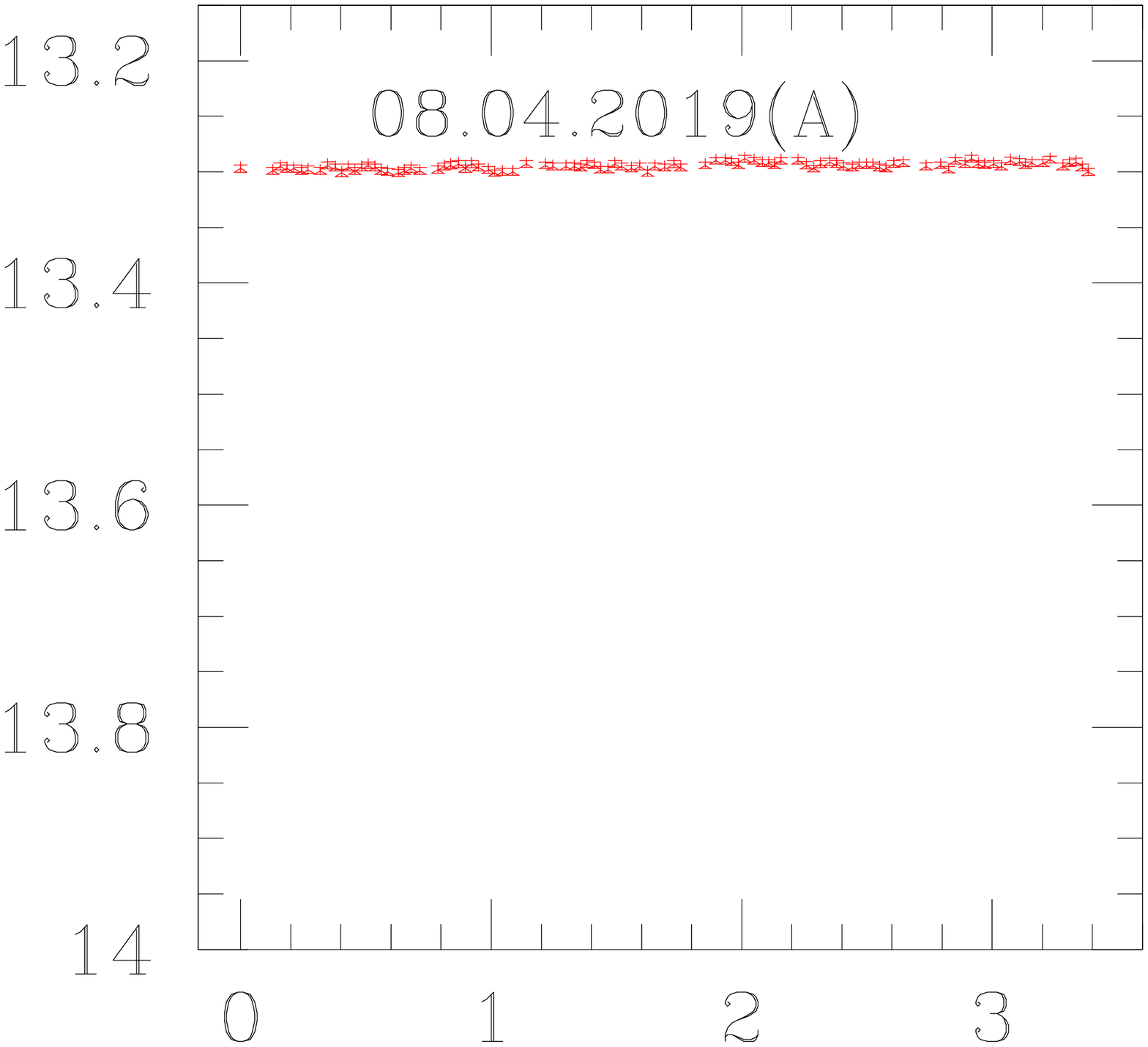,height=2.1in,width=2in,angle=0}
\epsfig{figure=  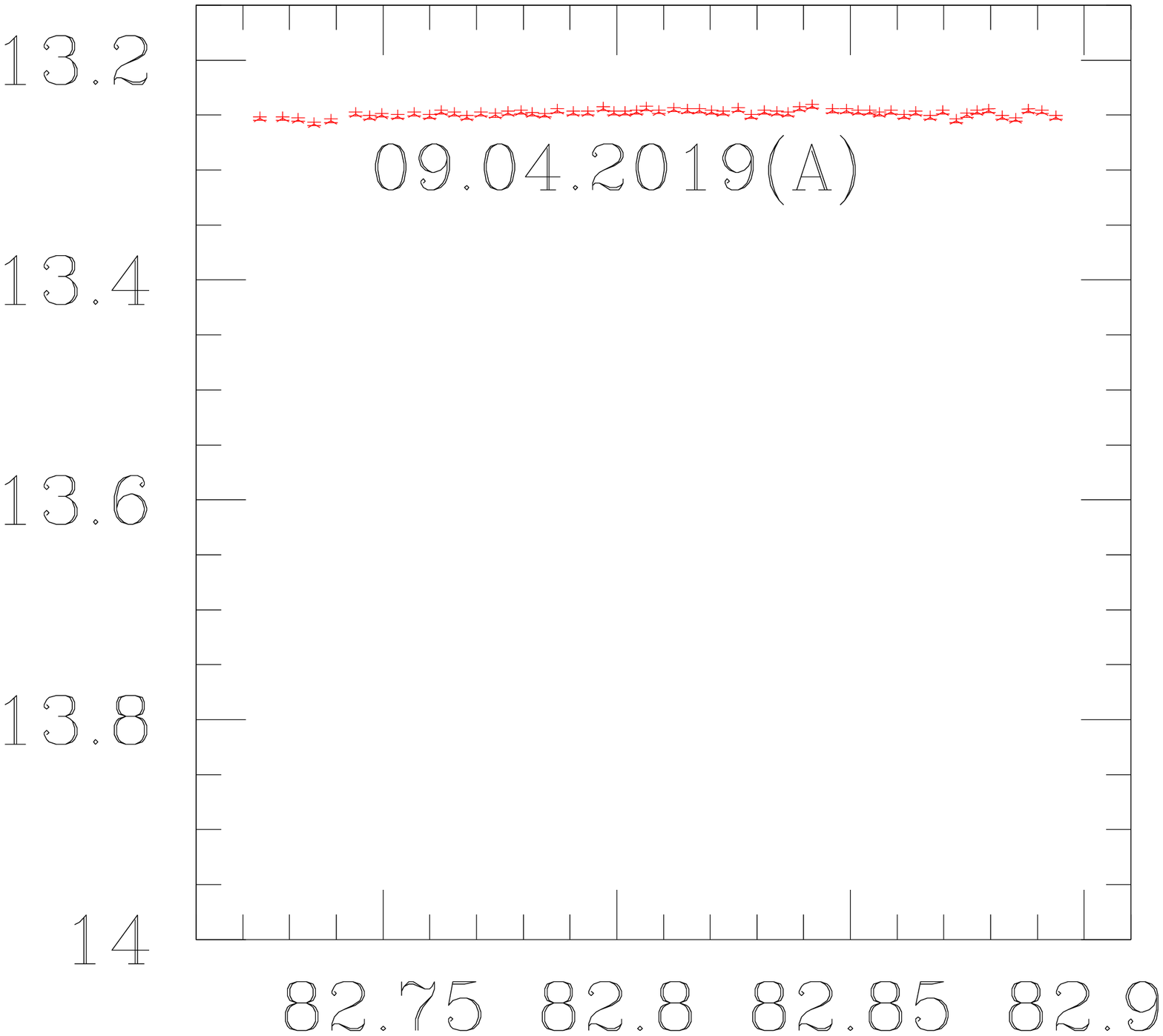,height=2.1in,width=2in,angle=0}
\epsfig{figure=  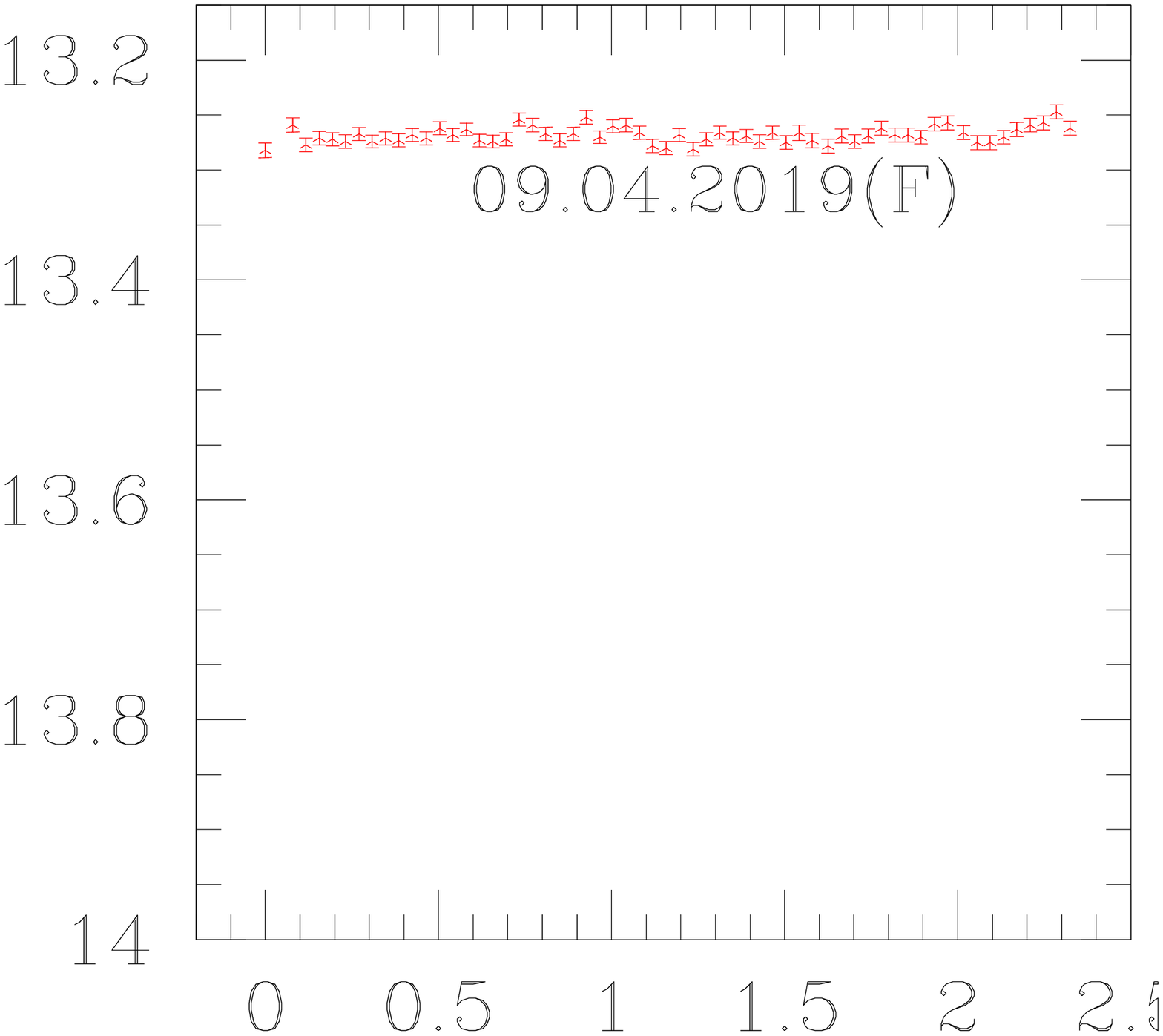,height=2.1in,width=2in,angle=0}
\epsfig{figure=  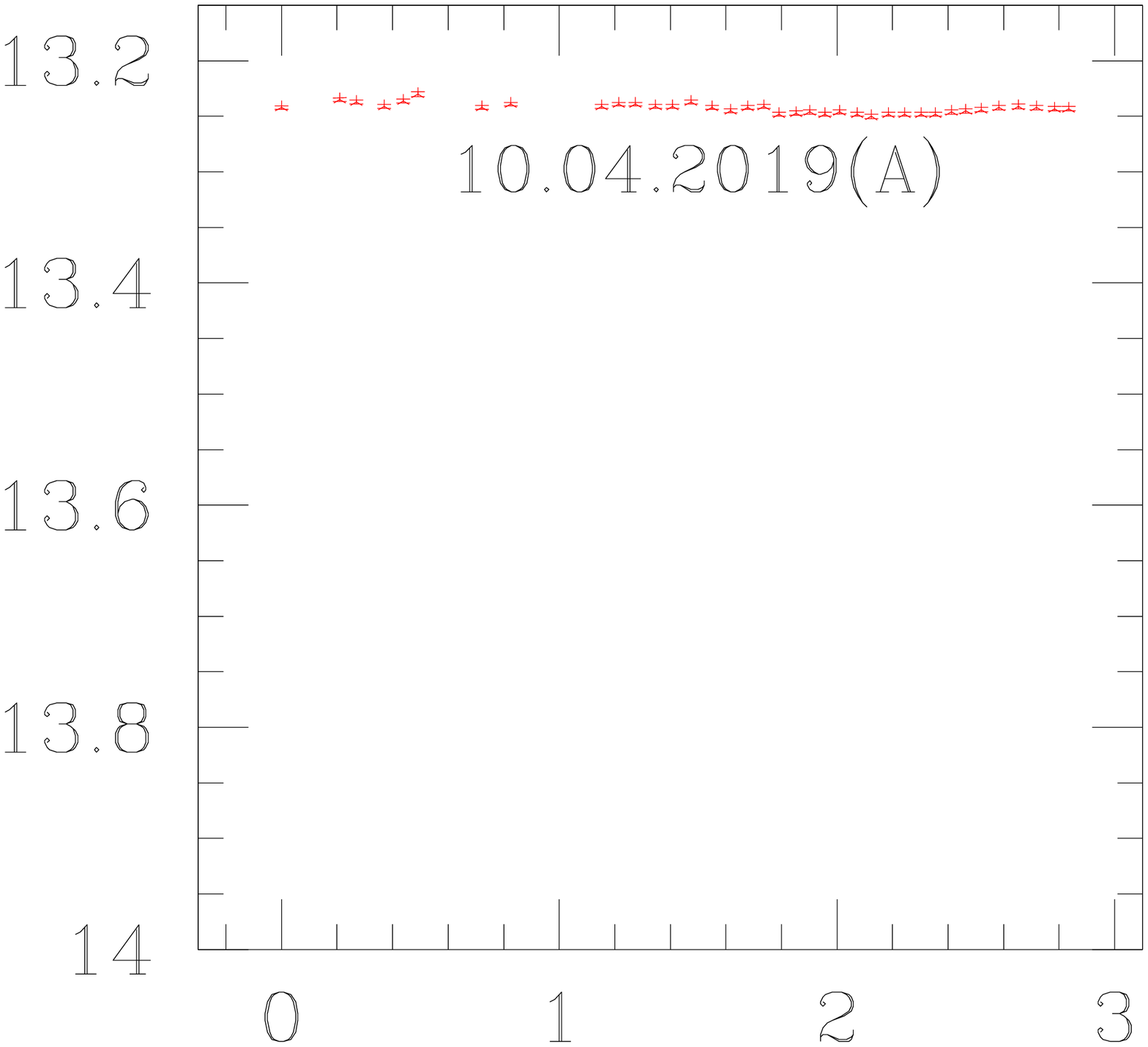,height=2.1in,width=2in,angle=0}
\epsfig{figure=  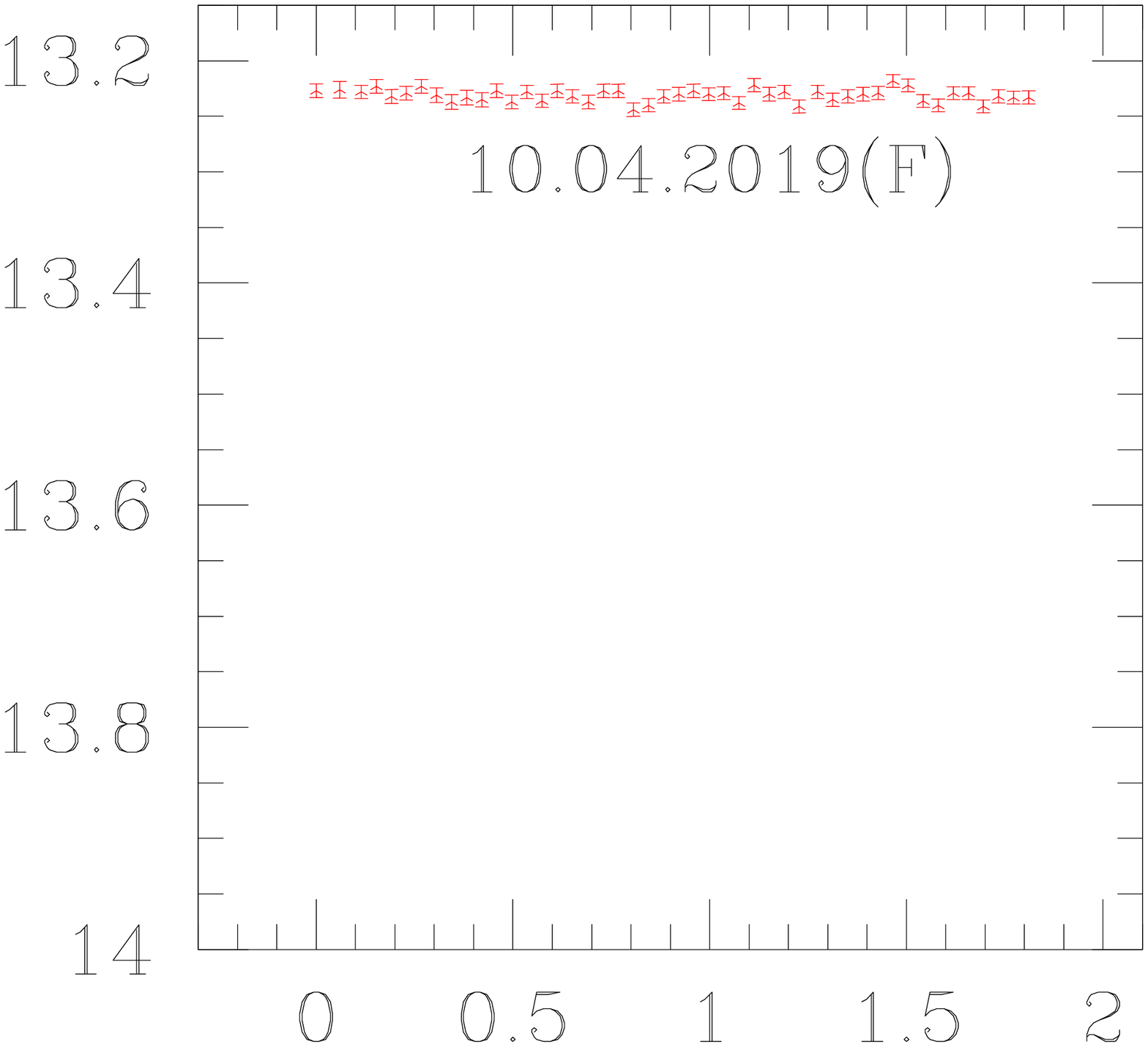,height=2.1in,width=2in,angle=0}
\epsfig{figure=  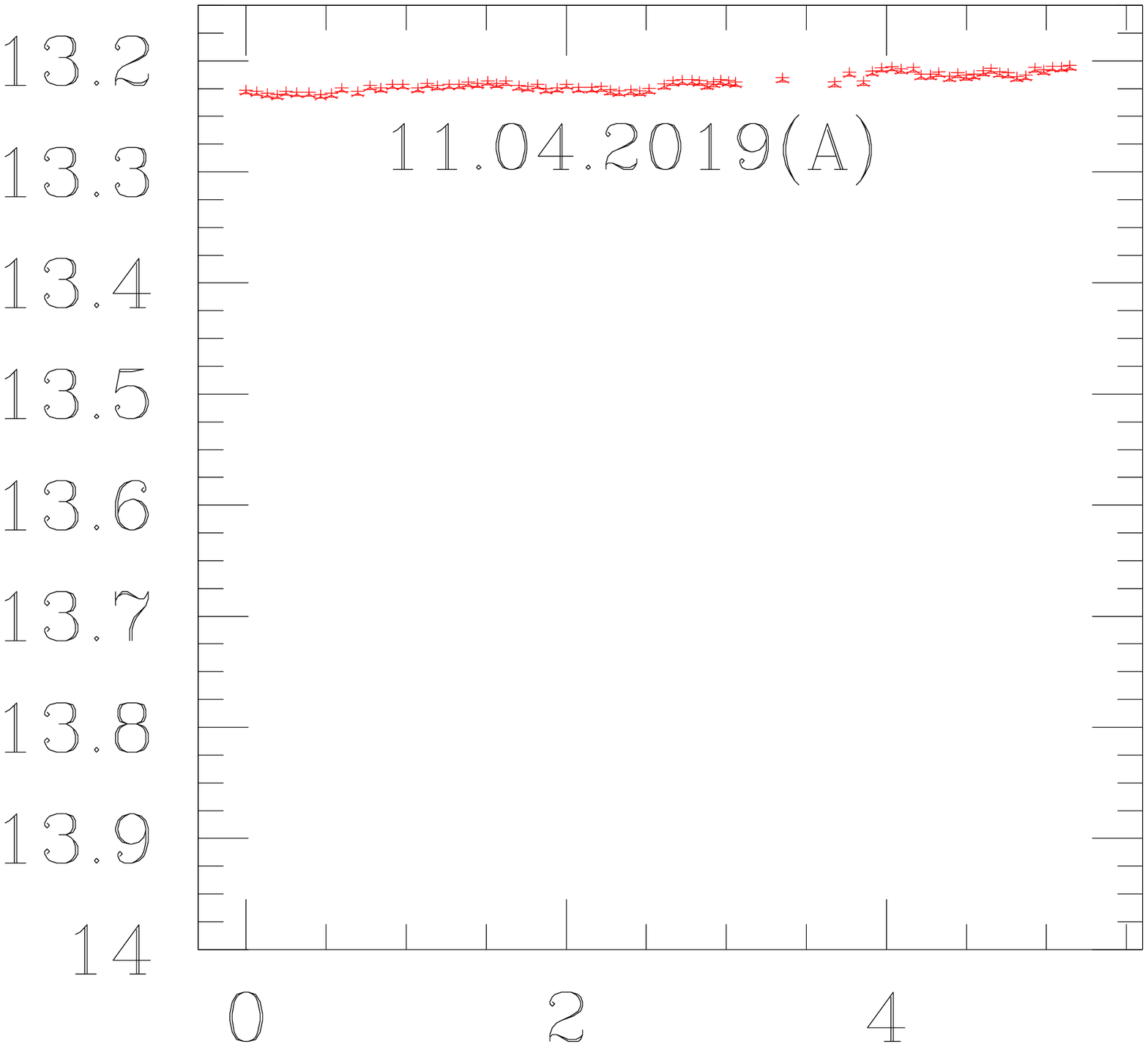,height=2.1in,width=2in,angle=0}
\hspace{0.55in}
\epsfig{figure=  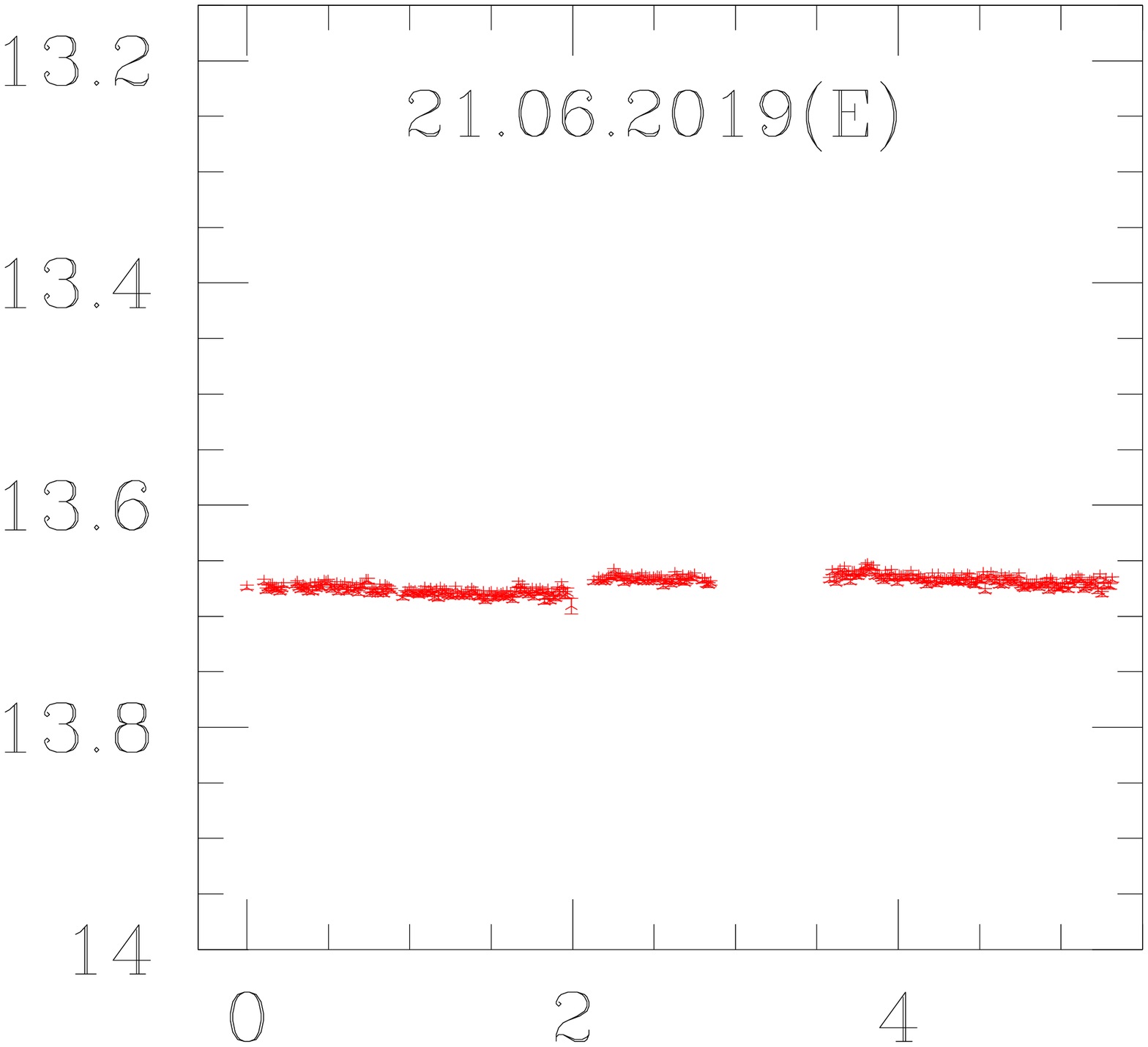,height=2.1in,width=2in,angle=0}
\hspace{0.55in}
\epsfig{figure=  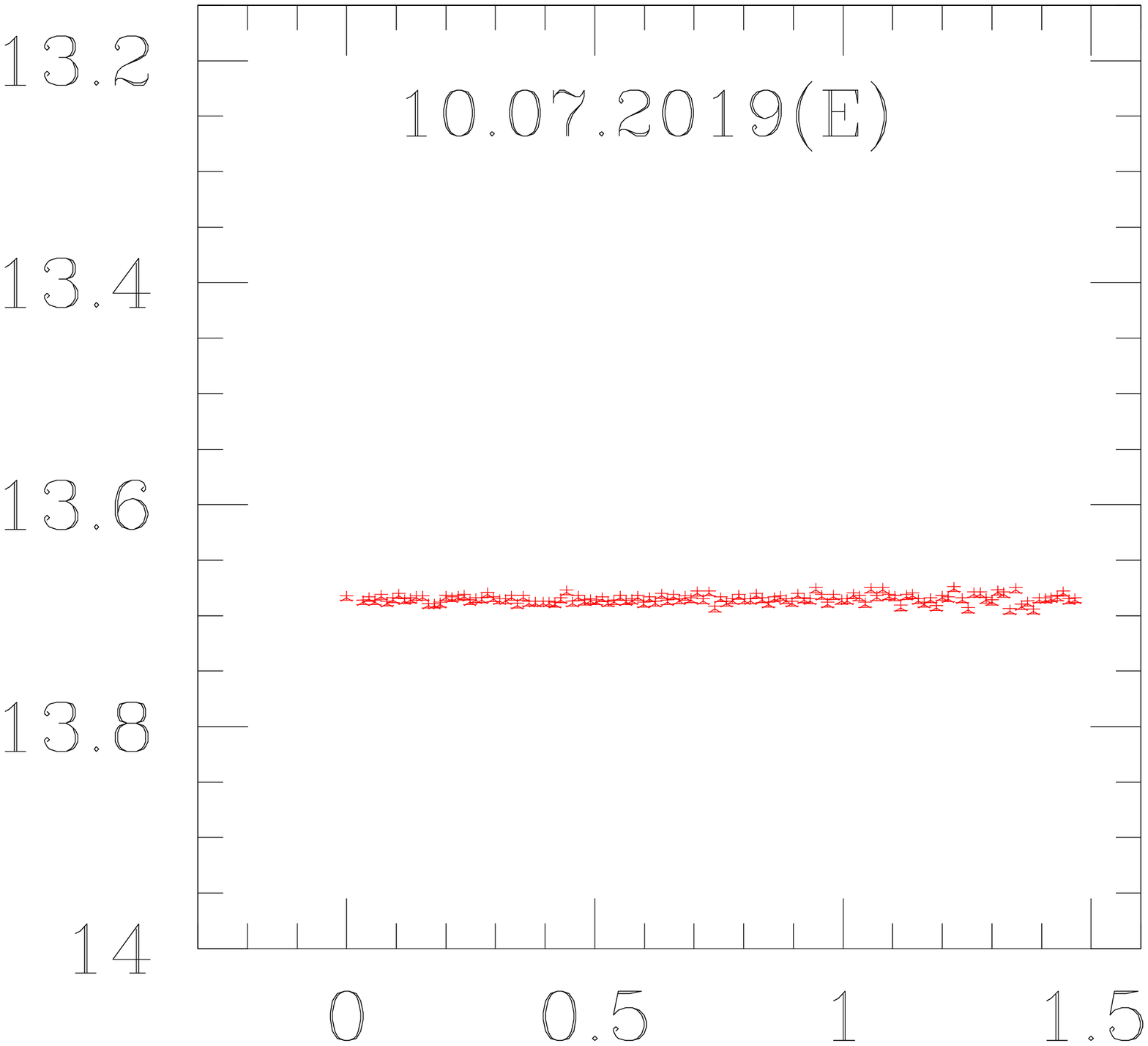,height=2.1in,width=2in,angle=0}

  \caption{Continued.}
\label{LC_BL2}
\end{figure*}

\addtocounter{figure}{-1}
\begin{figure*}[t]
\epsfig{figure=  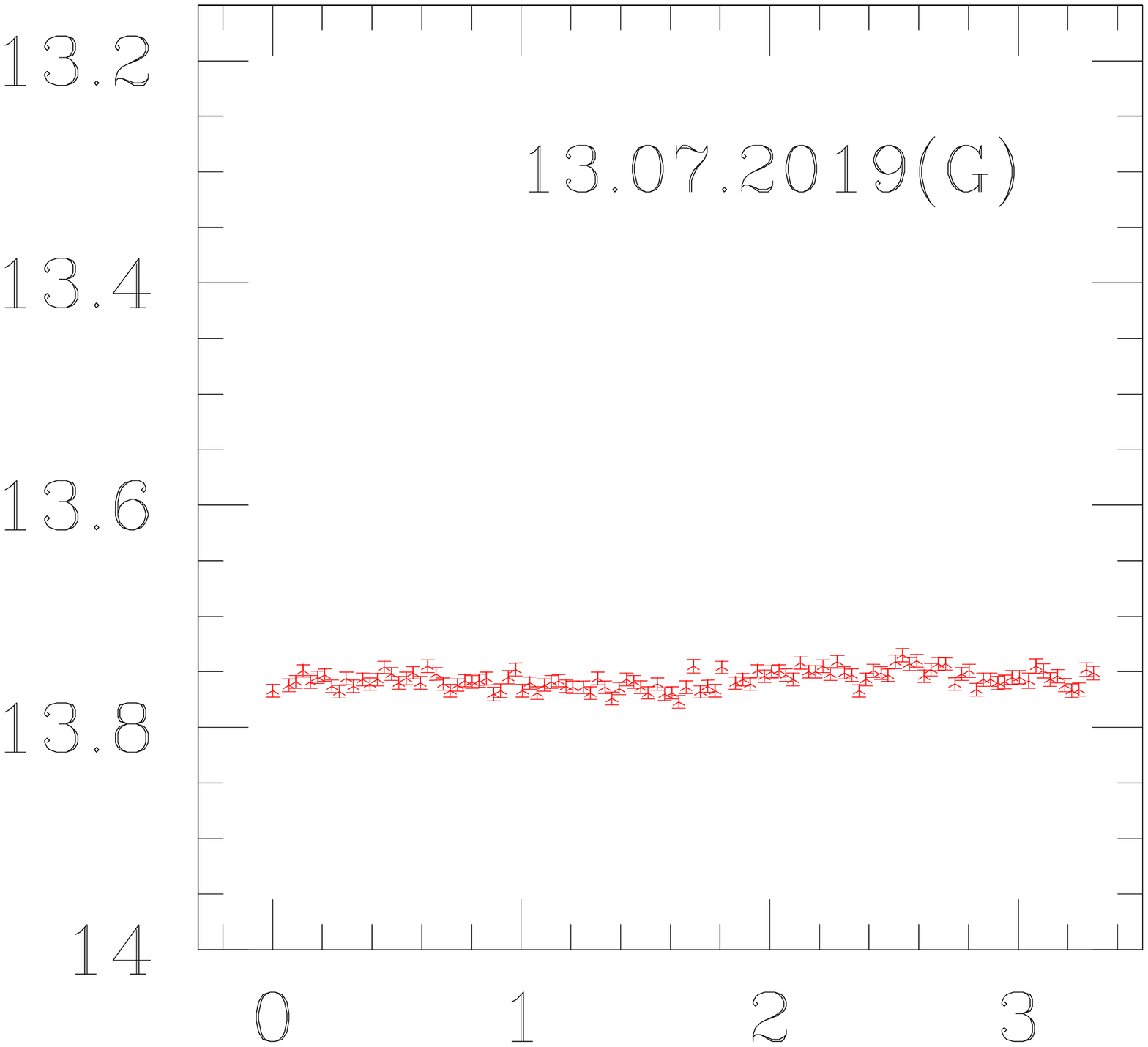,height=2.1in,width=2in,angle=0}
\epsfig{figure=  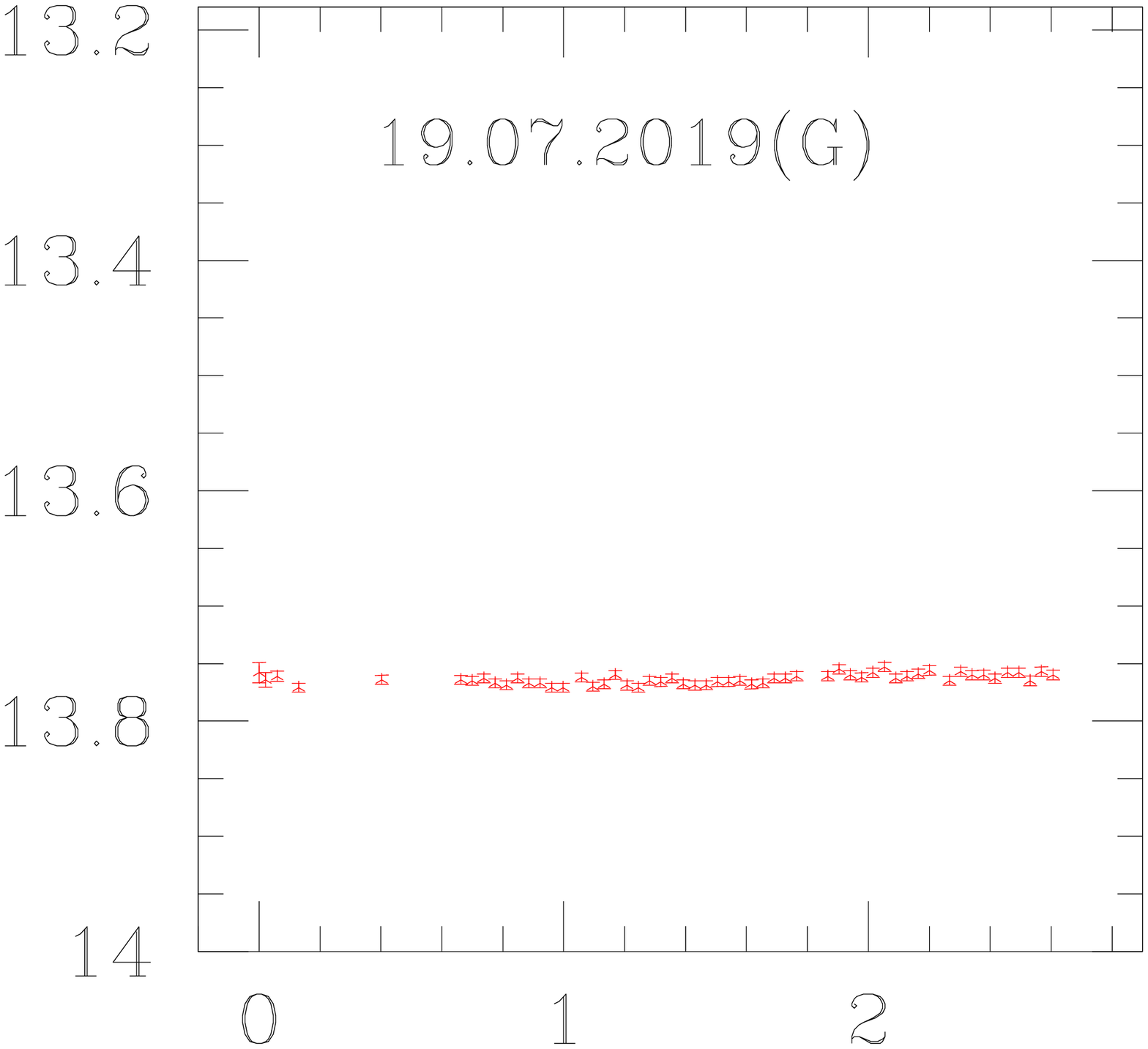,height=2.1in,width=2in,angle=0}
\epsfig{figure=  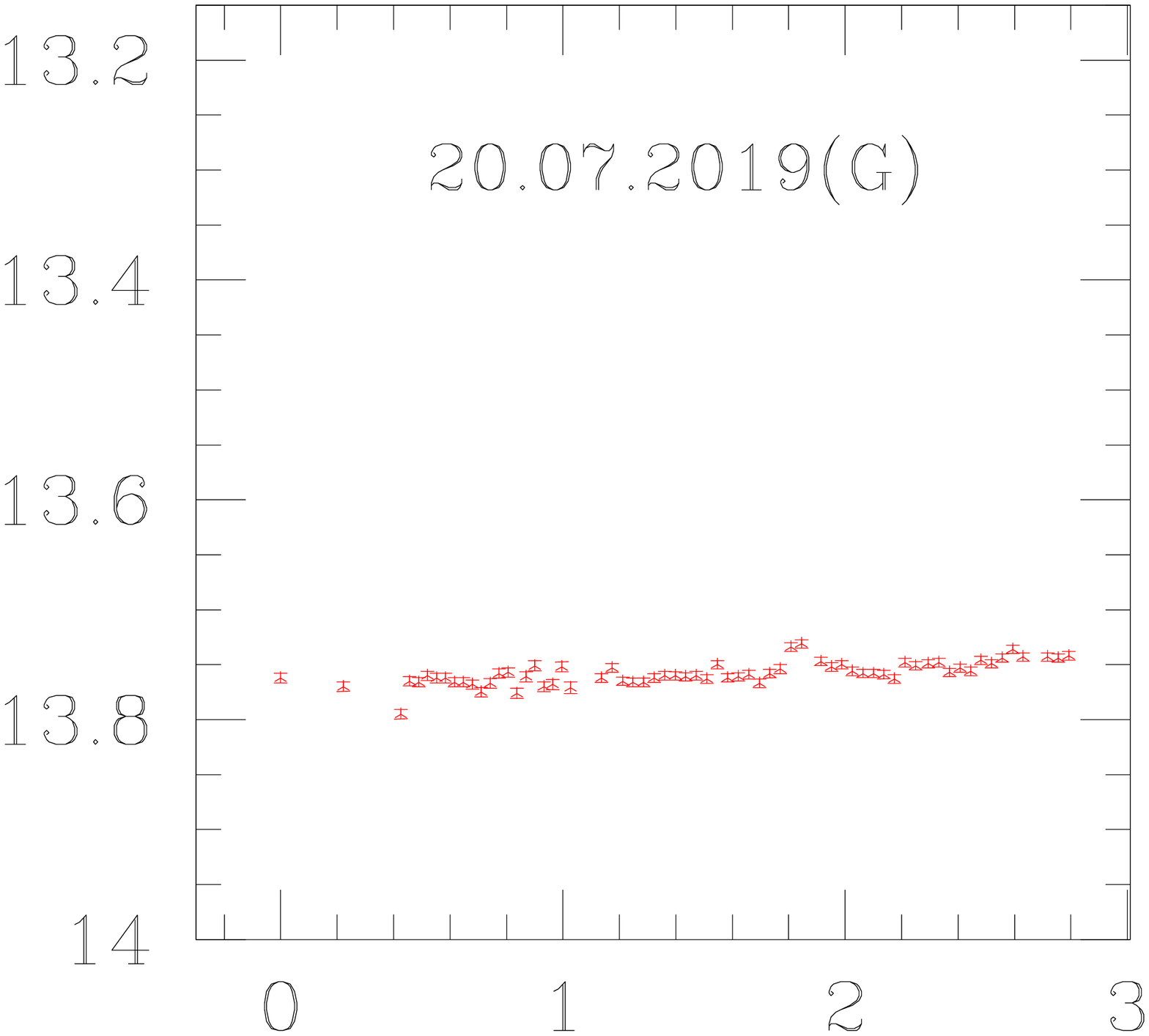,height=2.1in,width=2in,angle=0}
\hspace{0.55in}
\epsfig{figure=  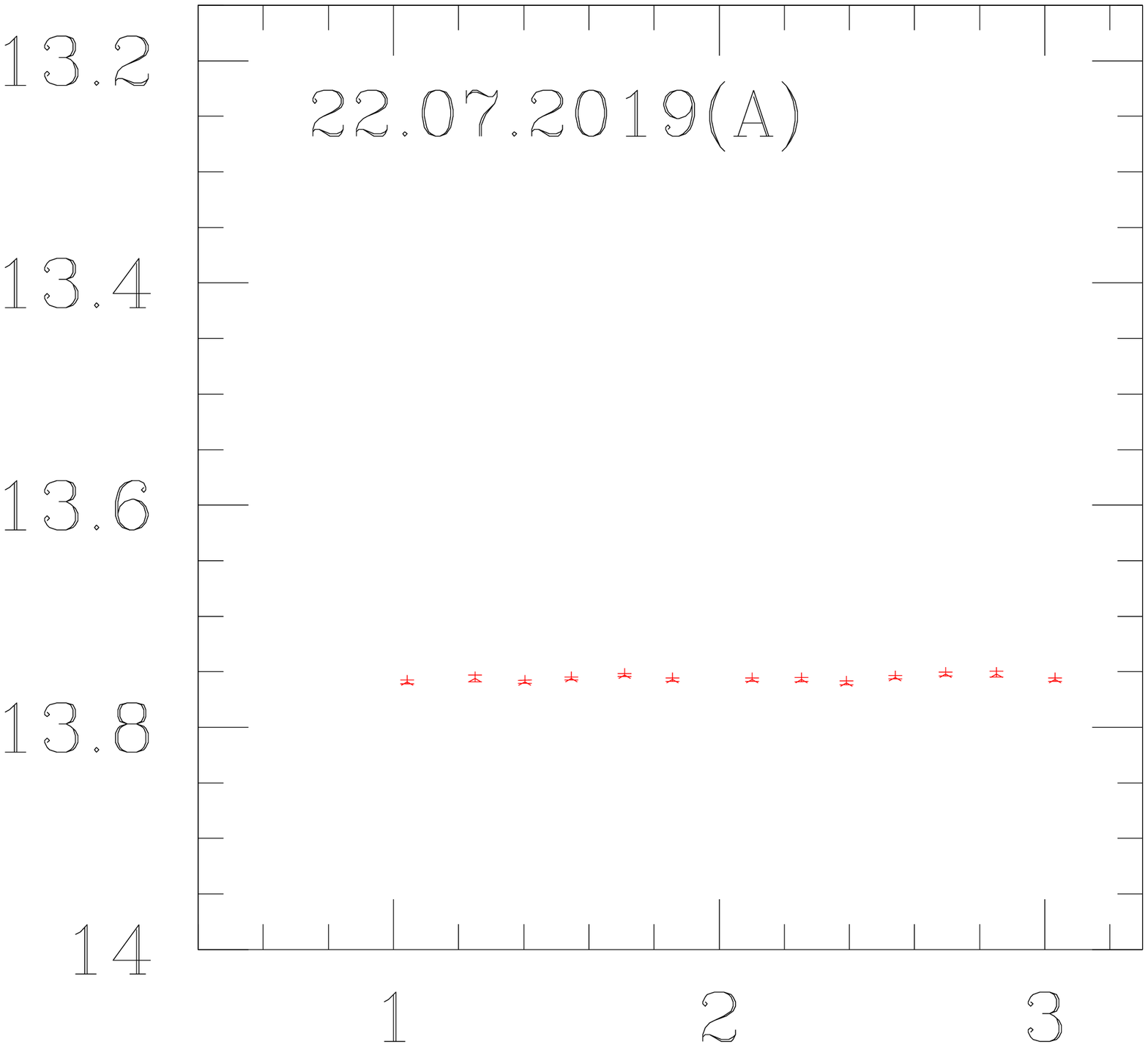,height=2.1in,width=2in,angle=0}

  \caption{Continued.}
\label{LC_BL3}
\end{figure*}

\clearpage

\section{Historical light curves}
\label{app:hist}

In order to study in detail the variability of \pg\ on long-term timescales, we constructed its historical LC. We used only major public data sets, so, this LC will not include all available photometric data on the blazar.

The first data set we considered was the one obtained at the Steward Observatory\footnote{http://james.as.arizona.edu/$\sim$psmith/Fermi/} of the University of Arizona. The observations were performed from 2012 to 2018 through $VR$ filters and were described in \citet{2009arXiv0912.3621S}. 
Comparing Steward photometry to ours, we found 7 $V$ and 9 $R$ data points that have been obtained either in one and the same night or in two consecutive nights~-- in all cases, the magnitude differences were less than the corresponding uncertainties. Therefore, there is no need the Steward photometry to be adjusted to ours and we merged the two data sets into a basic set to which the other archival data will be matched. After each successful matching the basic data set is updated accordingly.

The next major $VR$ data set comes from \citet{2016ApJ...833...77I}. The observations were performed from 2008 to 2014 using the Kanata telescope. This data set was adjusted and then added to the basic one. We have only one common observing night, so we were forced to use an interpolation onto the overlapping parts of the LCs to do the inter-calibration; this procedure was favoured by the fact that the source did not show significant variability during the period of interest.

The $V$-band LC was successively supplemented with the following data sets:
\begin{enumerate}
    \item Catalina Surveys \citep[][The Catalina Surveys Data Release 2]{2009ApJ...696..870D}. The observations have been obtained without a filter from 2005 to 2014 and the photometry has been transformed by the Catalina pipeline to $V$-band\footnote{http://nesssi.cacr.caltech.edu/DataRelease/FAQ2.html}. The inter-calibration with the updated basic data set was performed on the base of 33 observing nights in common.
    \item All-Sky Automated Survey for Supernovae\footnote{http://www.astronomy.ohio-state.edu/asassn/index.shtml} \citep[ASAS-SN,][]{2014ApJ...788...48S,2017PASP..129j4502K,2019MNRAS.485..961J}. We matched ASAS-SN photometry (taken from 2013 to 2018) to the updated basic data set using 21 nights in common.
    \item The Zwicky Transient Facility\footnote{https://www.ztf.caltech.edu/} \citep[ZTF,][]{2019PASP..131a8003M}. The data time span is from 2018 to 2019. Before adjustment to the updated basic data set, the ZTF $gr$ magnitudes were transformed to the Johnson-Cousins $VR$ ones. Then the matching was done using 12 nights in common.
\end{enumerate}

The following data sets were added to the $R$-band LC:
\begin{enumerate}
    \item Katzman Automatic Imaging Telescope\footnote{http://herculesii.astro.berkeley.edu/kait/agn/} \citep[KAIT,][]{2001ASPC..246..121F}. The KAIT unfiltered magnitudes roughly correspond to the $R$-band \citep[][we used the results since March 2019 from a new pipeline]{2003PASP..115..844L}. The matching was done on the base of 34 nights in common.
    We also added to the historical LC the KAIT results obtained with the old pipeline before March 2019 that are missing among the results obtained with the new pipeline. 
    The old photometry has been calibrated with respect to the `R2' magnitudes of USNO-B1 catalogue \citep{2003AJ....125..984M}. In this case the matching was done on the base of 15 nights in common. The KAIT photometry needs no adjustment before its joining to the basic data set. The time span of KAIT data is from 2010 to 2019.
    \item Tuorla Blazar Monitoring Program\footnote{http://users.utu.fi/kani/1m/} \citep{2008AIPC.1085..705T}. The observations were performed from 2005 to 2012 in the $R$-band \citep{2018A&A...620A.185N} and were adjusted to the updated basic data set on the base of 35 nights in common.
    \item Intermediate Palomar Transient Factory\footnote{https://www.ptf.caltech.edu/iptf} \citep[iPTF][]{2009PASP..121.1395L,2009PASP..121.1334R}. The observing time span is from 2009 to 2010 and the matching was done using four nights in common.
    \item ZTF (see above). The matching was done using 18 nights in common.
\end{enumerate}

\begin{figure}[t]
\centering
\includegraphics[width=\columnwidth,clip=true]{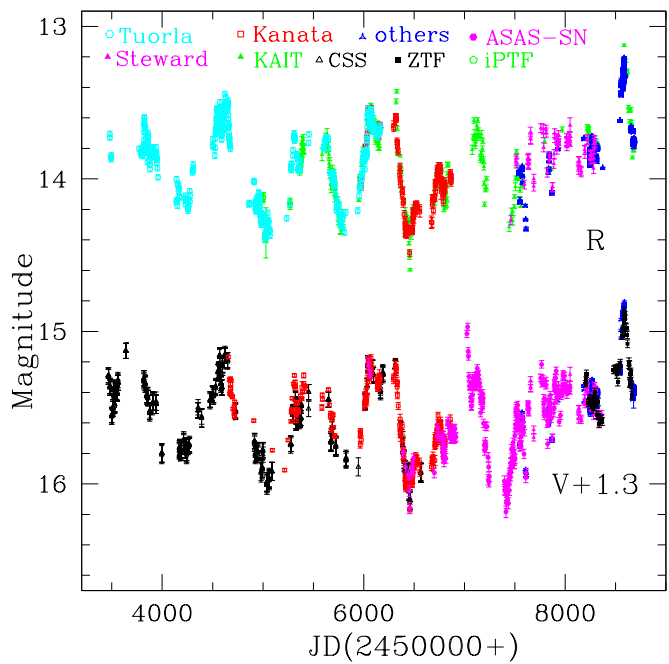}
\caption{Long-term $VR$-band LCs for \pg. Different colours and symbols denote data from different observatories and telescopes as indicated in the plot.
In particular, `others' denote the data collected by the nine telescopes mentioned in Sect.\,\ref{sect:obs}
The data plotted are not nightly binned.}
\label{fig:fullLC}
\end{figure}

After the different data sets were put together accounting for the offsets existing among them, nightly binning (and cleaning of the deviant data points) was performed as described in Sect.\,\ref{sect:res:ltv}. The so assembled historical $VR$ LCs for \pg\ are presented in Fig.\,\ref{fig:fullLC}. The variability amplitude of the historical $VR$-band LCs is $A\simeq1.4$\,mag.

\end{appendix}
\end{document}